\documentclass[twocolumn]{aastex631}

\usepackage{amsmath}
\usepackage{lipsum}
\usepackage{hyperref}
\usepackage{xcolor}
\usepackage{varwidth}
\usepackage{float}
\usepackage{array}

\newcommand\omicron{o}

\newcommand{\Mearth}{$\rm M_\oplus$}
\newcommand{\ms}{${\rm m\,s}^{-1}$}
\newcommand{\rvsearch}{{\texttt{RVSearch}}}
\newcommand{\secref}[1]{\S\ref{#1}}
\newcommand{\clf}{$\rm CL_{50}$}
\newcommand{\cln}{$\rm CL_{90}$}

\submitjournal{\aj}

\received{September 13, 2024}
\revised{September 5, 2025}
\accepted{September 23, 2025}

\shorttitle{SPORES-HWO. II. RV Limits on Companion Masses}
\shortauthors{Harada et al.}

\graphicspath{{./}{figures/}}

\begin{document}

\title{SPORES-HWO\footnote{SPORES-HWO: System Properties and Observational Reconnaissance for Exoplanet Studies with the Habitable Worlds Observatory}. II. Companion Mass Limits and Updated Planet Properties for 120 Future Exoplanet Imaging Targets from 35 Years of Precise Doppler Monitoring}

\author[0000-0001-5737-1687]{Caleb K.\ Harada}
\altaffiliation{NSF GRFP Fellow}
\affiliation{Department of Astronomy, 501 Campbell Hall \#3411, University of California, Berkeley, CA 94720, USA}

\author[0000-0001-8189-0233]{Courtney D.\ Dressing}
\affiliation{Department of Astronomy, 501 Campbell Hall \#3411, University of California, Berkeley, CA 94720, USA}

\author[0000-0002-1845-2617]{Emma V.\ Turtelboom}
\affiliation{McMaster University, Department of Physics \& Astronomy, 1280 Main St W, Hamilton, ON L8S 4L8, Canada}

\author[0000-0002-7084-0529]{Stephen R.\ Kane}
\affiliation{Department of Earth and Planetary Sciences, University of California, Riverside, CA 92521, USA}


\author[0000-0002-3199-2888]{Sarah Blunt}
\altaffiliation{NSF AAPF Fellow}
\affiliation{Department of Astronomy \& Astrophysics, University of California, Santa Cruz, CA, USA}
\affiliation{Center for Interdisciplinary Exploration and Research in Astrophysics (CIERA) and Department of Physics and Astronomy, Northwestern University, Evanston, IL 60208, USA}

\author[0000-0001-6320-7410]{Jamie Dietrich}
\affiliation{Arizona State University School of Earth and Space Exploration, 781 Terrace Mall, Tempe, AZ 85287}

\author[0000-0003-0595-5132]{Natalie R.\ Hinkel}
\affiliation{Louisiana State University, Department of Physics and Astronomy, 202 Nicholson Hall, Baton Rouge, LA 70803, USA}

\author[0000-0002-4860-7667]{Zhexing Li}
\affiliation{Department of Earth and Planetary Sciences, University of California, Riverside, CA 92521, USA}

\author[0000-0003-2008-1488]{Eric Mamajek}
\affiliation{Jet Propulsion Laboratory, California Institute of Technology, 4800 Oak Grove Drive, Pasadena, CA 91109, USA}

\author[0000-0002-7670-670X]{Malena Rice}
\affiliation{Department of Astronomy, Yale University, New Haven, CT 06520, USA}

\author[0000-0003-3989-5545]{Noah W.\ Tuchow}
\affiliation{Steward Observatory and Department of Astronomy, The University of Arizona, Tucson, AZ 85721, USA}
\affiliation{NASA Goddard Space Flight Center, Greenbelt, Maryland, USA}

\author[0000-0001-9957-9304]{Robert A.\ Wittenmyer}
\affiliation{University of Southern Queensland, Centre for Astrophysics, West Street, Toowoomba, QLD 4350 Australia}


\author{Christopher Chin}
\affiliation{Department of Astronomy, 501 Campbell Hall \#3411, University of California, Berkeley, CA 94720, USA}

\author{Aidan Fernandez}
\affiliation{University of Southern California, Los Angeles, CA 90089, USA}

\author{Shivani Kulkarni}
\affiliation{Department of Astronomy, 501 Campbell Hall \#3411, University of California, Berkeley, CA 94720, USA}

\author[0009-0006-5989-4899]{Emerald Lin}
\affiliation{Department of Astronomy, 501 Campbell Hall \#3411, University of California, Berkeley, CA 94720, USA}

\author{Nykole Liu}
\affiliation{Department of Astronomy, 501 Campbell Hall \#3411, University of California, Berkeley, CA 94720, USA}

\author{Remy Liu}
\affiliation{Department of Astronomy, 501 Campbell Hall \#3411, University of California, Berkeley, CA 94720, USA}

\author{Abhi Nathan}
\affiliation{Department of Astronomy, 501 Campbell Hall \#3411, University of California, Berkeley, CA 94720, USA}

\author{Adam Zbriger}
\affiliation{Department of Astronomy, 501 Campbell Hall \#3411, University of California, Berkeley, CA 94720, USA}

\correspondingauthor{Caleb K. Harada}
\email{charada@berkeley.edu}


\begin{abstract}
One goal of the future Habitable Worlds Observatory (HWO) is to directly image and spectroscopically characterize $\sim$25 true Earth analogs. However, if a large fraction of HWO target stars host unknown disruptive giant planets in their habitable zones (HZs), then additional targets that are farther away will need to be added to the survey, potentially requiring a larger-aperture telescope and a coronagraph with a smaller inner working angle (IWA). Therefore, the sooner we constrain the presence of massive planets orbiting potential HWO target stars, the easier and less costly it will be to adjust key aspects of HWO's architecture. In this work, we uniformly analyze over 153,000 public archival radial velocity (RV) observations of 120 potential HWO target stars to derive mass limits on planetary companions. The RVs were measured by 23 high-precision spectrographs located at 15 observatories around the world, with the first data going back to 1987. Based on empirical search completeness tests, we determine that undetected Jupiter-mass (Saturn-mass) planets may be hiding in up to 38\% (53\%) of the HZs of targets in the ExEP Mission Star List. The median mass sensitivity limit in the middle of the conservative HZ is approximately 48 \Mearth. We also provide updated orbital parameters for 53 known companions, and we detect 4 planet candidates and 26 signals corresponding to stellar activity. We note that 44 of the ExEP stars lack substantial RV monitoring history, and we advocate for community-coordinated observing campaigns of these stars using moderate-precision RV facilities.
\end{abstract}


\section{Introduction} \label{sec:intro}

To date, over 5,900 exoplanets have been discovered in at least 4,300 unique planetary systems\footnote{NASA Exoplanet Archive, accessed 20 August 2025; \url{https://exoplanetarchive.ipac.caltech.edu/cgi-bin/TblView/nph-tblView?app=ExoTbls&config=PS}}. Despite this growing number of planet discoveries, astronomers have yet to achieve a complete census of the exoplanets population. Nonetheless, the observed population of exoplanets clearly demonstrates that many planetary systems bear little resemblance to what we see in our own solar system \citep[e.g.,][]{Mayor+2011arXiv, Howard+2012ApJS, Wright+2012ApJ, Fressin+2013ApJ, Petigura+2013ApJ, Dressing+2015ApJ, Fulton+2017AJ, Petigura+2018AJ, Duck+2021AJ, Gan+2023AJ}. This staggering diversity has led to many outstanding questions concerning the uniqueness of our solar system in terms of its planetary properties, formation and evolutionary history, and habitability.

The current census of exoplanets has depended heavily on the workhorse transit and radial velocity (RV) detection methods, which are both biased toward large (massive) planets orbiting close in to their host stars. However, multi-decade RV searches have begun to reach the long observational baselines required to detect more distant giant planets, thus shedding light on the population of cooler outer planets \citep[e.g.,][]{Howard+2016PASP, Wittenmyer+2020MNRAS, Fulton+2021ApJS, 2021ApJS..255....8R, Rosenthal+2022ApJS, 2023AJ....165..176L, Fernandes+2019ApJ...874...81F}. With decades of observational baseline, we can start to piece together the full architectures of planetary systems and fill in our incomplete understanding of planetary demographics. These observations are also key to understanding the dynamical past, present, and future of planetary systems, and constraining whether additional planets may be present \citep[e.g.,][]{Barnes+2004ApJ, Kane+2015ApJ, Kane+2019AJ, Dietrich+2021AJ, Dietrich+2022AJ, Kane+2023AJ}.

Future observational facilities will continue to improve our sensitivity to ever smaller and longer-period planets. One of the fundamental goals driving these advancements is detecting and precisely measuring Earth-like planets orbiting in the habitable zones (HZs) of Sun-like stars, including those that may be habitable \citep{astro2020}. For example, new generations of extreme precision radial velocity (EPRV) instruments have already begun intensive searches for terrestrial-mass planets whose sub-meter-per-second signals were too small for previous generations of instruments \citep[e.g.,][]{Crass+2021arXiv, Newman+2023AJ, Gupta+2021AJ....161..130G, 2025AJ....169....1G, Brewer+2020AJ....160...67B}. Future detections and characterization of potentially Earth-like planets, when appropriately contextualized by a more complete census of exoplanets, better theories of planet formation, and robust knowledge of the planets' astrophysical environments, will revolutionize astronomy and the human condition. Such an undertaking will require synergy between demonstrated techniques that have formed our broader understanding of planetary astrophysics over last few decades and new innovations like EPRV.

However, while EPRV is an important pathway toward identifying potentially Earth-like exoplanets and measuring their masses, no single detection technique is yet capable of measuring planetary habitability. High-fidelity spectroscopic measurements of terrestrial planet atmospheres will be essential for robustly characterizing global biospheres and ruling out false-positive claims of inhabitance. Such observations will likely only be possible with future high-contrast direct imaging (DI) facilities capable of measuring the reflected-light spectra of Earth-size exoplanets, with complementary ground- and space-based observations providing the context clues necessary for interpretation.

The Habitable Worlds Observatory (HWO) is NASA's flagship mission concept for a large ultraviolet/optical/near-infrared (UV/O/NIR) space telescope with high-contrast ($\sim$10$^{-10}$) DI instrumentation. Slated for the 2040s, HWO will be capable of imaging and measuring the reflectance spectra of Earth-like exoplanets in the HZs of nearby bright stars. With its exquisite sensitivity, broad wavelength coverage, and starlight suppression technology, HWO will not only constrain the prevalence of potentially habitable and inhabited extrasolar worlds, but will enable revolutionary astrophysics in star formation, galactic evolution, cosmology, and the solar system (Dressing et al., \emph{in prep}). Key technologies and strategies for HWO are actively being developed \citep[e.g.,][]{Vaughan+2023MNRAS, Stark+2024JATIS, Stark+2024JATIS..10c4006S} and provisional target lists of bright nearby stars for the exo-Earth survey are being constructed \citep[][Tuchow et al. \emph{in prep}]{Mamajek+Stapelfeldt_2024, Harada+2024ApJS, Tuchow+2024AJ}. As the HWO mission concept and target lists continue to co-mature, a robust understanding of likely exo-Earth survey target stars is necessary to inform exoplanet yield calculations, which in turn enable reliable trade studies of potential HWO architecture options. Furthermore, precursor and preparatory science for the HWO target stars will be essential for accurately interpreting biosignature observations and placing worlds that HWO discovers in astrophysical context \citep[e.g.,][]{Batalha+2019ApJ...885L..25B, Damiano+2025AJ....169...97D}. 

Previously, in Paper I \citep{Harada+2024ApJS}, we outlined a precursor science project for HWO that aims to characterize the most promising targets for a direct-imaging search for exo-Earths. We compiled and analyzed archival data of 164 nearby FGKM stars from the ExEP Mission Star List\footnote{\url{https://exoplanetarchive.ipac.caltech.edu/docs/2645_NASA_ExEP_Target_List_HWO_Documentation_2023.pdf}} \citep{Mamajek+Stapelfeldt_2024}, then synthesized key target star information in the SPORES-HWO (System Properties and Observational Reconnaissance for Exoplanet Studies with HWO) Catalog. The catalog includes fundamental stellar properties, multi-band photometry, chemical abundances, optical flare rates, and photometric variability. In this work (Paper II), we provide a comprehensive review and analysis of precursor RV observations of potential HWO exo-Earth survey targets. Specifically, we combine and uniformly reanalyze all of the publicly-available precision RV measurements of the 164 SPORES-HWO stars from the past $\sim$35 years in order to: (1) search for new planet candidates, (2) refine orbits and planet parameters of known planets, (3) identify false positive signals, and (4) empirically measure the search completeness to derive meaningful limits on companions to potential HWO target stars. 

Previous work to quantify RV mass limits on companions for potential DI target stars has suggested that planets as large as Saturn or larger can be undetectable in many close by systems \citep[e.g.,][]{Howard+2016PASP, 2023AJ....165..176L}. Understanding the current RV landscape for potential HWO target stars is particularly important for prioritizing dynamically-viable target systems, as prior knowledge of any RV-detected massive planets in the HZ can rule out the long-term stability of any potential Earth-like worlds \citep[e.g.,][]{Kane+2024AJ....168..195K, Kane+2024AJ....168..279K}. Furthermore, sibling planets to exo-Earths revealed by RVs may themselves provide clues to habitability by way of their migration history and ability to scatter volatile-rich debris inward to the HZ from beyond the snow line \citep{Raymond+2008, Meadows+2018haex.book, Raymond+2004Icar, Kane+2024ApJ}. 

By searching for new planet candidates, refining orbits and planet properties of known planets, and measuring the RV search completeness, we can accurately simulate the types of planets HWO will detect, and place limits on their masses. For example, if planets above a certain mass can be ruled out with RV data obtained now, any planets imaged with HWO in the future must less massive and hence more likely to be habitable. Knowledge of the current RV sensitivity also informs the requirements for HWO's onboard astrometry, which will determine the dynamical masses of new planets discovered by the survey. Lastly, a precursor RV survey supports the broader exoplanet science themes of HWO as well as the general exoplanet community. In addition to exo-Earths, HWO will detect a statistical sample of cool Jovian and Neptune-like planets, and constraining the masses of these planets now prepares us for interpreting the atmospheric spectra that HWO will measure.

The rest of this paper is organized as follows. In \secref{sec:targets}, we provide an overview of the target sample considered in this study. In \secref{sec:data}, we summarize the RVs and stellar activity data we obtained for the target stars, in addition to the community science project that was created to handle the large amount of data. Then, in \secref{sec:analysis} we describe the analysis of the RVs and stellar activity data and the injection-recovery tests used to measure search completeness. In \secref{sec:results}, we present the RV search and completeness test results and highlight six systems with new signal detections. In \secref{sec:discussion}, we discuss how our results add to previous work, improve our sensitivity to planets both inside and outside the habitable zones of nearby stars, and set the stage for future observing efforts to better understand the planetary systems around potential HWO target stars. Finally, we summarize our main conclusions in \secref{sec:conclusion}.

\section{Target List} \label{sec:targets}

Several potential target lists for future space-based DI survey missions have recently been developed. For HWO, the NASA Exoplanet Exploration Program\footnote{\url{https://exoplanets.nasa.gov/exep/about/overview/}} (ExEP) recently published a provisional list of 164 target stars for the HWO exo-Earth survey \citep{Mamajek+Stapelfeldt_2024}. The ExEP Mission Star List (EMSL) was adapted and refined from the preliminary mission study target lists from the Large UV/Optical/IR Surveyor \citep[LUVOIR;][]{LUVOIR_2019} and Habitable Exoplanet Observatory \citep[HabEx;][]{HabEX_2020}. In brief, the EMSL contains 66 F-type stars, 55 G-type stars, 40 K-type stars, and 3 M-type stars. The majority of stars are on the main sequence, but a few subgiant stars are also included (4 stars have $\log g < 4.0$).

In addition to the EMSL, other much larger lists of potential HWO targets have been created. For example, the Habitable Worlds Observatory Preliminary Input Catalog (HPIC) contains $\sim$13,000 possible HWO targets \citep{Tuchow+2024AJ}. The HPIC is descended from the TESS and Gaia DR3 catalogs, and, in contrast to the EMSL, is more useful for yield calculations which require large input catalogs in order to simulate the science output of possible observatory architectures \citep{Stark+2024JATIS..10c4006S}. 

While both the EMSL and HPIC are useful in different aspects of precursor science for HWO, the stars in the shorter EMSL were selected to have the most easily-accessible HZs for a variety of HWO architectures. Therefore, these stars are considered a \emph{high-priority short list} of potential HWO targets\footnote{While the EMSL is an important starting point for HWO precursor science, we note that the scope of the list is limited by some assumptions regarding the notional observatory design \citep{Mamajek+Stapelfeldt_2024}. We caution that the EMSL and SPORES-HWO Catalog should not be interpreted as final target lists for the HWO exo-Earth survey, but rather as target lists for HWO precursor science studies.}. Previously, \citet{Harada+2024ApJS} synthesized information about these high-priority HWO targets in the SPORES-HWO Catalog\footnote{\url{https://sites.google.com/berkeley.edu/spores-hwo}}, which includes fundamental stellar properties, multi-band photometry, chemical abundances, optical variability, and high-energy emission. In this work, we continue to focus on the 164 EMSL/SPORES-HWO stars, as this sample represents the most likely stars to be observed HWO regardless of its final architecture.

\begin{figure*}[t!]
    \centering
    \includegraphics[width=0.99\textwidth]{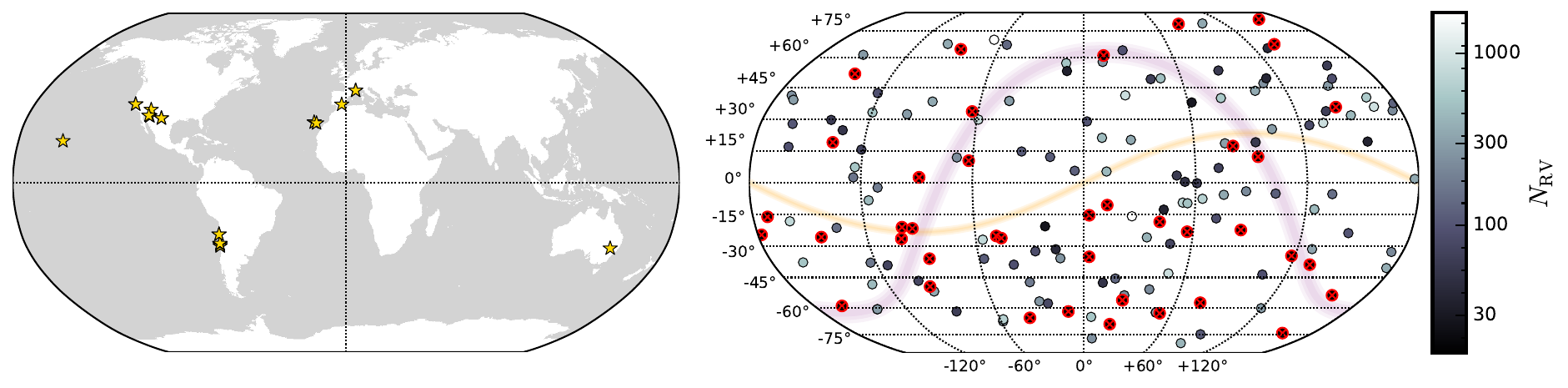}
    \caption{\textbf{Left:} Locations of the 20 telescopes at 15 different observatories used to measure the RV data analyzed in this work (see Table \ref{tab:observatories} in Appendix \ref{appx:figures_tables} for observatory details). \textbf{Right:} ICRS (J2000) coordinates of the target stars in this study with the number of public RV epochs indicated logarithmically by the colorbar. The 44 (of 164) targets with fewer than 20 RV epochs are indicated by red ``$\otimes$'' symbols---these targets were excluded from the subsequent RV analysis. The ecliptic and galactic planes are shown for reference in yellow and purple, respectively.}
    \label{fig:RV_coordinates}
\end{figure*}

\section{Data} \label{sec:data}

In this section we summarize the sources of the RV data and stellar activity indicators analyzed in this study and discuss how the data were collected. In brief, we acquired archival RV measurements and stellar activity indicators for all 164 SPORES-HWO stars, in all cases where such data were publicly accessible. For subsequent analysis, we required that each star have at least 20 unique nights of public RV observations. Of the 164 stars in the sample, we identified 120 stars (73\%) with at least 20 nights of public RV observations. An additional 21 stars have at least one RV observation, while the remaining 23 stars had no publicly-available RV data.

A total of 153,902 individual RV measurements (35,794 RVs binned to nightly intervals) were obtained from 40 literature sources. We also compiled 43,649 binned stellar activity indicator measurements, including $S$-index, $H$-index, $\rm R_{HK}^\prime$, $\rm H\alpha$ equivalent width, and the cross-correlation function FWHM and BIS. The data were originally collected by 23 different spectrographs located at 15 observatories across the globe. Considering all targets and all spectrographs, the observations span a total of over 36 years from June 1987 to December 2023, the product of the tremendous efforts of hundreds of observers over the entire history of the field of exoplanets. 

Figure \ref{fig:RV_coordinates} shows the positions of the targets on sky and the locations of the observatories used to collect the RVs analyzed in this study. A tabulated summary of the data from each spectrograph is given in Table \ref{tab:data_summary}, and a summary of the data collected for each target star is given in Table \ref{tab:targets} in Appendix \ref{appx:figures_tables}. The following subsections provide further details regarding the how these data were collected and processed for the purpose of this study. 

The final dataset for all target stars, including all unbinned RV and stellar activity indicator measurements, is published and available for download on Zenodo\footnote{\url{https://doi.org/10.5281/zenodo.17058228}}. In future work, we will also move the data to an online database where users can download data and upload new observations as they become available.

\begin{deluxetable*}{lllllll}
\tablecolumns{7}
\tablefontsize{\scriptsize}
\tablecaption{Summary of data}
\label{tab:data_summary}
\tablehead{
  \colhead{Instrument} & 
  \colhead{\# RVs} & 
  \colhead{\# SAIs} & 
  \colhead{First Obs.} & 
  \colhead{Last Obs.} & 
  \colhead{Span (yr)} & 
  \colhead{Reference(s)}
}
\startdata
AFOE & 65 & 0 & 1994-09-22 & 2001-09-07 & 7.0 & \begin{minipage}[t]{0.3\linewidth}\citet{2003AA...410.1051N,1999ApJ...526..916B}\end{minipage} \\
CARMENES & 316 & 0 & 2016-01-08 & 2020-01-10 & 4.0 & \begin{minipage}[t]{0.3\linewidth}\citet{2020AA...643A.112S,2023AA...671A..10S}\end{minipage} \\
CES-LC & 767 & 0 & 1992-11-03 & 1998-04-04 & 5.4 & \begin{minipage}[t]{0.3\linewidth}\citet{2013AA...552A..78Z,2001AA...374..675E}\end{minipage} \\
CES-VLC & 383 & 0 & 1999-11-21 & 2006-05-24 & 6.5 & \begin{minipage}[t]{0.3\linewidth}\citet{2013AA...552A..78Z}\end{minipage} \\
CHIRON & 181 & 0 & 2011-03-10 & 2013-05-25 & 2.2 & \begin{minipage}[t]{0.3\linewidth}\citet{2018AJ....155...24Z}\end{minipage} \\
CORALIE-07 & 98 & 48 & 2007-06-10 & 2014-07-17 & 7.1 & \begin{minipage}[t]{0.3\linewidth}\citet{2019AA...630A..50B,2023AA...674A.114B,2020AA...642A..31D,2013AA...551A..90M}\end{minipage} \\
CORALIE-14 & 250 & 0 & 2014-11-13 & 2021-09-09 & 6.8 & \begin{minipage}[t]{0.3\linewidth}\citet{2019AA...630A..50B,2023AA...674A.114B,2020AA...642A..31D}\end{minipage} \\
CORALIE-98 & 322 & 293 & 1998-07-27 & 2007-01-21 & 8.5 & \begin{minipage}[t]{0.3\linewidth}\citet{2019AA...630A..50B,2023AA...674A.114B,2001AA...375..205N,2004AA...415..391M,2020AA...642A..31D,2013AA...551A..90M}\end{minipage} \\
CTIO-ES & 286 & 0 & 2008-08-11 & 2010-05-18 & 1.8 & \begin{minipage}[t]{0.3\linewidth}\citet{2018AJ....155...24Z}\end{minipage} \\
ELODIE & 263 & 66 & 1994-04-24 & 2004-06-08 & 10.1 & \begin{minipage}[t]{0.3\linewidth}\citet{2003AA...410.1051N,2004PASP..116..693M,2004AA...414..351N}\end{minipage} \\
ESPRESSO & 108 & 148 & 2018-09-06 & 2023-03-25 & 4.5 & \begin{minipage}[t]{0.3\linewidth}\citet{2020AA...642A..31D,2025AA...693A.297N}\end{minipage} \\
EXPRES & 246 & 416 & 2018-04-28 & 2023-05-09 & 5.0 & \begin{minipage}[t]{0.3\linewidth}\citet{2022AJ....163..171Z,2022AJ....163...19R,2023AJ....166...46B,2021AJ....161...26C}\end{minipage} \\
HARPS-North & 23 & 91 & 2012-08-09 & 2014-03-29 & 1.6 & \begin{minipage}[t]{0.3\linewidth}\citet{2018AA...619A...1B,2015AA...584A..72M}\end{minipage} \\
HARPS-post & 1,804 & 6,963 & 2015-05-31 & 2021-12-29 & 6.6 & \begin{minipage}[t]{0.3\linewidth}\citet{2020AA...636A..74T}\end{minipage} \\
HARPS-pre & 5,348 & 20,128 & 2003-10-24 & 2015-05-16 & 11.6 & \begin{minipage}[t]{0.3\linewidth}\citet{2020AA...636A..74T,2019AA...632A..37B}\end{minipage} \\
HIRES-post & 5,444 & 9,422 & 2004-08-20 & 2020-03-10 & 15.6 & \begin{minipage}[t]{0.3\linewidth}\citet{2017AJ....153..208B,2021AJ....161..134H,2021ApJS..255....8R}\end{minipage} \\
HIRES-pre & 772 & 765 & 1996-07-11 & 2004-07-21 & 8.0 & \begin{minipage}[t]{0.3\linewidth}\citet{2017AJ....153..208B,2021ApJS..255....8R}\end{minipage} \\
HRS & 243 & 0 & 2003-10-15 & 2008-07-24 & 4.8 & \begin{minipage}[t]{0.3\linewidth}\citet{2009ApJS..182...97W,2010ApJ...715.1203M,2012ApJ...759...19E}\end{minipage} \\
Hamilton & 4,908 & 0 & 1987-06-11 & 2020-03-08 & 32.7 & \begin{minipage}[t]{0.3\linewidth}\citet{2014ApJS..210....5F,2021ApJS..255....8R}\end{minipage} \\
Levy & 7,514 & 147 & 2013-06-20 & 2021-04-27 & 7.9 & \begin{minipage}[t]{0.3\linewidth}\citet{2023AJ....165..176L,2021ApJS..255....8R,2021AJ....161..134H}\end{minipage} \\
NEID & 157 & 0 & 2021-01-03 & 2023-12-13 & 2.9 & \begin{minipage}[t]{0.3\linewidth}\citet{2025AJ....170...52G,2025AJ....169....1G}\end{minipage} \\
PFS-post & 113 & 108 & 2018-02-07 & 2021-05-26 & 3.3 & \begin{minipage}[t]{0.3\linewidth}\citet{2023AJ....165..176L}\end{minipage} \\
PFS-pre & 364 & 323 & 2011-01-26 & 2017-04-19 & 6.2 & \begin{minipage}[t]{0.3\linewidth}\citet{2023AJ....165..176L}\end{minipage} \\
SONG & 328 & 0 & 2015-04-03 & 2017-03-06 & 1.9 & \begin{minipage}[t]{0.3\linewidth}\citet{2023AA...671A..10S}\end{minipage} \\
SOPHIE & 2,358 & 2,790 & 2006-08-30 & 2023-05-31 & 16.8 & \begin{minipage}[t]{0.3\linewidth}\citet{2004PASP..116..693M,2019AA...625A..17D,2018AA...619A...1B}\end{minipage} \\
Tull & 336 & 126 & 1998-07-15 & 2014-02-08 & 15.6 & \begin{minipage}[t]{0.3\linewidth}\citet{2009ApJS..182...97W,2006AJ....132.2206B,2007ApJ...654..625W,2016ApJ...818...34E,2012ApJ...759...19E}\end{minipage} \\
UCLES & 2,794 & 1,815 & 1998-01-16 & 2015-11-22 & 17.8 & \begin{minipage}[t]{0.3\linewidth}\citet{2023AJ....165..176L,2006ApJ...646..505B}\end{minipage} \\
UVES & 3 & 0 & 2004-09-24 & 2017-08-28 & 12.9 & \begin{minipage}[t]{0.3\linewidth}\citet{2019MNRAS.490.5002F}\end{minipage} \\
\hline \\
\textbf{Total} & 35,794 & 43,649 & 1987-06-11 & 2023-12-13 & 36.5 & \\
\enddata
\tablecomments{Data have been binned to nightly intervals.}
\end{deluxetable*}

\subsection{Data Sources}

We compiled RV observations from a wide variety of sources in order to maximize the completeness of our search. We include observations beginning from the earliest generation of \'{e}chelle spectrographs that enabled some of the first detections of exoplanets around main-sequence stars (e.g., ELODIE on the 1.93-m Telescope at the Observatoire de Haute-Provence and Hamilton on the 3-m Shane Telescope at Lick Observatory), in addition to next-generation instruments with improved RV precision (e.g., HARPS on the ESO 3.6-m Telescope at La Silla Observatory and PFS on the 6.5-m Magellan II Telescope at Las Campanas Observatory), and current cutting-edge EPRV instruments with the highest available precision (e.g., EXPRES on the 4.3-m Lowell Discovery Telescope at Lowell Observatory and NEID on the 3.5-m WIYN Telescope at Kitt Peak National Observatory). A complete list of the 15 observatories and 23 spectrographs is provided in Table \ref{tab:observatories} in Appendix \ref{appx:figures_tables}.

Because each spectrograph was designed independently, often many years apart, each employs a unique set of characteristics that result in different levels of RV precision. For example, spectral resolving power, wavelength range, throughput, thermal stability, and wavelength calibration method are all factors that may vary from instrument to instrument. Some of the key spectrograph and observatory properties for the instruments used in this study are summarized in Table \ref{tab:observatories} in Appendix \ref{appx:figures_tables}, but we point the reader to the references therein for more detailed discussion of each spectrograph.

We note that some spectrographs have undergone one or more significant upgrades since becoming operational. For example, the HIRES instrument on the 10-m Keck I Telescope at Keck Observatory underwent a significant upgrade in 2004 when a new CCD was installed, improving the typical RV precision from $\sim$4-5~\ms\ to $\sim$1-2~\ms\ \citep{2017AJ....153..208B}. Similarly, HARPS underwent a significant upgrade in 2015 when the optical fibers were replaced \citep{LoCurto+2015Msngr.162....9L}. It is important to note that changes to the detector or fiber-optic cables can introduce systematic offsets between the pre- and post-upgrade RV data, biasing potential planet detections and orbit fits. We therefore treat data before and after such upgrades as independent datasets, as if the data were measured by separate instruments. We have indicated this by appending either ``-pre'' or ``-post'' or the year of the upgrade after the instrument name (e.g., ``HIRES-pre,'' ``HIRES-post,'' ``CORALIE-98,'' ``CORALIE-07.'')

In general, we gathered all publicly-available RV measurements that met the following basic criteria: (1) were timestamped with precision better than 0.5 days, (2) included an uncertainty value, and (3) were measured with a precision better than a few 10's of \ms. We only considered published or well-documented datasets that included the final reduced RV measurements (i.e., we did not reduce any raw spectra in this work). The next subsection details the steps we followed to locate and organize all the RV data.

\begin{figure*}[t!]
    \centering
    \includegraphics[width=0.9\textwidth]{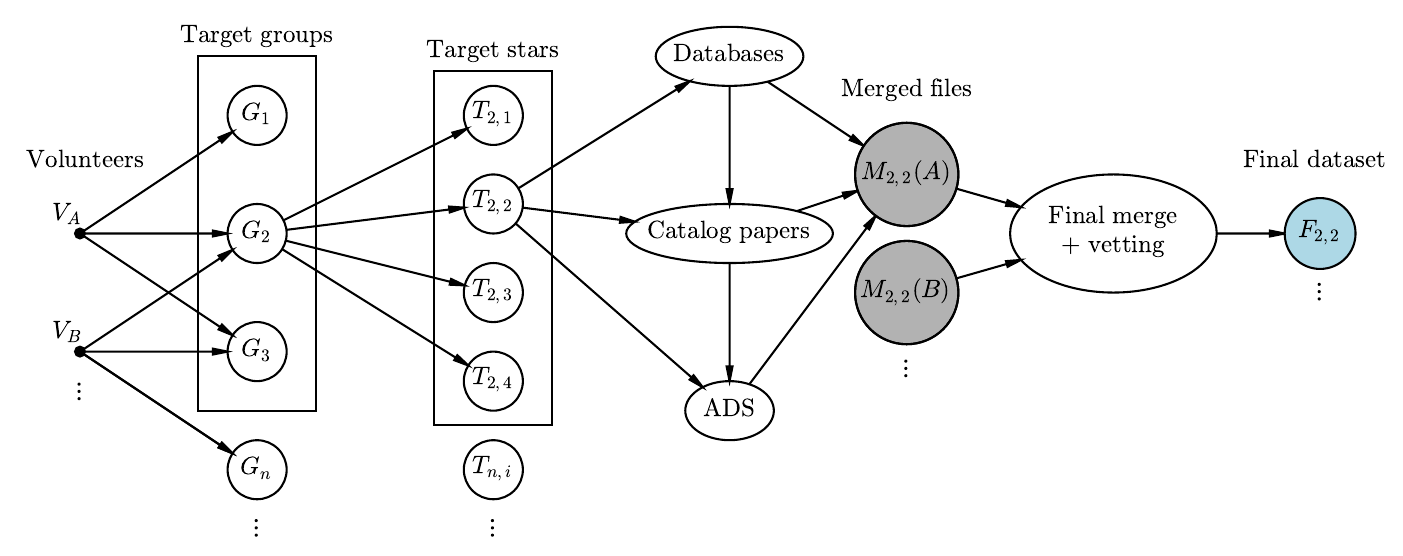}
    \caption{Flowchart summarizing the steps followed to compile the public RV datasets identified by the SPORES-HWO RV Hunters volunteers. The arrows in this example trace the data compilation progress for the second target star in Target Group 2 ($T_{2,2}$), from being assigned to volunteers to being completed as a final merged and vetted dataset. Each volunteer is randomly assigned three Target Groups, with each Target Group containing four target stars from the SPORES-HWO Catalog. Multiple volunteers are assigned to the same target star for redundancy. The volunteers then collect and merge data from various databases, catalog papers, and ADS, as described in \secref{sec:data:rvhunters}. Then, the merged data files from each volunteer are vetted and combined once more to create a final dataset for each target star.}
    \label{fig:flowchart}
\end{figure*}

\subsection{The SPORES-HWO RV Hunters Community Science Project}\label{sec:data:rvhunters}

Identifying and organizing all of the public RV measurements of the 164 SPORES-HWO target stars proved to be a nontrivial task. Completeness and consistency were two major challenges of this process. Because the targets are bright ($V<8$), many have historically been ideal targets for multiple RV exoplanet surveys. Fortunately, this has led to an extensive volume of RV data in the literature for 164 stars. However, this inherently leads to difficulty in determining just how much data exists and whether it is accessible. In other words, although we desire as many RV measurements from the literature as possible, we do not know \textit{a priori} how much data to expect any given target to have or where to find it. 

For example, RVs for some targets may be part of a large easily-accessible database, while for others the RVs may only be available as a formatted table within a single publication. Making matters more difficult is the fact that datasets from different sources are heterogeneous, often differing substantially in terms of which types of metadata are included, what units or conventions are assumed, the chosen file format (or lack thereof), and quality of documentation. These factors together mean that close attention to detail is required to accurately combine disparate RV datasets while striving for high completeness. This is meticulous work that is very difficult to automate for a large sample of stars.

This was the main impetus for the SPORES-HWO RV Hunters Community Science Project, a roughly 3-month long citizen-science-style project whose aim was to compile as many public RV measurements as possible for the SPORES-HWO target stars, and to do so in an efficient, consistent, and well-documented manner. Volunteers for SPORES RV Hunters were primarily undergraduate students at the University of California, Berkeley, whose primary majors were astrophysics or a related field. Through a combination of detailed written guidelines, online webinars, office hours, and digital collaboration, over 60 volunteers were trained and assigned prioritized groups of target stars to investigate\footnote{All volunteers were required to agree to a code of conduct and expectations based on the AAS Code of Ethics (\url{https://aas.org/policies/ethics}) prior to receiving their target assignments.}.

To evenly distribute the workload, the 164 target stars were divided into 41 random target groups (each containing 4 target stars), and each volunteer was assigned to three target groups for a total of 12 target stars per person. The target groups were assigned in a particular order as to maximize redundancy, should any volunteer only complete a single group. If all volunteers completed all three of their assigned groups then each target star would be investigated by at least 4 independent people, reducing the likelihood that any data were missed during the search. 

Once the target groups were assigned, volunteers followed a general procedure to search for public RV datasets, organize the relevant columns, and submit their findings through an online portal. This process is shown graphically in Figure \ref{fig:flowchart}. The following resources were used to identify the RV datasets:
\begin{itemize}
    \item{
        \textit{LCES HIRES Archive.} The most recent data release from the Lick-Carnegie Exoplanet Survey Team (LCES) HIRES/Keck RV exoplanet survey\footnote{\url{https://ebps.carnegiescience.edu/data/hireskeck-data}} \citep{2017AJ....153..208B} includes 75,062 RV measurements of 1,700 stars through 10 March 2020. These data were measured at a spectral resolving power of R$\sim$60,000 in the wavelength range 3700-8000\,\AA, using a gaseous iodine absorption cell located in the optical path for wavelength calibration \citep{Butler+1996ApJ}. The LCES HIRES Archive was the default source for HIRES/Keck RVs in this work because of its high coverage of the northern hemisphere sky.
    }
    \item{
        \textit{HARPS RVBank Archive.} The HARPS RVBank\footnote{\url{https://www2.mpia-hd.mpg.de/homes/trifonov/HARPS_RVBank.html}} is a public database of RVs from HARPS that have been corrected for systematic errors \citep{2020AA...636A..74T}. The latest release of the HARPS RVBank contains more than 200,000 unique observations of over 5,000 targets up to January 2022 \citep{Perdelwitz+2024AA}. These data were measured with a spectral resolving power of R$\sim$115,000 covering a wavelength range of approximately 3800-6900\,\AA, and a Th-Ar lamp with a dedicated fiber was used for wavelength calibration \citep{Pepe+2002Msngr}. All RVs from the HARPS RVBank were re-derived using the SpEctrum Radial Velocity AnaLyser (SERVAL) pipeline \citep{Zechmeister+2018AA,Zechmeister+2020ascl}. The HARPS RVBank was the default source for HARPS/ESO RVs in this work because of its extensive coverage of the southern hemisphere sky.
    }
    \item{
        \textit{California Legacy Survey (CLS).} The CLS catalog contains over 100,000 RV measurements of 719 stars from three different spectrographs: HIRES/Keck, Levy/APF, and Hamilton/Lick. More information regarding these data can be found in \citet{2021ApJS..255....8R}. For any targets with HIRES/Keck data, we preferentially used the LCES archive instead of CLS when possible because of more recent updates to the LCES data.
    }
    \item{
        \textit{Lick Planet Search (LPS).} The Twenty-five Year Lick Planet Search contains more than 14,000 RVs for 386 stars that were observed between 1987 and 2011 with the Hamilton spectrograph at Lick observatory. See \citet{2014ApJS..210....5F} for more information. While the Hamilton/Lick RV data were already included in the CLS catalog for many stars, some target stars were only included in LPS. 
    }
    \item{
        \textit{\citet{2023AJ....165..176L}.} The authors of this study examined archival RVs of potential targets for proposed direct imaging missions in the southern hemisphere. They compiled data from the HIRES/Keck, HARPS/ESO, Levy/APF, UCLES/AAT, and PFS/Magellan spectrographs. See \citet{2023AJ....165..176L} for additional information. We collected HIRES/Keck, HARPS/ESO, and Levy/APF observations from this source only if they were not included in the previously listed sources. However, much of the UCLES/AAT and PFS/Magellan data collected in this work were first published in \citet{2023AJ....165..176L}.
    } 
    \item{
        \textit{ELODIE/SOPHIE Archives.} The Observatoire de Haute-Provence (OHP) maintains archival spectra and RV measurements obtained from the ELODIE\footnote{\url{http://atlas.obs-hp.fr/elodie/}} and SOPHIE\footnote{\url{http://atlas.obs-hp.fr/sophie/}} spectrographs \citep{2004PASP..116..693M}. Although observations of some target stars are available on the ELODIE Archive, we chose not to include those data here due to the very large uncertainties derived from 1$\sigma$ width of the cross-correlation function (CCF) peak (on the order of a few k\ms). This was not an issue on the SOPHIE Archive, where the RV uncertainties were derived from either the CCF contrast and FWHM or the photon noise, depending on the signal-to-noise ratio of the observation. While we did obtain the ``SCI'' SOPHIE RVs for our target stars, we note that uncorrected instrumental drifts and nightly zero-point offsets were often observed in the data, which complicated the analysis. We therefore excluded the SOPHIE RVs from the RV search and orbital fitting procedure in any systems where the instrumental effects were significant. 
    }
    \item{
        \textit{Data \& Analysis Center for Exoplanets (DACE).} The DACE radial velocity archive\footnote{\url{https://dace.unige.ch/dashboard/}} is a useful tool for identifying previously-published RV datasets, especially those that are not part of large surveys or databases. We found additional RV data on DACE obtained from various spectrographs, including some that were not included in the previously listed sources. In cases where RVs were available on DACE, we went back to the original source paper in order to download and document the data.
    }
    \item{
        \textit{Astrophysics Data Systems (ADS) \& VizieR.} Once the previous sources had been checked for RVs, we manually searched ADS\footnote{\url{https://ui.adsabs.harvard.edu}} for individual articles or smaller catalogs with data for the target stars. These data were either downloaded from VizieR\footnote{\url{https://vizier.cds.unistra.fr/viz-bin/VizieR}} or accessed through the online journal. 
    }
\end{itemize}

\begin{figure}[t!]
    \centering
    \includegraphics[width=0.49\textwidth]{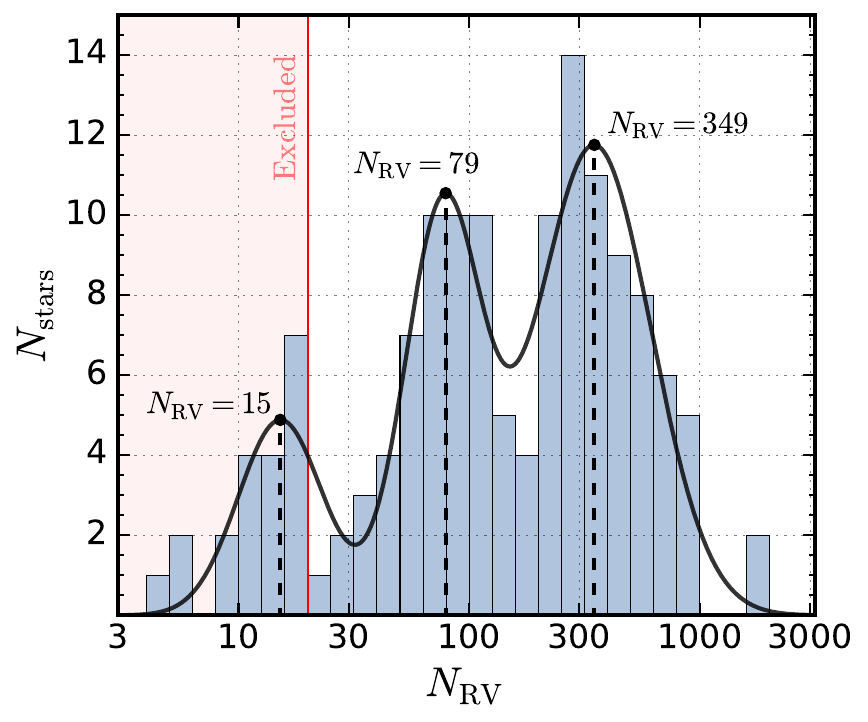}
    \caption{Histogram of the number of RV epochs obtained for each target star. Each histogram bin has a width of 0.1 dex. Interestingly, the stars appear to be clustered in roughly three groups based on how many times they have been observed, which we quantified using a best-fit 3-Gaussian model (shown in black). The least-observed stars with fewer than 20 RVs were excluded from the analysis in this work. A 2-D version of this histogram including the observational time span and the derived HZ sensitivity limits (\secref{sec:discussion:limits}) is shown in Figure \ref{fig:correlations} in Appendix \ref{appx:figures_tables}.}
    \label{fig:obs_histogram}
\end{figure}

Volunteers then merged the individual RV datasets for each of their targets into a single file with a standard pre-defined format. The individual datasets and merged file were then uploaded via an online portal for further processing. A total of over 1350 individual datasets and over 330 merged data files were submitted.

All 330 merged submissions were vetted semi-automatically for consistency and accuracy. The first step of the vetting process included checking that the proper units and column names were used, the telescopes were given consistent names, a reference was provided for each data point, null values were consistently formatted, there were no duplicates, and values were reasonable in both magnitude and sign. After making any appropriate corrections, we then manually cross-checked the multiple merged files for each target to check for completeness and correctness. We then took the union of the merged files to create a final vetted dataset for each target. As an example, a subset of the final dataset for the star HD 190360 is shown in Table \ref{tab:example_rvs} in Appendix \ref{appx:figures_tables}. 

A histogram of the final number of RV epochs obtained for each target star is shown in Figure \ref{fig:obs_histogram}. In total, the SPORES RV Hunters compiled 153,902 individual RV measurements of 141 stars, measured with 23 different spectrographs at 15 different observatories around the world over the last 36 years. We note, however, that not all stars have been observed equally, as Figure \ref{fig:obs_histogram} demonstrates. We found no public RV data for 23 stars, and fewer than 20 RVs were found for 21 stars. Interestingly, stars with more than 20 RV epochs appear to fall into roughly 2 categories: those with moderately good observational history ($\sim$80 RVs) and those with good or very good observational history ($\sim$350 RVs).

\subsection{Stellar Activity Indicators}

Inferring whether the gravitational effect of a planetary companion is responsible for periodic RV variations and avoiding false-positive detections requires knowledge of any intrinsic variations of the host star. There are several well-known sources of stellar activity that operate over a range of timescales and variability amplitudes. For example, acoustic oscillations in Sun-like stars, including p-mode oscillations, operate on timescales of several minutes and cause the stellar surface to pulsate with an amplitude of order several \ms\ \citep[e.g.,][]{Gelly+1986AA, Bouchy+2001AA}. Oscillations of similar magnitude are also induced by surface granulation, or the rising and falling of convective cells of plasma, on timescales up to a day \citep[e.g.,][]{Dumusque+2011AA_granulation, Meunier+2015AA}. 

Of particular concern for this study are stellar activity signals of Sun-like stars that can operate over timescales comparable to the orbital periods of potential planetary companions. In particular, magnetic activity-induced surface inhomogeneities such as star spots, plages, and faculae may be modulated by stellar rotation, leading to quasi-periodic oscillations in the observed RVs. These stellar activity signals can have amplitudes at the $\sim$10~\ms\ level for Sun-like stars with timescales on the order of the stellar rotation period \citep[e.g.,][]{Donahue+1997SoPh, Dumusque+2011AA_spots, Haywood+2016MNRAS}. Additional RV variations with periods comparable to the Solar magnetic cycle ($\sim$11~yr) may also be driven by a star's internal magnetic dynamos, with signal amplitudes exceeding $\sim$10~\ms\ \citep[e.g.,][]{Baliunas+1996ApJ, Dumusque+2011AA_magnetic, 2024ApJS..274...35I}. 

In order to vet any putative planet signals for false positives caused by stellar activity, we therefore also collected several well-documented stellar activity indicators during the search for RV measurements. One common stellar activity indicator, the Mount Wilson $S$-index, is a measure of the relative flux in the Ca~II H and K emission line cores at 3968~\AA~and 3934~\AA, respectively, compared to the local continuum flux \citep{Wilson_1968ApJ, Duncan+1991ApJS}. The $R_{\rm HK}'$ is another common stellar activity indicator based on the Ca~II H and K lines, which is closely related to the $S$-index with corrections for bolometric flux and photospheric contribution \citep{Middelkoop+1982AA, Rutten+1984AA, Hartmann+1984ApJ, Noyes+1984ApJ}. The Ca~II H and K lines are produced in active regions in the stellar chromosphere, where concentrated magnetic field lines are responsible for phenomena such as star spots. Therefore, time series observations of both the $S$-index and $R_{\rm HK}'$ are useful proxies for stellar activity on the timescales of stellar rotation and stellar magnetic cycles. 

Both $S$-index and $R_{\rm HK}'$ measurements were included for most stars with RVs in the LCES HIRES and HARPS RVBank archives. Other types of stellar activity indicators were available for other datasets. When available, we also collected $H$-index or H-$\alpha$ equivalent width (EW) measurements, which both trace active regions in stellar chromospheres, in addition to the full-width half-maximum (FWHM) and bisector inverse slope (BIS) of the cross-correlation function, which both trace line profile variations caused by stellar activity. Measurements of each of these stellar activity indicators were submitted and vetted along with the SPORES RV Hunters datasets. In total, 43,649 epochs of stellar activity indicator data were collected. These are summarized in Table \ref{tab:data_summary} and Table \ref{tab:targets} in Appendix \ref{appx:figures_tables}.

\section{Analysis} \label{sec:analysis}

In this section we describe the analysis of the RVs and stellar activity indicator data. First, we describe the search for Keplerian signals in the RVs and how we quantify the RV search completeness. Then, we describe the analysis of periodic signals in the stellar activity data and how we used those signals to vet false positives. The full figure sets from the analysis (e.g., Figures \ref{fig:HD190360_summary}-\ref{fig:HD190360_Sind_periodogram_plot} and Figures \ref{fig:HD190360_corner_plot}-\ref{fig:HD190360_derived_plot} in Appendix \ref{appx:figures_tables}) are available on Zenodo\footnote{\url{https://doi.org/10.5281/zenodo.17058228}}.

\subsection{Radial Velocity Search} \label{sec:analysis:rvsearch}

\begin{figure*}[t!]
    \centering
    \includegraphics[width=0.7\textwidth]{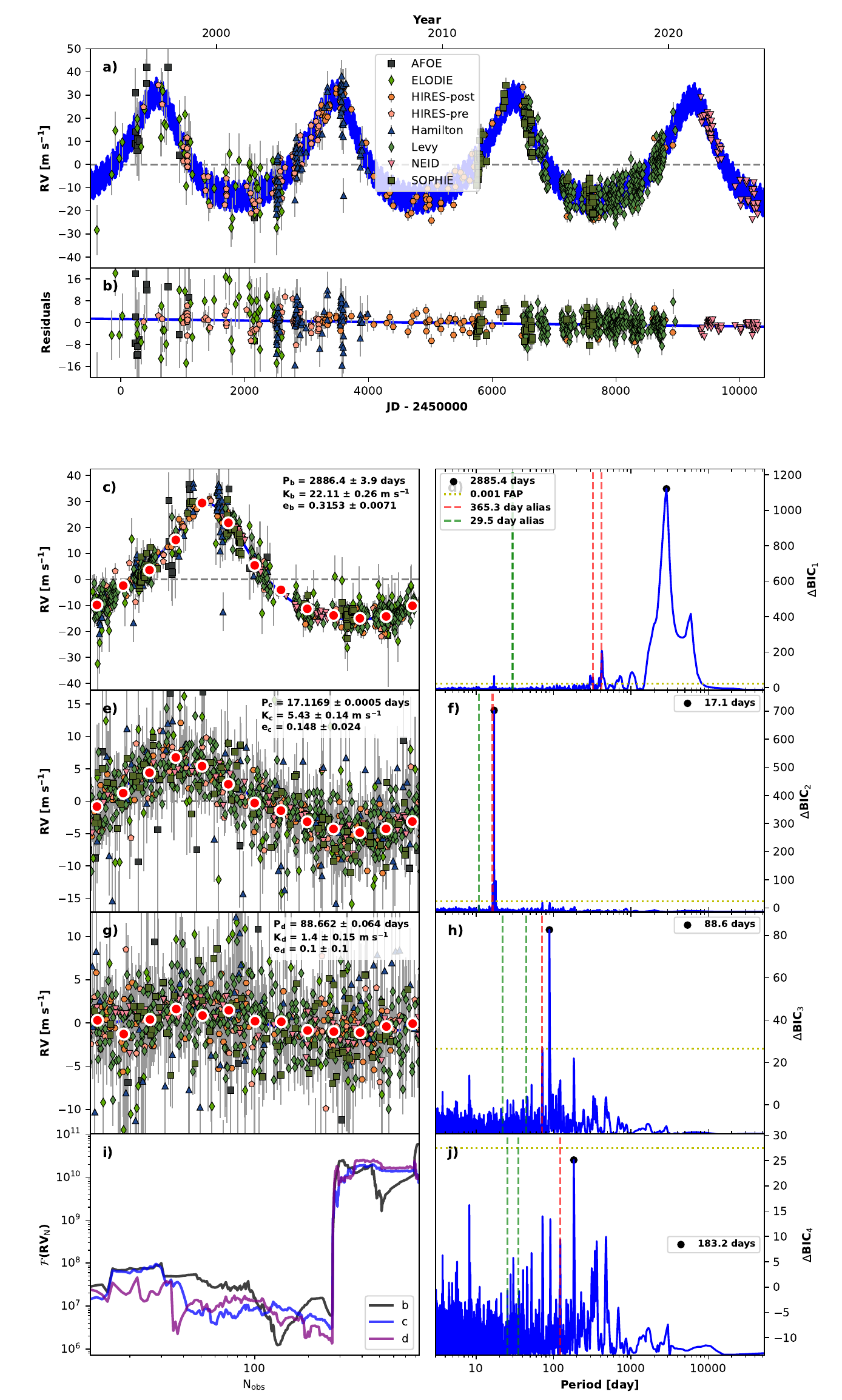}
    \caption{Example \rvsearch~summary plot for HD~190360, a $V=5.7$ G7IV-V star orbited by at least two known planets and one additional planet candidate. Panel (a) shows the full RV time series with the best-fit total model overplotted in blue, while panel (b) shows the residuals to the best-fit 3-planet Keplerian model. Panels (c), (e), and (g) show the phase-folded RVs associated with the $\Delta$BIC periodograms in panels (d), (f), and (h). Panel (i) shows the running Lomb-Scargle periodograms for each signal, and panel (j) shows the $\Delta$BIC periodogram of the final residuals. The complete figure set (120 images) is available.}
    \label{fig:HD190360_summary}
\end{figure*}

We used the iterative RV search algorithm implemented in the open-source \rvsearch\ package \citep{2021ApJS..255....8R} to identify periodic signals in the RV data and determine the best-fit Keplerian orbital solutions. Throughout the analysis, we binned the RVs to 0.5-day intervals to average over intra-night stellar variability caused by p-mode oscillations. For each target star, the search begins with an initial model consisting of the RV dataset and a likelihood function given by
\begin{equation}
    \ln (\mathcal{L}) = \sum_i \bigg[ \frac{\big(v_i - m(t_i) - \gamma_D\big)^2}{\sigma_i^2} + \ln(2\pi \sigma_i^2) \bigg]
\end{equation}
where $v_i$ is the $i$th RV measurement, $\sigma_i^2$ is the
quadrature sum of the instrumental error and a jitter term, and $\gamma_D$ is an instrumental velocity offset \citep{2021ApJS..255....8R}. The RV model $m(t_i)$ at time $t_i$ is given by 
\begin{equation}
    m(t) = \sum_n \nu(t | K_n, P_n, e_n, \omega_n, t_{c,n}) + \dot{\gamma}(t - t_0) + \Ddot{\gamma}(t - t_0)^2
\end{equation}
where $\nu$ is the $n$th Keplerian orbit signal at time $t$ given RV semi-amplitude $K$, period $P$, eccentricity $e$, argument of periastron $\omega$, and time of inferior conjunction $t_c$. Linear and quadratic trend terms are given by $\dot{\gamma}$ and $\Ddot{\gamma}$, respectively, and $t_0$ is the median time of observation. 

Our initial likelihood model contains a single-planet Keplerian model with uninformed orbital parameters. First, \rvsearch\ determines the trend terms $\dot{\gamma}$ and $\Ddot{\gamma}$ by fitting independent flat, linear, and quadratic models to the data, and then performing a statistical test to decide which trend model is preferred \citep{2021ApJS..255....8R}. For this step, the Bayesian Information Criterion (BIC) is calculated for each trend model, defined as
\begin{equation}
    {\rm BIC} = -2\ln(\mathcal{L}) + k \ln(N)
\end{equation}
where $\mathcal{L}$ is the likelihood of the given model, $k$ is the number of free model parameters, and $N$ is the total number of observations. A linear trend is preferred over a flat model if the difference in BIC between the models exceeds a threshold of $\Delta\text{BIC}=5$. Likewise, a quadratic trend is favored over the linear model if this $\Delta\text{BIC}$ threshold is exceeded.

Next, with the initial model in hand, \rvsearch\ defines an orbital period search grid in which the spacing between adjacent frequency grid points is $\Delta f = \frac{1}{2 \pi \tau}$, where $\tau$ is the observational baseline \citep{2021ApJS..255....8R}. For each target, we define the limits of the period grid to be 3 and 10,000 days, unless the system contains a known planet with an orbital period shorter than 3 days. A goodness-of-fit periodogram is then constructed by stepping through the period grid and fitting a sinusoid at a fixed period to the data. The goodness-of-fit is quantified at each grid point by the $\Delta\text{BIC}$ between the best-fit model with $n+1$ planets and the model with $n$ planets (i.e., a zero-planet model for the first iteration of the search).

The significance of each point in the $\Delta\text{BIC}$ periodogram is quantified by a false-alarm probability (FAP), which is determined empirically by extrapolating the log-scale histogram of periodogram power values \citep{Howard+2016PASP, 2021ApJS..255....8R}. We set a FAP threshold of 0.1\% for any periodic signal to be considered significant. For any periodic signal that meets this criterion, \rvsearch\ then refines the Keplerian model by performing a maximum \textit{a posteriori} (MAP) fit with all parameters free, and records the BIC of the best-fit model. At this stage, \rvsearch\ implements Keplerian orbit fitting using the \texttt{RadVel} package \citep{Fulton+2018PASP}.

The algorithm proceeds by adding one additional planet to the RV model, conducting another grid search, and calculating a new $\Delta\text{BIC}$ periodogram by comparing the BIC of the $n+1$-planet model to that of the best-fit $n$-planet model at each fixed period grid point. The empirical $\Delta\text{BIC}$ detection threshold corresponding to FAP=0.1\% is recomputed, and this process iterates until no further significant peaks are detected. The total number of parameters in the final model depends on both the number of planets detected and the number of instruments used to observe the system. For example, the model for a system with 2 planets observed by 4 instruments would include up to 20 parameters: 12 parameters shared across instruments (2 sets of 5 Keplerian orbit components plus 2, 1, or 0 trend terms) and 8 parameters unique to each instrument (4 sets of instrumental offset and jitter terms).

Once the search process is complete and the MAP estimates of the final model parameters are calculated, \rvsearch\ samples the posteriors using an affine-invariant Markov Chain Monte Carlo (MCMC) implemented in \texttt{emcee} \citep{emcee} and \texttt{RadVel} \citep{Fulton+2018PASP}. We sample the posteriors in the default \texttt{RadVel} basis $\{\log P, K, \sqrt{e} \sin \omega, \sqrt{e} \cos \omega, t_c\}$ assuming uniform priors on all parameters\footnote{We enforce two hard bounds constraining $K > 0$ and $0 \leq e < 1$ \citep{2021ApJS..255....8R}.}. For a comprehensive review of \texttt{RadVel}'s MCMC implementation, see \cite{Fulton+2018PASP}.

Finally, we derive $P$, $e$, and $\omega$ from the converged posteriors using the standard utility functions provided in \texttt{RadVel}. Using the stellar masses and uncertainties adopted from the SPORES-HWO Catalog \citep{Harada+2024ApJS}, we also derive the semi-major axis $a$ using Kepler's third law, and the projected planet mass according to 
\begin{equation}\label{eq:msini}
    \frac{M_{\rm p} \sin i}{{\rm M_{Jup}}} = \frac{K \sqrt{1 - e^2}}{28.4\ \text{m\,s}^{-1}} \bigg(\frac{M_\star + M_{\rm p}}{M_\odot}\bigg)^{2/3} \bigg(\frac{P}{\text{yr}}\bigg)^{1/3}
\end{equation}
where $M_\star$ is the stellar mass. 

Each unique Keplerian signal identified by \rvsearch\ is listed in Table \ref{tab:rvsearch_signals} in Appendix \ref{appx:figures_tables}. Any signals associated with known planets or companions are listed with the corresponding planet letter and classified as ``known companion'' (KC). Otherwise, each signal has been assigned an arbitrary decimal ID number in the order it was detected.  For each signal, Table \ref{tab:rvsearch_signals} provides the posterior median and 1$\sigma$ uncertainty values for $P$, $T_c$, $e$, $\omega$, $K$, $M_{\rm p} \sin i$, and $a$, alongside the FAP associated with the detection and a classification of the signal (see \secref{sec:analysis:activity}). For systems where we detected a linear or quadratic, we list the posterior median and 1$\sigma$ uncertainty values for $\dot{\gamma}$ and $\Ddot{\gamma}$ in Table \ref{tab:rvsearch_trends} in Appendix \ref{appx:figures_tables}.

An example \rvsearch\ summary plot for the HD~190360 system is shown in Figure \ref{fig:HD190360_summary}, which includes the full RV time series and best-fit model, $\Delta$BIC periodograms, and phase-folded orbit plots (the converged posterior distributions of the MCMC fit parameters and derived planet parameters are shown in Figures \ref{fig:HD190360_corner_plot} and \ref{fig:HD190360_derived_plot} in Appendix \ref{appx:figures_tables}). HD~190360 has at least two known planets, an inner Neptune-mass planet and an outer Jovian planet \citep[e.g.,][]{2003AA...410.1051N, Vogt+2005ApJ}, which were easily detected by \rvsearch. We also detect a third planet candidate in this system, a sub-Neptune mass planet on an 88.66-day orbit, which we discuss more in \secref{sec:hd190360}.

We note that \rvsearch\ sometimes identifies RV signals with periods greater than or comparable to the observational baseline. In these cases, the MCMC may converge to a nonphysical period that is much greater than the one corresponding to the periodogram peak. In such cases, the uncertainties in the posterior median period and amplitude may be overly inflated due to poor sampling. Following \citet{2023AJ....165..176L}, we label these detections as long-period signals (LPS) and record only the period corresponding to the peak of the $\Delta$BIC periodogram (Table \ref{tab:rvsearch_signals}).

\begin{figure}[t!]
    \centering
    \includegraphics[width=0.49\textwidth]{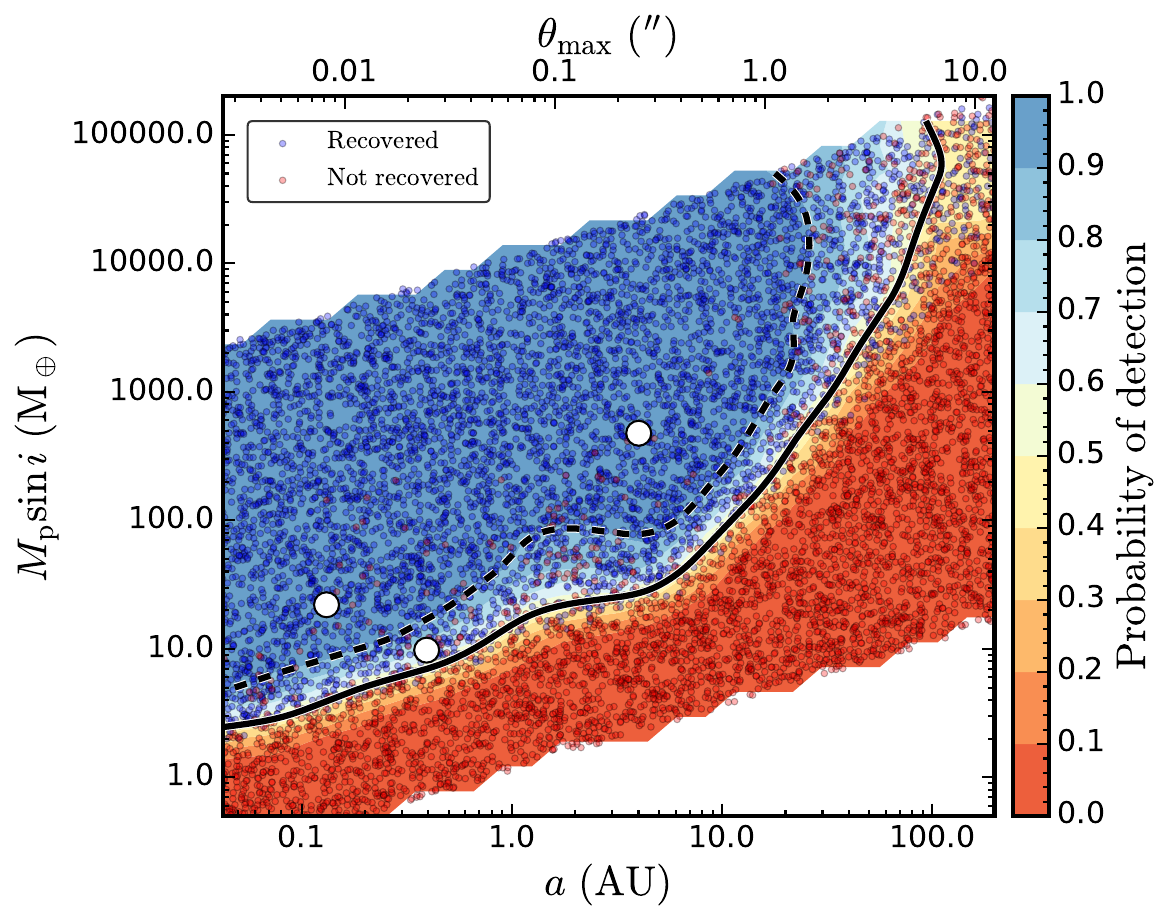}
    \caption{Example \rvsearch~completeness plot for HD~190360 computed from the RV residuals shown in Figure \ref{fig:HD190360_summary}. Each point represents a simulated Keplerian (circular) orbit in the $M_{\oplus} \sin i$ vs.~$a$ plane, where the blue points have been recovered by \rvsearch~and the red points have not. The top axis shows the converted maximum angular separation of the orbit given $a$ and the star's parallax. Detection probability contours are shown by the color scale. The 50\% completeness contour is indicated by the solid black line and the 90\% contour is shown by the dashed black line. The three Keplerian signals recovered by \rvsearch~are shown as white circles. The complete figure set (120 images) is available.}
    \label{fig:HD190360_recoveries}
\end{figure}

\begin{figure*}[t!]
    \centering
    \includegraphics[width=0.99\textwidth]{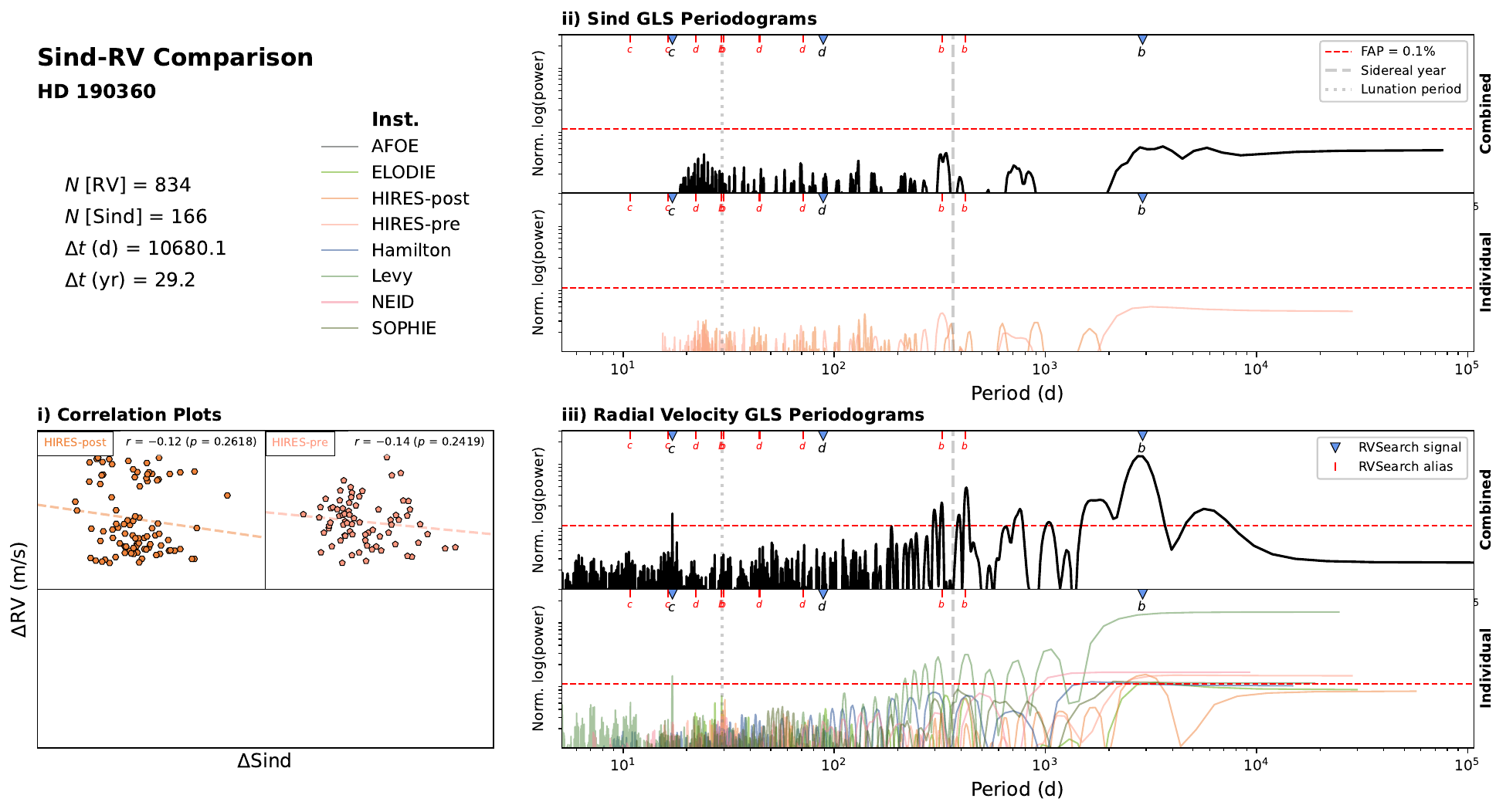}
    \caption{Example stellar activity diagnostic plot for the $S$-index measurements of HD~190360. Panel (i) shows the correlation between the RVs and $S$-index values (by instrument). The Pearson correlation coefficient and its corresponding $p$-value are shown in the inset. Panel (ii) shows the Generalized Lomb-Scargle (GLS) periodogram \citep{Zechmeister+2009AA} of the $S$-index measurements for the combined dataset (top) and individual instruments (bottom). The 0.1\% FAP level is indicated by the red dashed line. Keplerian RV signals identified by \texttt{RVSearch} are indicated by the blue triangles, and their alias periods with the sidereal year and the lunation period are indicated by vertical red ticks. Panel (iii) is the same, but for the RV measurements. The complete figure set (567 images) of stellar activity diagnostic plots for each stellar activity indicator and RV window function is available to download.}
\label{fig:HD190360_Sind_periodogram_plot}
\end{figure*}

\subsection{Injection and Recovery Tests} \label{sec:analysis:injrec}

An key goal of this study is to determine limits on undetected planets in each of the target systems given the existing public RV data. We empirically measure the search completeness for each system with at least 20 RV epochs as a function of semi-major axis and $M_{\rm p} \sin i$ using Keplerian injection and recovery tests. We used the injection and recovery functionality implemented in the \rvsearch\ package \citep{2021ApJS..255....8R}.

We inject a 10,000 individual synthetic Keplerian signals into the residuals from the best-fit RV model described in \secref{sec:analysis:rvsearch}. The periods and RV amplitudes of the synthetic planets are drawn from log-uniform distributions spanning from 3 to $10^6$ days and 0.1 to 1000 \ms, respectively. For simplicity, we assume all orbits are circular. The search algorithm described in \secref{sec:analysis:rvsearch} then attempts to recover each synthetic planet. The result is a map in semi-major axis vs.\ $M_{\rm p} \sin i$ space showing which hypothetical planets would or would not be detectable. 

The results of the injection and recovery tests are then used to compute the search completeness for each data set, shown as contours of detection probability in the $a$ vs.\ $M_{\rm p} \sin i$ plane. Figure \ref{fig:HD190360_recoveries} shows an example of the injection and recovery results for HD~190360 with the corresponding completeness contours. For this system, which has a total of 835 RV epochs from 8 different instruments spanning 29.2 years, we find that a hypothetical $M_{\rm p} \sin i\approx15$ \Mearth\ planet would have a detection probability of about 50\% at an orbital separation of 1~AU. Therefore, a Neptune-mass or more massive planet at this semi-major axis should be detectable, while any lower-mass planets would remain undetectable given the RV data. As expected, the detection sensitivity begins to drop off steeply at orbital separations $\gtrsim$10~AU due to the temporal baseline of the RV observations. All \rvsearch\ completeness plots are available for download.

\subsection{Stellar Activity Analysis} \label{sec:analysis:activity}

In order to help us identify potential stellar activity signals and vet any new RV signals for false positives, we compute the generalized Lomb-Scargle (GLS) periodogram \citep{Zechmeister+2009AA} of any available stellar activity indicators. Following the RV search outlined in \secref{sec:analysis:rvsearch}, we compute up to 6 sets of stellar activity GLS periodograms for each target: $S$-index, $H$-index, EW$_{\rm H\alpha}$, $\rm R_{HK}^\prime$, BIS, and FWHM. For comparison, we also compute GLS periodograms for the RVs and the observational window function.

For each stellar activity indicator, we first compute GLS periodograms one instrument at a time for any instruments with at least 10 measurements. We iteratively remove 5$\sigma$ outliers and bin the stellar activity measurements to nightly intervals before calculating each periodogram using the Lomb-Scargle \texttt{autopower} function implemented in \texttt{astropy} \citep{astropy_2013_AA, astropy_2018_AJ}. Computing instrument-by-instrument periodograms allows us to isolate any potential instrumental systematics could that introduce false-positive RV signals. After removing any instrumental offsets, we then stitch together the stellar activity data the compute the combined GLS periodogram. This process is also repeated for the RVs and the window function.

We then generate a set of diagnostic plots for each target comparing any signals detected by \texttt{RVSearch} with the stellar activity GLS periodograms. For each activity indicator, the diagnostic plot shows the combined and instrument-by-instrument periodograms for both the RVs and the stellar activity data. Any \texttt{RVSearch} signals and their yearly and lunation period aliases are overplotted for comparison. The diagnostic plots also show instrument-by-instrument scatter plots of the stellar activity indicator vs.\ RV and the corresponding Pearson correlation coefficient and $p$-value. An example diagnostic plot for the $S$-index data for HD~190360 is shown in Figure \ref{fig:HD190360_Sind_periodogram_plot}. For this system, the $S$-index diagnostic plot shows no evidence that any of the \texttt{RVSearch} signals are false alarms or false positives due to stellar activity. All diagnostic plots for all the systems we analyzed are available for download.

All of the diagnostic plots are then used to help classify any RV signals detected by \texttt{RVSearch} as either false alarm (FA), stellar activity (SA), planet candidate (PC), or source requiring confirmation (SRC). Alternatively, signals may be classified as a known companion (KC) or a poorly constrained long-period signal (LPS), as described in \secref{sec:analysis:rvsearch}.  The classifications are described on a case-by-case basis in \secref{sec:hd190360} to \ref{sec:hd219623} and in Appendices \ref{appx:signals_known_planets}, \ref{appx:undetected_KPs}, and \ref{appx:other_signals}.

\section{Results} \label{sec:results}

Here we present the main results of the RV analysis. We start by summarizing the potential Keplerian signals identified by \texttt{RVSearch} in \secref{sec:results:rvsearch}, highlighting six systems with novel detections in \secref{sec:hd190360} to \ref{sec:hd219623}. We then summarize the search completeness of all the stars in the sample in \secref{sec:results:completeness}.

\subsection{\texttt{RVSearch} Results} \label{sec:results:rvsearch}

We analyzed 120 of the 164 provisional HWO target stars identified in the EMSL \citep{Mamajek+Stapelfeldt_2024} and SPORES-HWO Catalog \citep{Harada+2024ApJS} for which we had at least 20 epochs of RV observations. While we also compiled public RV data for 21 additional target stars, these stars were excluded from the analysis because they had fewer than 20 nights of observations. For the 120 target stars we analyzed, we used \rvsearch\ \citep{2021ApJS..255....8R} to find a total of 109 potential Keplerian signals in 49 unique systems (Table \ref{tab:rvsearch_signals} in Appendix \ref{appx:figures_tables}), which could either produced by gravitationally-bound companions (stellar or planetary), stellar activity (magnetic cycles or rotationally-modulated surface inhomogeneities), or instrumental systematics or aliases.

Of the 120 systems, we found that 74 also show evidence of a long-term trend in the RVs (Table \ref{tab:rvsearch_trends} in Appendix \ref{appx:figures_tables}), which could be fit by either a first- or second-order polynomial. Many of these trends are associated with known stellar companions on wide orbits\footnote{We note that for stars with imaging and/or astrometry observations, it may be possible to use non-detections and sensitivity limits from those measurements to place limits on the luminosities and separations of stellar companions that might be responsible for the observed RV trends. While this is beyond the scope of the present analysis, we will investigate these observations in future work.}, while others are likely produced by distant planetary companions or stellar activity cycles with periods longer than the observational baseline.

Of the 164 target stars, we note that 30 have at least one previous claim of a planet detection\footnote{\url{https://exoplanetarchive.ipac.caltech.edu/cgi-bin/TblView/nph-tblView?app=ExoTbls&config=PS}}, totaling up to 70 known planets in the whole sample. We recovered 51 known planetary signals, whose updated orbital parameters are shown in Table \ref{tab:rvsearch_signals} and discussed in Appendix \ref{appx:signals_known_planets}. For the other 19 known planets, we summarize their non-detections in Table \ref{tab:undetected_planet_claims} and in Appendix \ref{appx:undetected_KPs}. We found that these undetected planets were often either disputed as planet candidates in the literature, or required additional observations such as astrometry or imaging to be detected. In other cases, the planetary signal simply did not reach the statistical significance required to be considered detections in this work, or the discovery RV data simply were not publicly available.
\begin{deluxetable}{lll}
\tablecolumns{3}
\tablecaption{RV signal classifications}
\label{tab:rvsearch_class_counts}
\tablehead{
  \colhead{Class.} & 
  \colhead{Definition} & 
  \colhead{Count}
}
\startdata
KC & Known Companion & 51 \\
KC-S & Known Companion -Stellar & 2 \\
SA & Stellar Activity & 17 \\
SA-R & Stellar Activity -Rotation & 9 \\
LPS & Long-period Signal & 5 \\
FA & False Alarm & 19 \\
PC & Planet Candidate & 4 \\
SRC & Source Requiring Confirmation & 2 \\
\hline \\
 & \textbf{Total} & 109 \\
\enddata
\end{deluxetable}

We identified an additional 2 signals originating from known stellar binaries, in cases where approximately one full orbit of the stellar companion was measured: HD 165499 (\secref{app:sec:hd165499}) and HD 212330 A (\secref{app:sec:hd212330}). We caution that HD 165499 is not currently listed in the Washington Double Star Catalog\footnote{\url{https://www.astro.gsu.edu/wds/}} \citep[WDS;][]{wds}, and therefore there is no indication in the EMSL \citep{Mamajek+Stapelfeldt_2024} that this star has a stellar companion. More alarmingly, HD 212330 A \textit{is} shown as a binary in the EMSL, but the reported WDS angular separation (23.6$^{\prime\prime}$) is much larger than maximum angular separation of the companion measured in the RV data at $\sim$16,000 d. The companion properties shown in the EMSL could correspond to a wider-separation stellar companion, given that we observe a long-term trend in the CORALIE RVs, but it is also possible that the information is simply inaccurate or out-of-date. Both of these stars are ranked in Tier B of the EMSL. However, \citet{Mamajek+Stapelfeldt_2024} intended to exclud binaries with stellar companions separated by less than 3$^{\prime\prime}$. Based on this criterion, HD 165499 and HD 212330 A should not have made the cut for the EMSL to begin. This raises concerns about the reliability of the WDS catalog in informing exoplanet DI target lists, considering that close stellar companions pose a major challenge for coronagraphic starlight suppression.

We classified each of the remaining 56 RV signals based on the stellar activity diagnostic plots (e.g., Figure \ref{fig:HD190360_Sind_periodogram_plot}) and previous results from the literature. We determined that 26 signals were caused by stellar activity either in the form of magnetic activity cycles or rotationally-modulated activity (SA or SA-R). Another 19 of the signals were classified as false alarms (FA) due to instrumental systematics or aliasing, and 5 signals were classified as long-period signal (LPS) as defined in \secref{sec:analysis:rvsearch}. The last 6 signals were classified as either planet candidates (PC) or sources requiring classification (SRC). These signals were not obviously associated with any known planets, stellar activity, instrumental effects, or aliases. Following \citet{2023AJ....165..176L}, we classified a signal as PC if the periodogram peak was well defined and the RV signal strength increased roughly monotonically with the number of RV observations. In more ambiguous cases (e.g., multiple periodogram peaks clustered around a narrow range of periods) we labeled the signal as SRC, indicating that more data or more sophisticated modeling is needed to robustly classify the signal. We classified 4 RV signals as PC and 2 RV signals as SRC.

All of the signal classifications are summarized in Table \ref{tab:rvsearch_class_counts}. We elaborate on these classifications in 6 systems with newly-detected signals (PC or SRC) in \secref{sec:hd190360} to \ref{sec:hd219623}. More detailed discussion of the classifications in systems with updated parameters of \emph{known planets} can be found in Appendix \ref{appx:signals_known_planets}, and discussion of \emph{non-detections} in systems with previously reported planet can be found in Appendix \ref{appx:undetected_KPs}. Lastly, all other non-companion classifications are discussed in Appendix \ref{appx:other_signals}.

\subsubsection{HD 190360}\label{sec:hd190360}
HD 190360 (HIP 98767; HR 7670; GJ 777 A; TIC 105999792) is a G7IV-V star \citep{Gray+2006AJ} at a distance of $d = 16.003 \pm 0.009 \,{\rm pc}$ \citep{GaiaDR3_2022yCat}. This system hosts an eccentric Jovian planet and an inner Neptune planet, which were first detected with RV observations from ELODIE/OHP, AFOE/Whipple, and HIRES/Keck \citep{2003AA...410.1051N, Vogt+2005ApJ}. The properties of these two planets have since been revised by further ground-based RVs \citep{2009ApJ...693.1084W, Courcol+2015AA, 2021ApJS..255....8R} and astrometry from Gaia and Hipparcos \citep{Feng+2021MNRAS}. In this work, we detected both previously-known planets orbiting HD~190360 using 835 public RVs from the AFOE/Whipple, ELODIE/OHP, HIRES/Keck, Hamilton/Shane, Levy/APF, NEID/WIYN, and SOPHIE/OHP spectrographs, improving the precision on both period and semi-amplitude by at least a factor of two compared to previous literature values \citep[see Table \ref{tab:rvsearch_signals} and ][]{Feng+2021MNRAS, 2009ApJ...693.1084W}. We also find evidence of a linear trend in the RVs (Table \ref{tab:rvsearch_trends}), which is likely produced by HD 190360's faint stellar companion.

In addition to the two previously-confirmed planets, we also detected a third planet candidate, HD 190360 d ($P_{\rm d} = 88.662^{+0.065}_{-0.064}$\,d, $M_{\rm d}\sin i = 10.0 \pm 1.0$\,\Mearth), as shown in Figure \ref{fig:HD190360_summary}. This signal was first identified as a planet candidate in an analysis by \citet{2021AJ....161..134H}. Interestingly, \citet{2021ApJS..255....8R} also found a signal at 90.34 days which they classified as a false-positive, likely due to how close the period is the the 1/4 annual harmonic. More recently, \citet{2025AJ....170...52G} found a similar periodicity at 90 days using newer RV data from NEID. The period we recovered is not consistent with the 1/4 annual harmonic and is not obviously associated with any stellar activity signals, hence we classified it as a planet candidate. During the revision stage of this paper, the confirmation of HD 190360 d was announced by \citet{Giovinazzi+2025arXiv250818169G}, who conducted a more thorough investigation of the system using both RVs and astrometry from Gaia and Hipparcos. Our results for HD 190360 d are consistent with their orbital solution, albeit with slightly higher uncertainties. 

\subsubsection{HD 192310}\label{sec:hd192310}
HD 192310 (HIP 99825; HR 7722; GJ 785; TIC 326096771) is a K2+V star \citep{Gray+2006AJ} at a distance of $d = 8.812 \pm 0.004 \,{\rm pc}$ \citep{GaiaDR3_2022yCat}. This system hosts two known planets with orbital periods of about 74 days and 526 days \citep{Howard+2011ApJ, 2011AA...534A..58P}. Using 768 RVs from HARPS/ESO, HIRES/Keck, PFS/Magellan, and UCLES/AAT, we detected both known planets at periods of $P_{\rm b} = 74.262 \pm 0.032 \,{\rm d}$ and $P_{\rm c} = 546.5 \pm 3.4 \,{\rm d}$, consistent with recent results for this system \citep{2023AJ....165..176L}. 

We also identified four additional signals not associated with any known planets in this system (Table \ref{tab:rvsearch_signals}). Two of the signals at periods of $P=43.623^{+0.025}_{-0.024}$ d and $P=39.44^{+0.05}_{-0.03}$ d have been attributed to differential stellar rotation \citep{2023AJ....165..176L, 2011arXiv1107.5325L}, and another signal at $P=5650.0^{+4000.0}_{-1600.0}$ d is consistent with a previously reported stellar activity cycle \citep{2011arXiv1107.5325L}. The sixth signal we detected at $P=24.567^{+0.02}_{-0.018}$ d does not match with any known planets or stellar activity. The signal has an amplitude of $K=0.632^{+0.1}_{-0.099}$\,\ms, corresponding to a planet mass of $M_{\rm p} \sin i = 2.3^{+0.38}_{-0.37}$. \citet{2023AJ....165..176L} also reported a significant signal with the same periodicity and noted a well-defined periodogram peak and well-constrained posteriors. We found very similar results for this system using the same data but with an additional 10 nights of HIRES/Keck data. Following \citet{2023AJ....165..176L}, we classified this signal as a planet candidate. Finally, we note that the best-fit model for this system prefers a quadratic trend term (Table \ref{tab:rvsearch_trends}), which may hint at a long-period companion or stellar activity signal. 

\subsubsection{HD 210302 ($\tau$ PsA)}
HD 210302 ($\tau$ PsA; HIP 109422; HR 8447; GJ 9770; TIC 97402436) is an F6V star \citep{Gray+2006AJ} at a distance of $d = 18.46 \pm 0.03$~pc \citep{GaiaDR3_2022yCat} with no known companions. Analyzing 35 RVs from HIRES/Keck and 43 RVs from HARPS/ESO, we detected a single periodic signal at $P=115.52^{+0.19}_{-0.27}$ d with an amplitude of $K = 8.3^{+1.6}_{-1.4}$\,\ms. This signal was only weakly detected with an FAP of 0.046\%, and we note a high level of scatter in the residuals and incomplete orbital phase coverage. A small peak in the GLS periodogram of $\rm R_{HK}^\prime$ measurements from HARPS/ESO and a correlation between $\rm R_{HK}^\prime$ and the RVs (Pearson correlation $r=0.65$ with $p\ll0.05$) suggest that this signal may be associated with stellar activity. However, we currently do not have enough data to confirm whether the $\rm R_{HK}^\prime$ periodicity is statistically significant and rule out the possibility that the signal is planetary in nature. We therefore classify the signal as SRC until further observations can be obtained.

\subsubsection{HD 217987 (Lacaille 9352)}
HD 217987 (Lacaille 9352; HIP 114046; GJ 887; TIC 155315739) is an M1.0V star \citep{Dieterich+2012AJ} at a distance of $d = 3.2871 \pm 0.0005 \,{\rm pc}$ \citep{Gaia_DR2_2018AA}. This system was recently reported to have two super-Earths and a possible third planet in comprehensive study of RVs from HARPS/ESO, HIRES/Keck, PFS/Magellan, UCLES/AAT, and photometry from TESS \citep{2020Sci...368.1477J}. Using the publicly-available HARPS/ESO and HIRES/Keck data, we found consistent orbits for the two known planets, GJ~887~b and GJ~887~c, with periods of $P_{\rm b} = 9.26177^{+0.00055}_{-0.00053}$ d and $P_{\rm c} = 21.7952^{+0.0025}_{-0.0024}$ d. However, we did not detect the putative $\sim$50-day signal identified by \citet{2020Sci...368.1477J}, instead finding a potential new signal at $P=39.223^{+0.016}_{-0.015}$ d with an amplitude of $K=2.28^{+0.34}_{-0.32}$\,\ms. \citet{2020Sci...368.1477J} noted that GJ 887 is an unusually quiet M dwarf, and therefore were unable to determine a rotation period for the star based on its TESS light curve. Without knowing the stellar rotation period, we cannot confidently classify the 39-day signal detected here as either stellar activity or planetary in nature. Therefore, we report this signal as SRC. 

\subsubsection{HD 219134} \label{sec:hd219134}
HD 219134 (HIP 114622; HR 8832; GJ 892; TIC 283722336) is a K3V star \citep{Keenan+1989ApJS} at a distance of $d = 6.542 \pm 0.002 \,{\rm pc}$ \citep{GaiaDR3_2022yCat}. This star hosts a multi-planet system with at least five known planets with masses ranging from $M_{\rm p} \sin i =4.36$\,\Mearth\ to 108\,\Mearth\ and with orbital periods ranging from about 3 days to 6 years \citep{2015ApJ...814...12V}. At least two of the planets, HD~219134~b and HD~219134~c, have been detected in transit by TESS and \textit{Spitzer} \citep{2015AA...584A..72M, Gillon+2017NatAs, Seager+2021AJ}. In this work, we detected five known planets (b, c, d, f, and h) using 543 RVs from HARPS-North/TNG, HIRES/Keck, Hamilton/Shane, and Levy/APF (Table \ref{tab:rvsearch_signals}). Our derived planet properties and orbits are consistent with previously published results with improvements in orbital period precision of up to a factor of 3 \citep{Gillon+2017NatAs, 2015ApJ...814...12V}. 

In addition to the five known planets, we also detected a long-period signal (LPS) with a periodogram peak at 19,241.3 days ($\sim$52 years) and the star's rotation period at $P=43.475^{+0.024}_{-0.026}$ d \citep{2015AA...584A..72M}. Notably, we did not detect the putative planet g at 94.2 days that was published in the original study \citep{2015ApJ...814...12V}. In fact, there have been no subsequent detections of planet g in the literature, despite independent analyses of the system by multiple authors \citep[e.g.,][]{Gillon+2017NatAs, Rosenthal+2022ApJS}. We therefore consider planet g to be a disputed planet detection. However, we did detect one additional signal in this system with a period of $P=192.32^{+0.45}_{-0.39}$ d and an amplitude of $K=2.03 \pm 0.17$\,\ms\ (which we refer to as HD 219134.04). Interestingly, \citet{Rosenthal+2022ApJS} also recovered this 192-day signal and reported it as a systematic, while \citet{2021AJ....161..134H} suggested that it may be the actual period of planet g (94 days is close to the 1/2 harmonic of the signal). We classified HD 219134.04 as PC, but note that a closer investigation of the data for this system is needed to robustly determine its origin. Finally, we also find evidence of a long-term quadratic trend in the RVs (Table \ref{tab:rvsearch_trends}), which is likely associated with the orbit of HD 219134's faint stellar companion.

\subsubsection{HD 219623}\label{sec:hd219623}
HD 219623 (HIP 114924; HR 8853; GJ 4324; TIC 24467943) is an F8V star \citep{Gray+2001AJ} with no known planets at a distance of $d = 20.61 \pm 0.02 \,{\rm pc}$ \citep{GaiaDR3_2022yCat}. Measurements from \textit{Spitzer} show that this star has an infrared excess ($\geq 3\sigma$ at 70\,$\mu$m), suggesting the presence of a debris disk \citep{Beichman+2006ApJ}. Our analysis of 36 RVs from HIRES/Keck spanning 11.5 years revealed a significant signal (FAP = 0.02\%) at $P = 48.481^{+0.046}_{-0.038} \,{\rm d}$ with an amplitude of $K = 9.4 \pm 1.3 \,{\rm m\,s}^{-1}$, corresponding to planet mass of $M_{\rm p} \sin i = 49.3^{+7.5}_{-7.3} \,{\rm M_\oplus}$. 

We found that the peak in the $\Delta$BIC periodogram was well defined and the strength of the signal increased roughly monotonically with the number of observations. We also note that HD~219623 is a relatively inactive star with $R^\prime_{\rm HK} = -4.993$ \citep{Isaacson+2010ApJ}, ranking in the 16th percentile of all $R^\prime_{\rm HK}$ values in the sample. There were no significant peaks in the stellar activity periodograms, and there were no significant correlations between the RVs and stellar activity indicators. Additionally, HD~219623's measured rotation of $v\sin i = 5.50 \,{\rm km\,s}^{-1}$ \citep{Mishenina+2012AA}, combined with its radius of $1.29 \,{\rm R_\oplus}$ \citep{Harada+2024ApJS}, corresponds to a maximum rotation period of about 11.9 days. HD~219623 is also an earlier spectral type than the Sun and is near the lower end of the Kraft break \citep{Kraft_1967ApJ} with an effective temperature of $T_{\rm eff} = 6080 \pm 43 \,{\rm K}$ \citep{Harada+2024ApJS}, so we would not expect this star to rotate slowly. Therefore, the 48-day RV signal is unlikely associated with stellar rotation. Provided these factors, we cannot rule out a planetary origin for the 48-day signal given the 36 available RVs, and hence classify this signal as a PC requiring additional data and analysis.

\begin{figure*}[t!]
    \centering
    \includegraphics[width=0.99\textwidth]{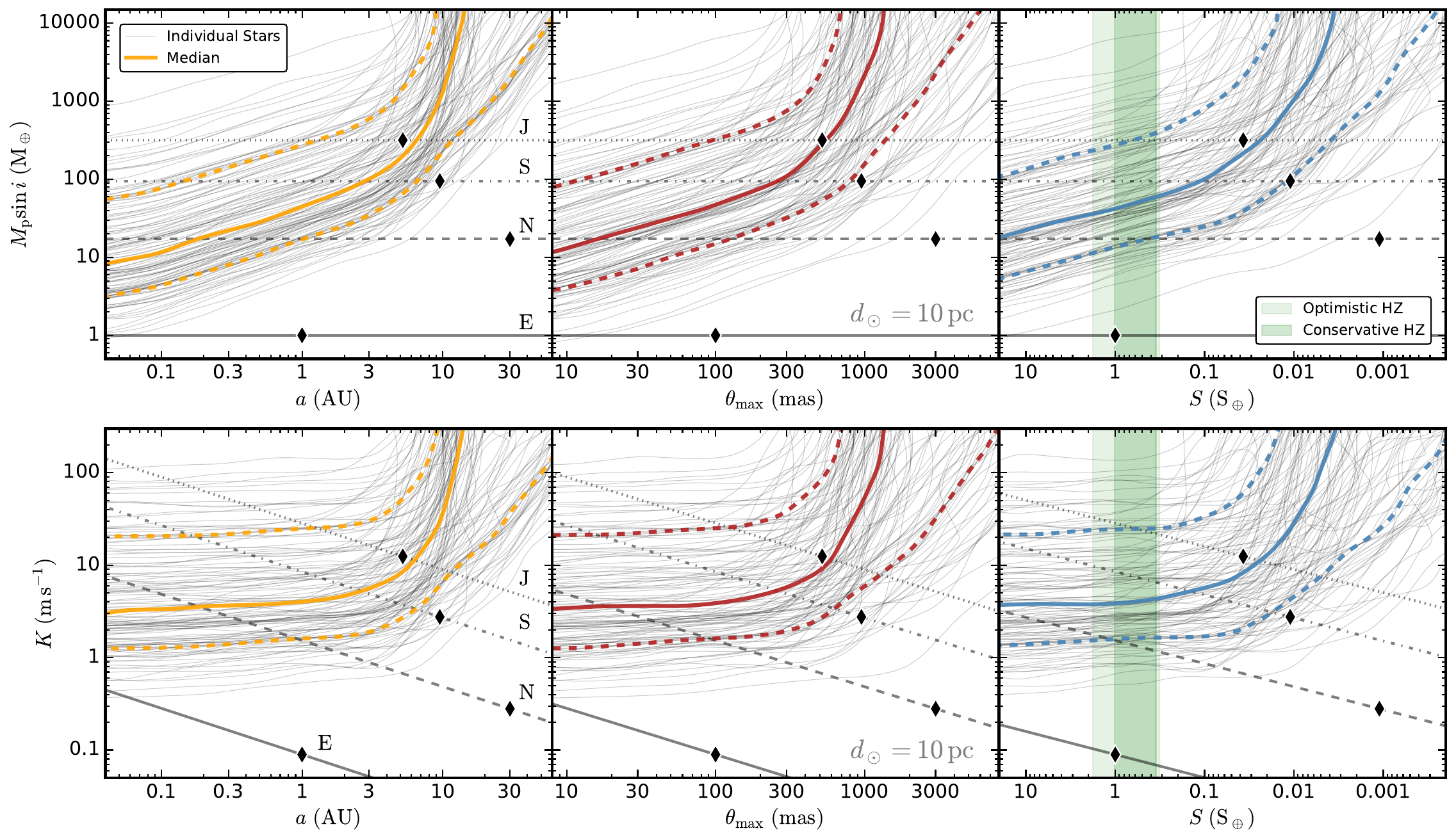}
    \caption{Contours of 50\% search completeness for the 120 stars with at least 20 nights of RVs (black lines). \textbf{Top:} From left to right, each panel shows the same \clf\ lines as a function of $M_{\rm p}\sin i$ and semi-major axis, maximum projected separation (assuming circular orbits), and stellar insolation flux. The median \clf\ in each panel is shown by the solid colored line in either orange, red, or blue. The middle 68\% of \clf\ lines are between the dashed colored lines. The conservative and optimistic HZs for a solar twin \citep{Kasting+1993Icar, Kopparapu+2013ApJ} are indicated by the green shaded regions in the right-most panel. \textbf{Bottom:} Same as the top row, only with RV semi-amplitude $K$ shown on the vertical axis. Earth, Neptune, Saturn, and Jupiter values are indicated by black diamonds in each panel, assuming $e=0$, $i=90^\circ$, and $d=10$~pc for a solar twin.}
    \label{fig:completeness_contours}
\end{figure*}

\subsection{Search Completeness} \label{sec:results:completeness}

We quantified the RV search completeness for each target star with at least 20 RV epochs by running injection and recovery simulations as described in \secref{sec:analysis:injrec}. Search completeness varies significantly between different targets and can depend strongly on stellar spectral type, instrument precision and systematics, number of observations, survey cadence and baseline, and gaps between observations. Therefore, to assess and compare the achieved search completeness between different targets across the sample, we computed the the value of $M_{\rm p} \sin i$ as a function of $a$ at which 50\% of injected planet signals are recovered for a given system. To achieve this, we fit a cubic B-spline smoothing function with regularization parameter $\lambda=1$ to the contour of 50\% detection probability from the completeness plot of each system (e.g., Figure \ref{fig:HD190360_recoveries}). Hereafter, we refer to this as the ``50\% completeness level'' (\clf).

Figure \ref{fig:completeness_contours} shows all 120 \clf\ lines for targets with $\geq$20 epochs of data.. Using the stellar parallaxes compiled in the SPORES-HWO Catalog \citep{Harada+2024ApJS} and EMSL \citep{Mamajek+Stapelfeldt_2024}, we also calculated the \clf\ values as a function of maximum angular separation, $\theta_{\rm max}$, assuming circular orbits. We converted semi-major axis to $\theta_{\rm max}$ using $\theta_{\rm max} = \varpi(a / \text{au})$, where $\varpi$ is the stellar parallax in arcseconds. To assess the search completeness in relation to the conservative and optimistic HZ limits \citep{Kasting+1993Icar, Kopparapu+2013ApJ}, we then also calculated the \clf\ values as a function of instellation flux using $S / {\rm S_\oplus} = (L_\star / {\rm L_\odot}) \cdot (a / {\rm au})^{-2}$, where $L_\star$ are the stellar luminosities in the SPORES-HWO Catalog. Figure \ref{fig:completeness_contours} shows the \clf\ of each star as functions of $M_{\rm p} \sin i$ (and $K$) and $a$, $\theta_{\rm max}$, and $S$.

The \clf\ lines in Figure \ref{fig:completeness_contours} follow the general expected behavior from analytic arguments: minimum detectable mass increases roughly as $M_{\rm p} \sin i \propto a^{1/2}$ (see Equation \ref{eq:msini}) for $a \lesssim (M_\star / {\rm M_\odot})^{1/3} (t_\text{span} / {\rm{yr}} )^{2/3}\,{\rm AU}$, where $t_\text{span}$ is the temporal baseline of the RV observations. At larger values of $a$, the minimum detectable mass increases steeply as we rapidly lose sensitivity to planets whose orbital periods are $\gtrsim t_\text{span}$. Deviations from this behavior on a system-by-system basis result from differences in stellar spectral type, instrument characteristics, survey cadence and baseline, and gaps between observations.

For a typical star in the sample (i.e., the median \clf; solid orange line in Figure \ref{fig:completeness_contours}), we could expect to detect planets with masses\footnote{Throughout this discussion we sometimes use ``planet mass'' and $M_{\rm p} \sin i$ interchangeably, despite these being different quantities. Unless otherwise specified, ``planet mass'' here actually refers to $M_{\rm p} \sin i$---the completeness maps were calculated in $K$-space, but of course we gain the $\sin i$ degeneracy factor when we convert $K$ to a mass (Equation \ref{eq:msini}).} as small as about 10\,\Mearth\ at a distance of about $a = 0.05$ AU with at least 50\% probability. The minimum detectable planet mass increases to approximately 50\,\Mearth\ at $a = 1$ AU, and to 200\,\Mearth\ at $a = 5$ AU. At farther orbital distances the typical minimum detectable planet mass increases more rapidly, with no planets less massive than $\sim$5 $\rm M_J$ detectable at $a \gtrsim 10$ AU for the median \clf. 

At a fixed semi-major axis of $a = 0.1$ AU, the middle 68\% of \clf\ lines have $M_{\rm p} \sin i$ limits between about 5 \Mearth\ and 80 \Mearth\ (dashed orange lines in Figure \ref{fig:completeness_contours}). At $a = 0.3$ AU, the middle 68\% of \clf\ lines are between about 8 \Mearth\ and 150 \Mearth, and at $a = 1$ AU, the middle 68\% of lines are between about 1 Neptune mass (17 \Mearth) and about 1 Jupiter mass (318 \Mearth). In the worst case (for the stars with at least 20 RVs), we could not expect to detect any planets less massive than a few $\rm M_J$. On the other hand, in the best case scenario, a $\sim$1\,\Mearth\ planet could be detected within $a \leq 0.05$ AU (assuming stellar effects do not limit planet sensitivity at this level) and a $\sim$6\,\Mearth\ planet could be detectable at $a = 1$ AU. 

Further discussion of the search completeness results and implications can be found in \secref{sec:discussion}.

\section{Discussion} \label{sec:discussion}

In this section, we discuss the results from our analysis of archival RVs. First, we discuss the context of these results in relation to past studies of Doppler sensitivity to planets around nearby stars (\secref{sec:discussion:past_work}). We then take a closer look at planet sensitivity limits in the HZs of the target stars (\secref{sec:discussion:limits}) and the stars that currently lack sufficient RV observations (\secref{sec:discussion:neglectedstars}). Lastly, we discuss what we currently know or do not know about the architectures of some of the most well-studied nearby planetary systems (\secref{sec:discussion:architectures}) and how this may inform future precursor science studies for the HWO exo-Earth survey.

\subsection{Comparison to Past Work} \label{sec:discussion:past_work}

This study showcases the planet-detecting capability of over 35 years of precision RV monitoring. Compared to past research analyzing Doppler constraints on planetary companions to nearby stars, this work shows an overall improvement in RV search sensitivity thanks to a broader scope of global, multi-decade RV monitoring efforts \citep{Howard+2016PASP, 2023AJ....165..176L}. In this work, we have studied a larger, more complete sample of 120 nearby stars that are likely to be observed by future high-contrast imaging mission like HWO. We have also included data from significantly more instruments by compiling over 150,000 RV measurements that were publicly available from a wide variety of resources. 

For example, \citet{Howard+2016PASP} previously analyzed a sample of 76 stars with RVs from HIRES/Keck and the Hamilton/Shane, which were observed by the California Planet Survey between 1987 and 2014. They selected stars with spectral types no earlier than F8 from the Exo-S, Exo-C, and WFIRST-AFTA study preliminary target lists \citep{Spergel+2013arXiv, Stapelfeldt+2014SPIE, Seager+2015SPIE}, and found that for a typical star they were sensitive to approximately Saturn-mass planets inside of $a = 1$ AU and Jupiter-mass planets inside of $a \approx 3$ AU. With our significantly expanded dataset and target list, we have improved the sensitivity for a typical star by at least factor of 2, reaching down to about 0.5 $\rm M_S$ at 1 AU and about 1 $\rm M_S$ at 3 AU. Moreover, in the best-case scenario, \citet{Howard+2016PASP} could detect Neptune-mass planets orbiting at $a = 3$ AU, while in the best-case scenario in this study we could reach planets at the same semi-major axis down to $\sim$10\,\Mearth. We note that while the target lists analyzed in this study and in \citet{Howard+2016PASP} are not identical, both target lists were compiled with similar criteria and have many targets in common. The improvement in search completeness in this work is predominantly driven by the increase in HIRES/Keck observations since 2014 and the inclusion of archival RVs from additional spectrographs.

While \citet{Howard+2016PASP} focused on stars in the northern hemisphere sky, a subsequent study by \citet{2023AJ....165..176L} conducted a similar analysis of southern stars. They considered archival Doppler measurements of 52 potential DI mission target stars from the NASA/NSF EPRV Working Group's list of high-priority stars for future DI missions \citep{Crass+2021arXiv}. Using RV data from HARPS/ESO, HIRES/Keck, UCLES/AAT, PFS/Magellan, and Levy/APF, \citet{2023AJ....165..176L} demonstrated a lack of sensitivity to Saturn-mass planets in the HZ for most of their targets. In this study, we have shown that a Saturn-mass planet could be detected in the HZ for over half of the targets. This improvement in sensitivity is again driven by the addition of archival RVs from more spectrographs in this study.

\begin{figure*}[t!]
    \centering
    \includegraphics[width=0.95\textwidth]{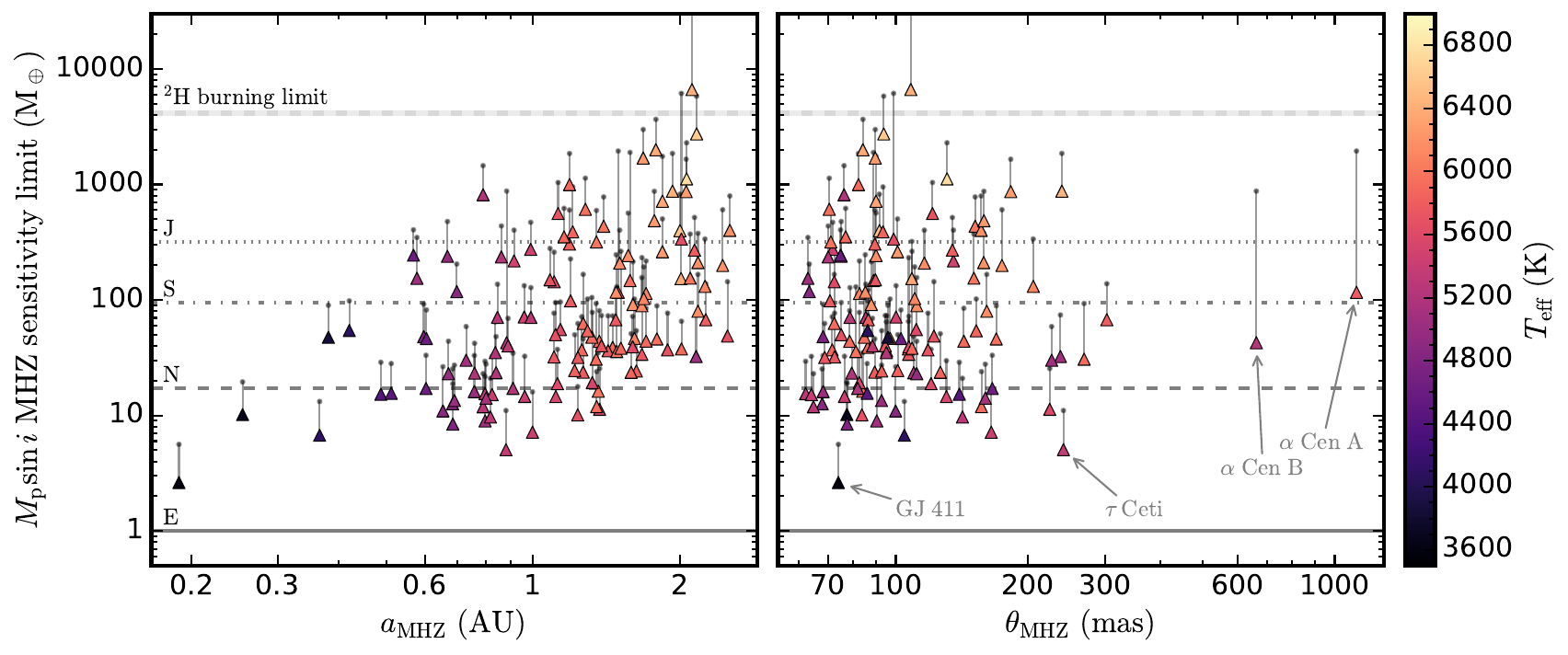}
    \caption{Lower $M_{\rm p} \sin i$ sensitivity limits in the middle of the conservative HZ \citep{Kasting+1993Icar, Kopparapu+2013ApJ} derived from the \clf\ lines shown in Figure \ref{fig:completeness_contours}. More conservative sensitivity estimates are shown by the black points and vertical lines, which have been derived from the \cln\ contours. The sensitivity limits are shown as a function of the physical separation (left) and maximum angular separation (right) of the middle of the conservative HZ defined in instellation space. The points are colored by stellar effective temperature. The sensitivity limit values shown here, in addition to sensitivity limits computed for the IHZ and OHZ are listed in Table \ref{tab:sensitivity_limits} in Appendix \ref{appx:figures_tables}.}
    \label{fig:sensitivity_limits}
\end{figure*}

\subsection{Habitable Zone RV Sensitivity Limits} \label{sec:discussion:limits}

Understanding the current sensitivity of RV observations of nearby stars is critical for the early development of the HWO target list and overall mission architecture. For the HWO exo-Earth survey in particular, we must know if potential target stars have disruptive giant planets in the HZ or if we can rule them out. Even planets as small as 5\,\Mearth\ in the HZ can lead to dynamically unstable conditions for any potentially Earth-like planets \citep{Kane+Fetherolf_2023AJ}. If a large fraction of otherwise viable HWO targets host disruptive HZ planets, more target stars that are farther away will need to be surveyed, requiring a larger aperture telescope with a more aggressive coronagraph inner working angle (IWA). Therefore, the more targets with dynamically nonviable HZs we can identify now, the easier it will be to implement engineering changes that will enable HWO to characterize a robust sample of potentially habitable planets. 

Using the results from the RV search completeness analysis (\secref{sec:analysis:injrec} and \secref{sec:results:completeness}), we assessed what planet masses we are sensitive to in the HZs of 120 potential HWO targets given the current state of RVs. From the \clf\ contours (Figure \ref{fig:completeness_contours}) we calculated the $M_\text{p} \sin i$ sensitivity limits at the inner edge, middle, and outer edge of the conservative habitable zone for each star, assuming the ``moist greenhouse'' and ``maximum greenhouse'' limits from classic 1D radiative-convective climate models \citep{Kasting+1993Icar, Kopparapu+2013ApJ}.

For Earth-like planets with H$_2$O- and CO$_2$-dominated atmospheres, \citet{Kopparapu+2013ApJ} derived the following parameterization for the HZ flux limits around main-sequence stars:
\begin{equation}
    S_\text{eff} = S_{\text{eff},\odot} + aT_* + bT_*^2 + cT_*^3 + dT_*^4
\end{equation}
where $S_{\text{eff},\odot}$ is the effective incident flux corresponding to a particular HZ definition for a solar twin (i.e., moist greenhouse or maximum greenhouse), $T_* \equiv T_\text{eff} - 5780$ K, and $a$, $b$, $c$, $d$ are empirically-determined coefficients. 

We calculated $S_{\text{eff}}$ at the middle of each star's HZ (MHZ) by averaging the $S_{\text{eff}}$ values at the inner edge of the HZ (IHZ; moist greenhouse limit) and the outer edge of the HZ (OHZ; maximum greenhouse limit). Using stellar properties from the SPORES-HWO Catalog \citep{Harada+2024ApJS}, the corresponding IHZ, MHZ, and OHZ distances were then computed from the stellar luminosity as
\begin{equation}
    a_{\rm HZ} = \sqrt{\frac{L_\star / {\rm L_\odot}}{S_{\rm eff} / {\rm S_\oplus}}}\,\text{AU}
\end{equation}

For each star, we then recorded the \clf\ sensitivity limits in both $M_\text{p} \sin i$ (\Mearth) and RV semi-amplitude $K$ (\ms) corresponding to the IHZ, MHZ, and OHZ distances. As a more conservative estimate of the sensitivity limits, we also recorded the \cln\ sensitivity limits (i.e., those corresponding to the search completeness contour of 90\% detection probability)\footnote{We caution that the \cln\ limits tend to be more sensitive to the number of injection trials than the \clf\ limits. In the limit of a very large number of injection trials, we expect the \cln\ limits to decrease slightly and become closer to \clf. Therefore, we default to using \clf\ in discussing mass sensitivity limits.}. The IHZ, MHZ, and OHZ distances, corresponding orbital periods, \clf\ sensitivity limits, and \cln\ sensitivity limits are listed for each star in Appendix \ref{appx:figures_tables} Table \ref{tab:sensitivity_limits}.

\begin{figure*}[t!]
    \centering
    \includegraphics[width=0.95\textwidth]{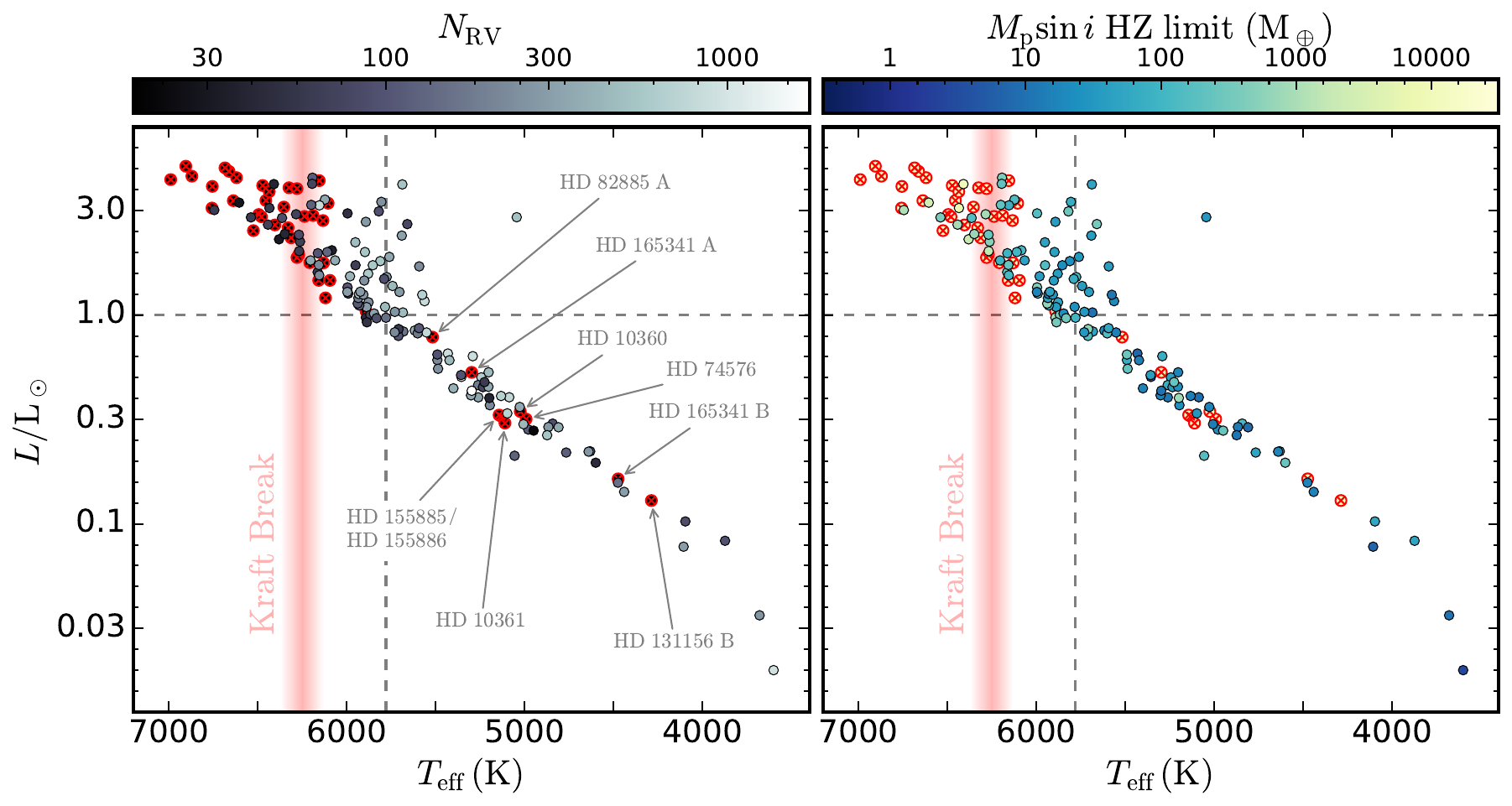}
    \caption{H-R diagrams of the 164 target stars, where color indicates the number of RV epochs (left) or the lower mass sensitivity limit in the MHZ (right). Stars with fewer than 20 epochs of RV observations are marked with a red ``$\otimes$'' in both panels. Solar values are shown by the thin dashed lines, and the approximate effective temperature corresponding to the Kraft Break \citep[$\sim$6250\,K;][]{Kraft_1967ApJ} is indicated by the transparent red line. The stellar properties plotted here were taken from the SPORES-HWO Catalog \citep{Harada+2024ApJS}.}
    \label{fig:RV_HRD}
\end{figure*}

Figure \ref{fig:sensitivity_limits} shows the \clf\ and \cln\ sensitivity limits for each star as a function of the MHZ distance and maximum angular separation. Each marker indicates the minimum $M_\text{p} \sin i$ that could be detected for a circular orbit with 50\% probability in a given system's MHZ, and the corresponding point connected by a vertical line indicates the minimum $M_\text{p} \sin i$ that could be detected with 90\% probability. The median sensitivity limits for the IHZ, MHZ, and OHZ are 42 \Mearth, 48 \Mearth, and 61 \Mearth, respectively (or, in terms of RV semi-amplitude, 3.9 \ms, 4.0 \ms, and 4.3 \ms). For the more conservative estimates (\cln), the median sensitivity limits are 90 \Mearth, 104 \Mearth, and 144 \Mearth\ for the IHZ, MHZ, and OHZ, respectively (or 8.1 \ms, 8.2 \ms, and 9.0 \ms\ in RV semi-amplitude). 

Ignoring dynamical constraints and factors of $\sin i$, the RVs analyzed in this work can rule out non-eccentric Jupiter-mass planets in the MHZ for 95 of the 114 systems (83.3\%) without known, undisputed planets in their conservative HZs (we discuss systems with known HZ planets in \secref{sec:discussion:architectures}). We can similarly rule out Saturn-mass and Neptune-mass planets in the MHZ for 71 (62.3\%) and 24 (21.1\%) systems, respectively\footnote{Taking the more conservative \cln\ sensitivity limits in the MHZ, we can rule out Jupiter-mass planets for 76 systems, Saturn-mass planets for 55 systems, and Neptune-mass planets for 3 systems.}. We are unable to rule out Earth-mass or smaller planets in any of the stars' habitable zones. Considering the 44 additional targets that were not analyzed in this work due to insufficient data, this means that up to 63 (38.4\%) of the 164 potential HWO target stars could have hidden Jovian planets in their HZs. Moreover, up to 87 (53\%) could have undetected Saturns and up to 134 (81.7\%) could have undetected Neptunes in their HZs.

Figure \ref{fig:RV_HRD} shows HR-diagrams of the target stars, demonstrating how the number of RV epochs and MHZ sensitivity limits are related to stellar properties. It is clear that stars hotter than $\sim$6250 K and stars with stellar companions tend to have far fewer RV observations than cooler stars, and therefore tend to have worse (or absent) constraints on planets in the HZ. Cooler stars, in addition to being more amenable to high-precision RV observations, have the added benefit of closer-in HZs where planets are easier to detect. We discuss how stellar properties relate to the sensitivity limits in more detail in \secref{sec:discussion:neglectedstars}. In addition to the stellar properties and the number of RV observations, the RV sensitivity can also depend on observing strategy parameters such as the survey cadence and baseline. For example, Figure \ref{fig:correlations} in Appendix \ref{appx:figures_tables} shows the correlations between the number of RV measurements, the total observing duration, and the MHZ sensitivity.

The two systems whose habitable zone planets we would be most sensitive to are HD 10700 ($\tau$ Ceti) and HD 95735 (Lalande 21185; GJ 411). The star $\tau$ Ceti is a Sun-like G8V star which has been studied extensively over the last several decades (see \secref{sec:tauCeti})---with 1,692 nights of RVs, this was the most well-observed star in the sample. In $\tau$ Ceti's habitable zone (0.72-1.25 AU) we can rule out planets more massive than 4.4-6.5 \Mearth\ (or, more conservatively, planets more massive than 9.5-13.9 \Mearth). Lalande 21185, an M2V star, is the coolest star in the sample and has also been extensively observed (see \secref{sec:Lalande21185}). The fact that Lalande 21185 is a cool star means that its HZ is closer in, where RV observations are more sensitive to lower-mass planets. With 921 RV epochs, including 297 from CARMENES, we are sensitive to planets more massive than 2.3-3.6 \Mearth\ (more conservatively, 4.8-7.6 \Mearth) in Lalande 21185's habitable zone (0.15-0.28 AU).
 
As stated at the beginning of this section, these companion mass sensitivity limits are crucial for informing trade studies of HWO's architecture and guiding future precursor science observations of potential target stars for HWO's exo-Earth survey. Additionally, these mass constraints provide key priors on any HZ planets that HWO will discover in the future, significantly aiding the atmospheric modeling required to interpret spectra that HWO measures \citep[e.g.,][]{Damiano+2025AJ....169...97D}. While providing specific recommendations for future DI missions like HWO is beyond the scope of this work, we encourage the community to apply the sensitivity limits provided in Table \ref{tab:sensitivity_limits} to coordinate future observing efforts to better understand the planetary systems of potential HWO targets.

\subsection{Neglected Target Stars}\label{sec:discussion:neglectedstars}

While the majority of target stars ($\sim$73\%) had at least 20 epochs of public RV data, we emphasize that 21 stars had $0 < N_{\rm RV} < 20$ (Figure \ref{fig:obs_histogram}), and we were unable to obtain any public RVs for 23 stars (see Appendix \ref{appx:figures_tables} Table \ref{tab:tab_neglected_stars}). We refer to these 44 under-observed stars as the neglected sample\footnote{We note that the middle cluster of ``moderately'' observed stars in Figure \ref{fig:obs_histogram} would also benefit from additional RV observations.}. All of these stars fall into at least one of the following categories: (1) hot stars, (2) highly active stars, and (3) binary stars. We speculate that previous RV surveys tended to exclude or de-prioritize stars belonging to these categories because they were less favorable targets for the science goals they were aiming to achieve. We note there was also at least one case of a star having been observed, but those RVs were not accessible apart from a single figure in the discovery paper (HD 33564; \secref{sec:HD33564}).

\begin{figure}[t!]
    \centering
    \includegraphics[width=0.49\textwidth]{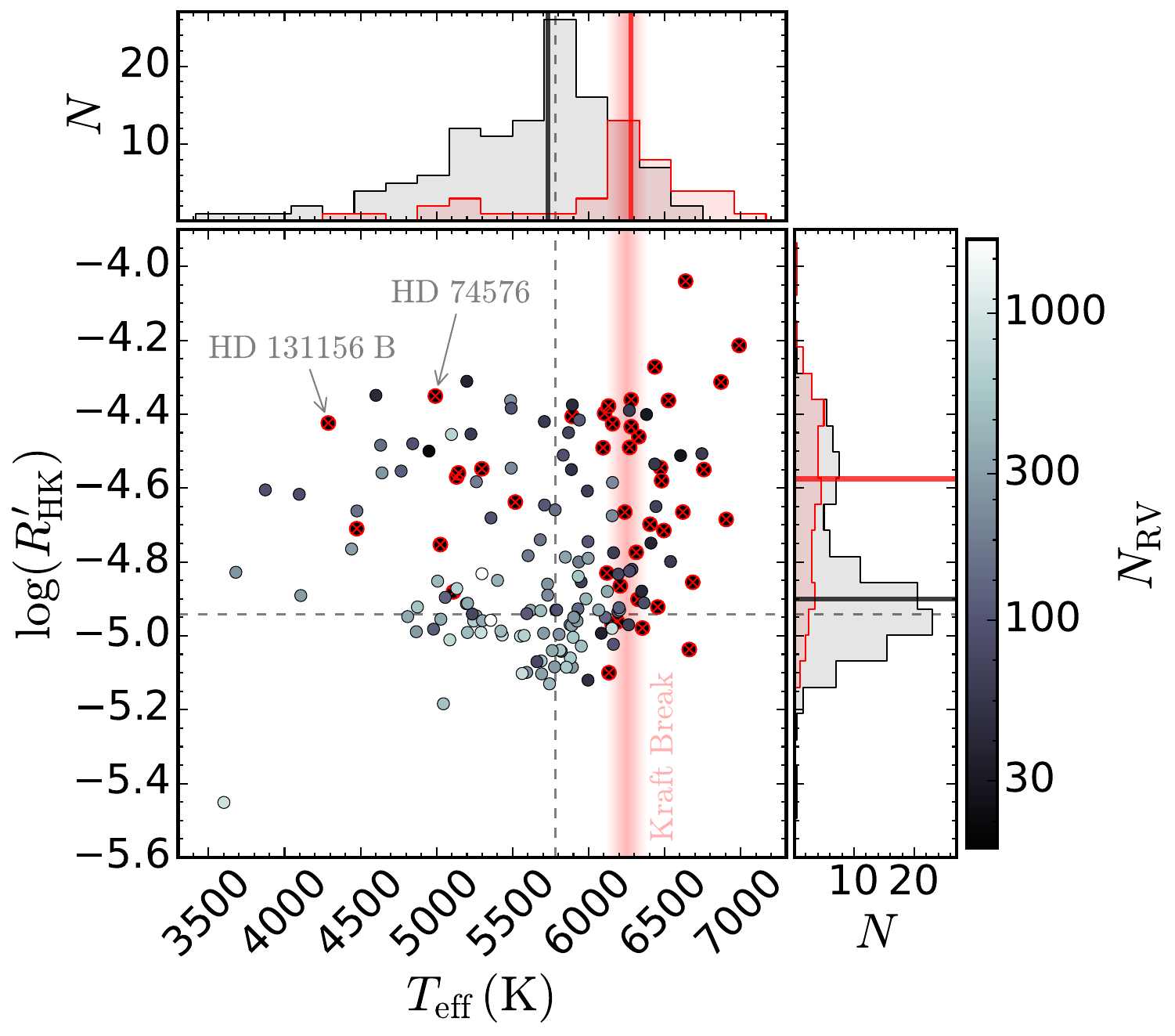} 
    \caption{Scatter plot and histograms of stellar effective temperature and $\log R^\prime_{\rm HK}$ \citep[from][]{Harada+2024ApJS}. The color scale corresponds to the number of RV observation epochs. Targets with fewer than 20 RV epochs are marked by a red ``$\otimes$'' symbol. The shaded 1D histograms above and to the right distinguish between targets with at least 20 RV epochs (black) and those with fewer than 20 RV epochs (red). The median 1D histogram values are indicated by solid lines. The approximate location of the Kraft Break temperature is shown by the transparent red line, and solar values are shown by the dashed lines \citep[$\log R^\prime_{\rm HK, \odot} = -4.943$;][]{Egeland+2017ApJ}.}
    \label{fig:RV_rhk_vs_teff}
\end{figure}

In Figure \ref{fig:RV_HRD}, the left HR-diagram shows that hot stars ($T_{\rm eff} \gtrsim 6250$ K) tend to have few or zero RV observations, which leads to worse or non-existent mass sensitivity limits in the right HR-diagram. Cooler stars with few or no observations tend to be in multi-star systems (e.g., 36 Oph, 70 Oph, p Eri), or have unusually high levels of stellar activity, indicated by $R_{\rm HK}^\prime$ (e.g., HD 74576). Figure \ref{fig:RV_rhk_vs_teff} shows the joint distribution of $R_{\rm HK}^\prime$ and $T_{\rm eff}$, as well as the number of RV epochs for each target star. Stars with at least 20 nights of RVs have a median effective temperature of $5731$ K and median $R_{\rm HK}^\prime$ index of $-4.9$, which are quite similar to solar values. On the other hand, stars with fewer than 20 RVs have a higher median effective temperature of $6280$ K and a higher median $R_{\rm HK}^\prime$ index of $-4.575$. 

We have already discussed how high levels of stellar activity can complicate RV searches for planets (\secref{sec:analysis:activity}), which partly explains why stars with high $R_{\rm HK}^\prime$ have not been studied as closely as less active stars. Certain binary star systems can also be challenging for RV surveys because of slit placement. For binary systems in which both stars are similar in brightness, special attention is required to isolate the spectrum of one component of the star system from the other to avoid blending. 

It is also common that RV surveys exclude stars with effective temperatures near or above the Kraft Break \citep[$\sim$6250 K;][]{Kraft_1967ApJ}, as these hot stars tend to rotate faster, resulting in more strongly Doppler-broadened spectral line profiles and hence lower precision RV measurements \citep[e.g.,][]{Beatty+2015PASP..127.1240B}. We note that the precise definition of the Kraft Break varies in the literature\footnote{The original \citet{Kraft_1967ApJ} paper does not actually provide a specific value of $T_{\rm eff}$, despite that paper is often cited in the literature for a specific effective temperature.} and is not strictly a single value \citep[e.g.,][]{Beyer+2024ApJ...973...28B}, but it is typically quoted at $T_{\rm Kraft} = 6250\,{\rm K}$ \citep[e.g.,][]{Santos+2021ApJS, Avallone+2022ApJ}. This is the value we adopted for Figures \ref{fig:RV_HRD} and \ref{fig:RV_rhk_vs_teff}. The median number of observations for stars hotter than this definition of the Kraft Break is 14 RVs, while the median number of observations for stars cooler than the Kraft Break is 221 RVs. 

\begin{figure*}[t!]
    \centering
    \includegraphics[width=0.95\textwidth]{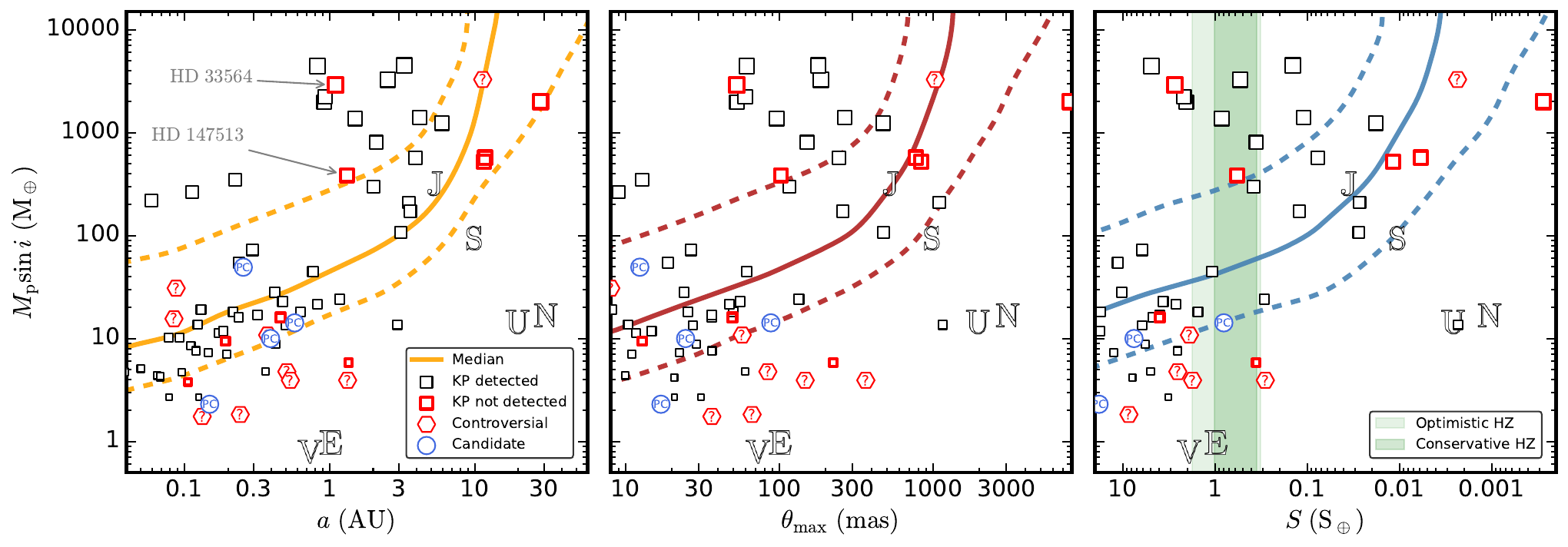}
    \caption{Known exoplanets in the target sample as a function of $M_{\rm p} \sin i$, semi-major axis, maximum angular separation (assuming circular orbits), and incident stellar flux. The sizes of the squares also correspond to the relative $M_{\rm p} \sin i$. The median \clf\ limits (Figure \ref{fig:completeness_contours}) are shown by the solid colored lines, and the middle 68\% of \clf\ limits are bounded by the dashed colored lines. Previously-detected planets that were not detected in this study are indicated by red squares, and the locations of the Solar System planets are labeled by the first letter of each planet. The green shaded regions in the right panel show the conservative and optimistic habitable zone limits in instellation space.}
    \label{fig:known_planets}
\end{figure*}

While the neglected sample were not included in previous RV surveys, we advocate for future RV programs to include these stars in coordinated multi-year observing campaigns. We also advocate for additional community-coordinated RV observations of the stars with only ``moderate'' observing history (i.e., those in the middle cluster of Figure \ref{fig:obs_histogram}). Even if observations of these stars are not precise enough to detect the lowest-mass planets, long-term RV monitoring will still be useful for constraining massive long-period companions and measuring stellar magnetic activity cycles. 

If we continue to only monitor the ``best'' RV targets, we will not only miss out on potentially new and interesting science, but also critical precursor limits on the dynamical viability of the HZs of potential HWO targets. Having no prior knowledge of the planetary systems around roughly a quarter of the HWO target stars would be a significant risk to the mission. Moreover, because $\sim$\ms\ precision may not be possible for many of the neglected stars, they are ideal targets for lower-precision instruments that are less highly oversubscribed. While state-of-the-art EPRV spectrographs are necessary to achieve 10-30 ${\rm cm\,s}^{-1}$ precision for detecting very small planets \citep[e.g., ESPRESSO, KPF, EXPRES, NEID;][]{2007AA...462..769P, Gibson+2016SPIE, Jurgenson+2016SPIE, Schwab+2016SPIE}, such high precision is not needed to achieve the science goals for the neglected sample stated here.

\subsection{Architectures of Potential Exo-Earth Hosting Systems} \label{sec:discussion:architectures}

Although the defining goal of HWO is to detect and characterize exo-Earths as living worlds, HWO is being designed to be a versatile facility that will revolutionize many aspects of astrophysics. In exoplanetary science, HWO will give unprecedented insights into the diversity of planetary system architectures and help to place the Solar System in context. For example, the exquisite sensitivity, IWA, and contrast requirements for imaging Earth-analogs will make HWO a premier facility for spectroscopically characterizing hundreds of cool outer planets with a wide range of masses, orbital characteristics, atmospheric compositions, dynamical histories, and other properties. Opening this new parameter space will change our understanding of cool planet demographics, planet formation, and dynamical evolution. Because HWO will also discover Earth-analog exoplanets in these systems, it will enable us to start addressing questions like ``Is the particular architecture and dynamical history of the Solar System a prerequisite for habitability?'' or ``How common are Earth-like planets in systems with architectures significantly different from our Solar System?''

RV surveys have thus far already demonstrated a rich diversity of planets and planetary system architectures among potential targets stars for HWO. However, the knowledge we can gain of these systems from RVs is still incomplete. Not only are many potential HWO targets lacking RV observations (\secref{sec:discussion:neglectedstars}), but there are many cases of disputed planetary signals or new RV planet candidates that need confirmation. 

Figure \ref{fig:known_planets} shows the $M_{\rm p} \sin i$ of the 70 known planets in the sample (in 30 different systems) as a function of $a$, $\theta_{\rm max}$, and $S$. We detected and updated the properties of 51 of these known planets in addition to 4 planet candidates (and 2 known stellar companions) with the RVs analyzed in this study (\secref{sec:results:rvsearch}). We note that at least 9 previous planet claims have been disputed (labeled as controversial in Figure \ref{fig:known_planets}). The disputed planets and the 10 other previously-known planets that were not detected in this work are summarized in Appendix \ref{appx:figures_tables} Table \ref{tab:undetected_planet_claims}.

The known planets in the sample represent a wide range of masses, semi-major axes, and radiation environments. Although knowing the true masses of these planets requires knowledge of the orbital inclination (which RVs alone are not sensitive to), if we assume that most systems have inclinations close to $i = 90^\circ$, then sub-Neptunes ($M_p < 17\,{\rm M_\oplus}$) account for 28 (46\%) of the undisputed known planets\footnote{Though most known planets do not have measured inclinations, we can estimate how the $\sin i$ degeneracy effects true planet masses in a statistical sense. Assuming orbits are oriented randomly, the inclination of any given system has a probability density function of $dp/di = \sin i$. Integrating $dp/di$ over $30^\circ < i \leq 90^\circ$, we would expect about 86.6\% of systems to have inclinations within that range. Therefore, the $\sin i$ degeneracy correction factor should be $<2$ ($=1/\sin30^\circ$) for about 86.6\% of systems shown in Figure \ref{fig:known_planets}.}, of which 15 are super-Earths ($2\,{\rm M_\oplus} < M_p < 8\,{\rm M_\oplus}$). Those planets receive between 0.0023 and 2665 times as much stellar flux as the Earth and their semi-major axes range 0.015 AU to 2.94 AU.

The next largest group of planets is super-Jupiters ($M_p > 318\,{\rm M_\oplus}$), which account for 17 (28\%) of the undisputed planets. These planets experience $1.9\times10^{-6}$ to 36 times as much stellar flux as the Earth and orbit their stars at distances ranging from 0.225 AU to 773 AU. Super-Neptunes/sub-Saturns ($17\,{\rm M_\oplus} < M_p < 95\,{\rm M_\oplus}$; $N=10$) and Jovian planets ($95\,{\rm M_\oplus} < M_p < 318\,{\rm M_\oplus}$; $N=6$) make up the rest of the undisputed planets. The super-Neptunes/sub-Saturns receive 0.29 to 68.5 times the flux received by Earth and orbit at semi-major axes of 0.13 AU to 1.18 AU, while the Jovians receive 0.027 to 956 times the flux at Earth and orbit their stars at 0.059 AU to 3.6 AU. Most of the controversial or disputed planets (7 out of 9) are smaller than Neptune, and the smallest ($\tau$ Ceti g; \secref{sec:tauCeti}) has a putative mass of 1.75 \Mearth\ \citep{2017AJ....154..135F}.

\begin{figure}[t!]
    \centering
    \includegraphics[width=0.45\textwidth]{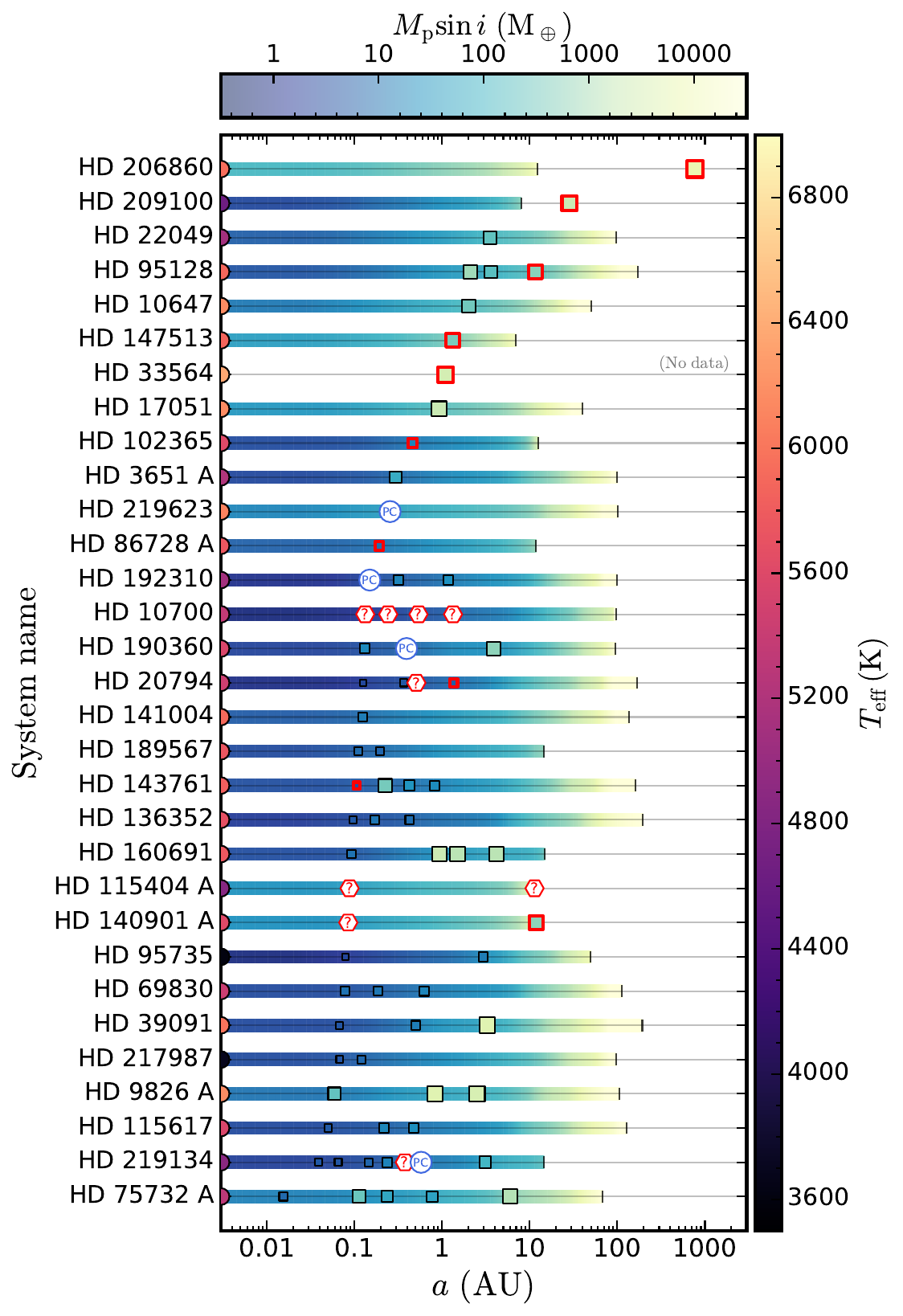}
    \caption{Architectures of systems in our sample with at least one known planet or planet candidate. Both the color and size of each marker correspond to the planet $M_{\rm p} \sin i$. Markers that are outlined in red show the known planets that were not detected in this study. Disputed planet detections are shown as red ``?''s inscribed in hexagons, and planet candidates are indicated by blue circles with the text ``PC.'' Each colored horizontal line shows the minimum $M_{\rm p} \sin i$ sensitivity limit (from the \clf) as a function of semi-major axis, and the colored circles on the left-hand side show stellar effective temperature. Note that HD 33564 has no sensitivity limits because no public RV observations were available.}
    \label{fig:system_architectures}
\end{figure}

Figure \ref{fig:system_architectures} shows the architectures of all target systems with at least one known planet or planet candidate, in addition to the \clf\ sensitivity limits as a function of semi-major axis. More than half of the target stars (16 out of 30) with known planets are multi-planet systems with at least 2 undisputed planets. At least 11 of those systems have a planet multiplicity of 3 or more, and the highest-multiplicity systems each have at least 5 planets---HD~219134 (\secref{sec:hd219134}) and 55 Cancri (\secref{sec:55Cnc}). None of the architectures of these systems are consistent with the Solar System.

While none of the known planets are exact analogs of any of the Solar System planets, an interesting exercise is to see whether it would be possible to detect Solar System analogs in any of these systems given our sensitivity limits. Using the \clf\ contours in Figure \ref{fig:completeness_contours}, we determined that we could detect a Jupiter analog (in terms of semi-major axis) in about 69 of the 120 systems (57.5\%), and a Saturn analog in about 7 systems (5.8\%). No other Solar System planets would be detectable. If we assume the more conservative sensitivity limits (i.e., using the \cln\ instead of the \clf), then only 30 Jupiters and no Saturns would be detected. 

If instead we define the Solar System analogs in terms of instellation flux to account for differences in stellar spectral across the sample, we find similar but perhaps more optimistic results. In terms of $S$, we could detect a Jupiter in about 68 systems (56.7\%) and a Saturn in about 18 systems (15\%). Using the more conservative limits, we could detect Jupiter in about 45 systems (37.5\%) and Saturn in about 4 systems (3\%). Again, no Solar System planets smaller than Saturn would be detectable. 

Finally, we note that 6 systems have at least one undisputed planet in the conservative HZ (assuming circular orbits), and 8 have at least one undisputed planet in the optimistic HZ \citep[the ``early Mars'' and ``late Venus'' limits][]{Kasting+1993Icar, Kopparapu+2013ApJ}. The HZ planets include HD 69830 d (\secref{sec:hd69830}), 55 Cnc f (\secref{sec:55Cnc}), HD 147513 b (\secref{sec:hd147513}), 82 Eri f (\secref{sec:82Eri}), $\mu$ Ara b (\secref{sec:muAra}), q1 Eri b (\secref{sec:q1Eri}), 47 UMa b (\secref{sec:47Uma}), and $\upsilon$ And d (\secref{sec:upsAnd}), as well as the planet candidate HD 219134.04 (\secref{sec:hd219134}). All of these planets are more massive than the Earth, ranging in $M_{\rm p} \sin i$ from 5.8 \Mearth\ to 1370 \Mearth, and therefore decrease the dynamical viability of the HZ for hosting potentially habitable Earth-like planets. 

While dynamical simulations of these systems are beyond the scope of this paper, previous dynamical studies of the 164 stars considered in this work found that the HZs of 11 planet-hosting are $<$50\% dynamically viable due to giant planets whose orbits pass near or through the HZ \citep{Kane+2024AJ....168..195K}. The impact of stellar and planetary uncertainties on the dynamical influence of HZ perturbers is also discussed in \citet{Kane+2024AJ....168..279K}. Alarmingly, as discussed in \secref{sec:discussion:limits}, as many as 63 (38.4\%) of the 164 EMSL target stars could have undetected Jovian planets in their HZs. Saturn- and Neptune-mass planets, which also negatively impact the dynamical viability of the HZ, could also be hiding in up to 53\% and 82\% of the stars' HZs given the current RV sensitivity limits. This poses a significant challenge to HWO that must be addressed through additional precursor RV observations of potential exoplanet DI target stars. Such observational efforts must begin as soon as possible to build up the RV sensitivity to giant HZ planets before a final architecture for HWO is decided.

\section{Conclusions} \label{sec:conclusion}

In this work, we have presented a uniform analysis of over 35 years of archival RV observations of 120 potential target stars for future exoplanet DI missions such as HWO. We analyzed over 150,000 RV measurements to search for planetary signals, refine the orbits and properties of known companions, identify false positives, and determine sensitivity limits on companion masses. Our main conclusions are as follows.

\begin{enumerate}
    \item{
        Over 35,000 epochs of high-precision ($\lesssim$10 \ms) RV observations (153,902 total unbinned measurements) have been made of 141 potential HWO target stars from the SPORES-HWO Catalog \citep{Harada+2024ApJS} and ExEP Mission Star List \citep{Mamajek+Stapelfeldt_2024}. Public RV observations of these stars span over 36 years, going as far back as 1987, and were observed by 23 different high-resolution spectrographs located across the globe. To obtain the most complete and accurate dataset possible, we created the SPORES-HWO RV Hunters community science project to redundantly collect, organize, and vet public RVs and stellar activity indicator data from 40 different public sources. 
    }
    \item{
        Analyzing the periodograms of the RVs and stellar activity data for 120 stars (with $N_{\rm RV} \geq 20$), we identified and modeled 109 periodic RV signals in 49 unique systems. We associate 51 signals with previously-known planets and 2 signals with known stellar companions, hence providing revised orbital parameters for those companions. Of the other signals, 26 are associated with known stellar activity and 19 are likely systematic false alarms. Additional observations or analyses are needed to confirm the nature of the remaining 11 signals, of which 4 are likely planet candidates. We also fit either linear or quadratic trends to the RVs in 74 systems, which may be caused by known stellar binaries or indicate the presence of unseen long-period companions.
    }
    \item{
        We derived sensitivity limits on companion masses for 120 stars from the results of planet injection-recovery simulations. The lower mass sensitivity limit for a typical star in the sample is approximately 50 \Mearth\ for a 1 AU circular orbit. For most of the sample (the middle 68\% of stars), the lower mass sensitivity limit at 1 AU is between about 1 Neptune mass (17 \Mearth) and 1 Jupiter mass (318 \Mearth). In the best case, a $\sim$6 \Mearth\ planet would be detectable at 1 AU around a quiet host star, while in the worst case, even planets larger than 10 $\rm M_J$ would not be detectable at 1 AU. These mass limits are a significant improvement over limits derived in previous studies \citep[e.g.,][]{Howard+2016PASP, 2023AJ....165..176L} in terms of the typical sensitivity achieved, the number of targets analyzed, and the scope of RV and stellar activity data.
    }
    \item{
        The mass sensitivity limits for the conservative HZ, which depends on stellar spectral type \citep{Kasting+1993Icar, Kopparapu+2013ApJ}, are about 42 \Mearth\ at the IHZ and 61 \Mearth\ at the OHZ for a typical star in the sample. Including the stars that did not have enough observations to be analyzed, the HZ sensitivity limits suggest that up to 63 (38.4\%) of the 164 potential HWO target stars on the EMSL \citep{Mamajek+Stapelfeldt_2024} could be hiding undetected disruptive Jovian planets in their HZs. The outlook is worse for lower-mass HZ perturbers. We cannot rule out Saturn-mass planets in the HZs of up to 87 (53\%) of the EMSL targets, or Neptune-mass planets in the HZs of up to 134 (81.7\%) of the targets.
    }
    \item{
        While some of the EMSL target stars have already been thoroughly studied by past RV surveys (see Figure \ref{fig:obs_histogram}), as many as 21 stars have fewer than 20 RV observations, and as many as 23 have no publicly-available RV data. These neglected stars are typically either hot (near or above the Kraft Break), highly active, or belong to multi-star systems. Because stellar noise is likely the limiting factor of RV precision for these targets, we advocate for community-coordinated observing campaigns to monitor the Doppler motions of these stars using moderate- to high- precision RV facilities ($\sigma_{\rm RV}\gtrsim1$ \ms). Such instruments are well suited for the long-baseline observations required to measure giant planet masses in the HZ, and are far less likely to be oversubscribed than EPRV resources.
    }
    \item{
        We identified 6 RV signals that are either planet candidates or sources requiring confirmation. During the revision stage of this paper, one of these candidates (HD 190360 d; \secref{sec:hd190360}) was confirmed to be planetary in nature \citep{Giovinazzi+2025arXiv250818169G}. Further observations and analysis are needed to determine the origin of the putative signals in HD 192310, $\tau$ PsA, Lacaille 9352, HD 219134, and HD 219623 (\secref{sec:hd192310}-\ref{sec:hd219623}).
    }
    \item{
        While 70 planets have previously been reported in 30 of the target systems, we note that at least 9 reported planets have been disputed or were not detailed in the original discovery papers (Table \ref{tab:undetected_planet_claims}). We advise caution in assuming that these are in fact bona fide planet detections, and recommend follow-up studies to properly classify the disputed signals. This is critical because some of the disputed planets have orbits in or near the HZ, which could preclude those systems from hosting Earth-analogs if the planets turn out to be real.
    }
    \item{
        We emphasize that HD 165499 ($\iota$ Pav; \secref{app:sec:hd165499}) and HD 212330 A (\secref{app:sec:hd212330}) have unresolved stellar companions at $<1^{\prime\prime}$ of angular separation \citep[e.g.,][]{2023AA...674A.114B}. However, $\iota$ Pav was not flagged as a binary star in the EMSL, and HD 212330 A was listed in the EMSL with inaccurate WDS companion properties. In fact, neither system should have been included in the EMSL based on the criteria for binary stars set by \citet{Mamajek+Stapelfeldt_2024}, as they excluded any binary stars separated by $<$3$^{\prime\prime}$. We therefore advise caution in including these systems in future studies of potential HWO targets.
    }
    \item{
        Lastly, we have created a set of 120 human-readable PDF \LaTeX\ summary reports for each system we analyzed. Each file contains the \rvsearch\ summary and completeness plots, stellar activity GLS periodogram plots, and MCMC posterior corner plots, in addition to several tables summarizing the data, significant \rvsearch\ signals, MCMC posteriors, and derived planet properties. The full set of PDF \LaTeX\ reports for each system are published along with the complete figure sets, machine-readable tables, and all RV and stellar activity data on Zenodo\footnote{\url{https://doi.org/10.5281/zenodo.17058228}}.
    }
\end{enumerate}

The aggregation of decades of RV data and determination of HZ sensitivity limits presented in this work are an important step towards realizing a high-contrast DI space telescope capable of resolving Earth-like planets in nearby habitable zones. However, there is a long road ahead to the first direct detection of a habitable world beyond the Solar System, and understanding the planetary system architectures of potential exo-Earth hosts is just one important element of the precursor science needed to make HWO successful.

We make two final recommendations based on the findings of this work. First, we recommend the creation and ongoing maintenance of a public repository for RV measurements of future exoplanet DI mission targets. This would be a extremely powerful resource for the exoplanet community to coordinate new RV observations (whether related to HWO precursor science or not) and utilize limited observing resources as efficiently as possible. We encourage observers to publish and document their RV datasets, especially for targets with non-detections, and add them to database.

Second, we advocate for community-coordinated RV observations of the neglected sample of potential HWO target stars. All of the data and analysis products presented in this work are available to download and are intended to provided guidance for future observing campaigns. Historically under-subscribed RV facilities will be especially valuable for precursor HWO studies because extremely high-precision measurements are not needed to rule out long-period giant planets in the HZs of these systems.

\bigskip

The authors thank the anonymous referee for constructive feedback that improved the quality of this manuscript. We thank Arvind Gupta for thoroughly reviewing this manuscript and providing useful feedback that substantially improved this work. We are grateful to the hundreds of scientists who proposed for and collected the observations that enabled this project. We respectfully acknowledge all the Indigenous Peoples whose sacred lands---upon which many of the world’s most influential astronomical observatories have been built---made these observations possible. We thank all of the following volunteers from the SPORES-HWO RV Hunters community project whose efforts to aggregate and document these observations from around the world made this research possible: Lubna Alrubayyi, Nouran Alyousif, Aaliyah-Tiffany Anderson, Jeremy Baden, Griffin Beckman, Paul Beroza, Kobe Bilstad, Diego Bonilla, Shep Brooke, Linus Buchholz, Kyle Chuon, Jose Farias-Ramirez, Celeste Gomez, Chase Graham, Paige Haworth, Sarah Jauregui, Krish Jhurani, Sean Johnson, Nathanael Kanter, Myra Khieu, Pranathi Kolla, Kaitlyn Le, Allison Lew, Melanie Linares, Leslie Lopez, Yugandhara Luthra, Brayden Matejka, Rosendo Medina-Uribe, Austin Mei, Trevor Nguyen, Raiki Omori, Arjun Patel, Halim Perez Melendez, Brandon Pestoni, Ava Phillips, Aakarsh Real, Colin Smith, Max Steiman, Chienyu Sun, Gavin Tamasi, Vivian Tamayo, Caitlin Tran, Alaina Tripp, David Trujillo, Flora Wang, Alana Waterman, Dylan Yamzon, and Bobby Yuan. 

We acknowledge support from the NASA Astrophysics Decadal Survey Precursor Science (ADSPS) program under Grant Number 80NSSC23K1476.
This research has made use of the Astrophysics Data System, funded by NASA under Cooperative Agreement 80NSSC25M7105. 
This research has made use of the SIMBAD database and the VizieR catalogue access tool, CDS, Strasbourg Astronomical Observatory, France (DOI: 10.26093/cds/vizier).
This publication makes use of The Data \& Analysis Center for Exoplanets (DACE), which is a facility based at the University of Geneva (CH) dedicated to extrasolar planets data visualization, exchange and analysis. DACE is a platform of the Swiss National Centre of Competence in Research (NCCR), federating the Swiss expertise in Exoplanet research. The DACE platform is available at \url{https://dace.unige.ch}.
This research has made use of the NASA Exoplanet Archive, which is operated by the California Institute of Technology, under contract with the National Aeronautics and Space Administration under the Exoplanet Exploration Program.
This research used the Savio computational cluster resource provided by the Berkeley Research Computing program at the University of California, Berkeley (supported by the UC Berkeley Chancellor, Vice Chancellor for Research, and Chief Information Officer).

We gratefully acknowledge the HARPS team and all the PIs and observers associated with the many ESO programs and data sets used in this work. This work is in part based on observations collected at the European Organization for Astronomical Research in the Southern Hemisphere under the ESO programs listed in the acknowledgments of \citet{2020AA...636A..74T}.
Some of the data presented herein were obtained at Keck Observatory, which is a private 501(c)3 non-profit organization operated as a scientific partnership among the California Institute of Technology, the University of California, and the National Aeronautics and Space Administration. The Observatory was made possible by the generous financial support of the W.\ M.\ Keck Foundation. We respectfully recognize and acknowledge the significant cultural role and reverence that the summit of Maunakea has always had within the Native Hawaiian community. We are most fortunate to have the opportunity to conduct observations from this sacred mountain belonging to the Native Hawaiian People. This research has also made use of the Keck Observatory Archive (KOA), which is operated by the W.\ M.\ Keck Observatory and the NASA Exoplanet Science Institute (NExScI), under contract with the National Aeronautics and Space Administration. We gratefully acknowledge the PIs and observers of many data sets that were obtained from the KOA \citep[for a complete list of names, please refer to][]{2017AJ....153..208B}.

C.K.H.\ acknowledges support from the National Science Foundation (NSF) Graduate Research Fellowship Program (GRFP) under Grant No.~DGE 2146752.
M.R.\ gratefully acknowledges support from Heising-Simons Grant \#2023-4478 and NASA Exoplanets Research Program NNH23ZDA001N-XRP (Grant No.~80NSSC24K0153).
N.W.T.\ was supported by an appointment with the NASA Postdoctoral Program at the NASA Goddard Space Flight Center, administered by Oak Ridge Associated Universities under contract with NASA.

\vspace{5mm}
\facilities{
    AAT (UCLES),
    ADS,
    APF (Levy),
    BRC (Savio),
    Calar Alto 3.5-m (CARMENES),
    DACE,
    ESO 3.6-m (CES-VLC, HARPS),
    Exoplanet Archive,
    HET (HRS),
    HJS (Tull),
    Keck I (HIRES),
    La Silla CAT (CES-LC),
    LDT (EXPRES),
    Leonhard Euler Telescope (CORALIE),
    Magellan II (PFS),
    OHP 1.93-m (ELODIE, SOPHIE),
    Shane 3-m (Hamilton),
    SMARTS (CHIRON, ES),
    Teide 1-m (SONG),
    Tillinghast 1.5-m (AFOE),
    TNG (HARPS-North),
    VizieR,
    VLT (ESPRESSO, UVES),
    WIYN (NEID)
}
\software{
    \texttt{Astropy} \citep{astropy_2013_AA, astropy_2018_AJ},
    \texttt{corner} \citep{corner},
    \texttt{emcee} \citep{emcee},
    \texttt{IPython} \citep{IPython},
    \texttt{matplotlib} \citep{matplotlib},
    \texttt{NumPy} \citep{numpy_2011, numpy_2020},
    \texttt{Pandas} \citep{pandas, mckinney-proc-scipy-2010},
    \texttt{RadVel} \citep{Fulton+2018PASP},
    \texttt{RVSearch} \citep{2021ApJS..255....8R},
    \texttt{SciPy} \citep{2020SciPy-NMeth}
}

\bibliography{main_R1}{}
\bibliographystyle{aasjournal}

\appendix


\section{Additional Figures \& Tables}\label{appx:figures_tables}

\begin{figure*}[th!]
    \centering
    \includegraphics[width=0.99\textwidth]{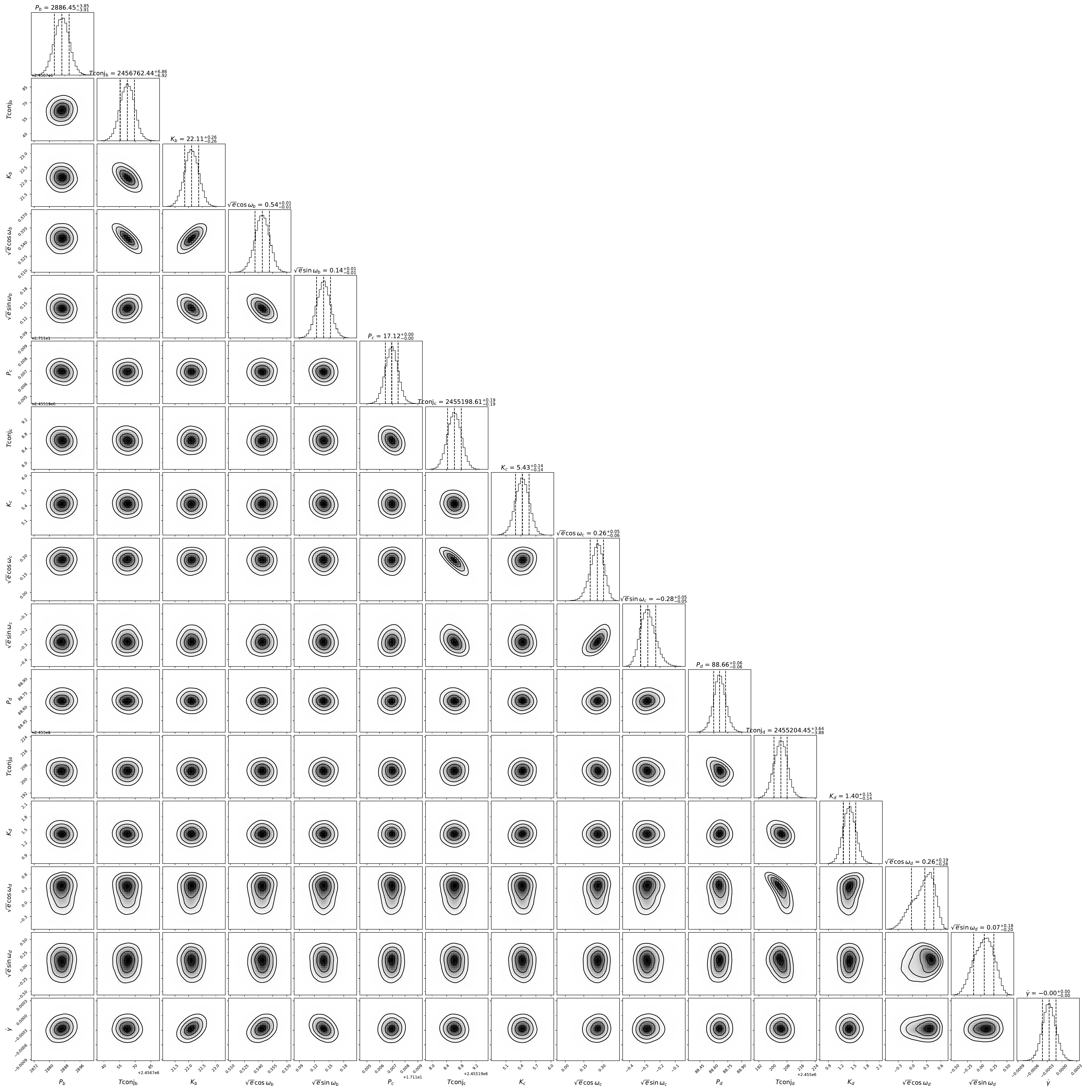}
    \caption{Example MCMC posterior corner plot demonstrating convergence of the 3-planet Keplerian model for HD~190360 shown in Figure \ref{fig:HD190360_summary}. The complete figure set (74 images) is available for all systems with RV detections and/or RV trends.}
    \label{fig:HD190360_corner_plot}
\end{figure*}

\begin{figure*}[th!]
    \centering
    \includegraphics[width=0.99\textwidth]{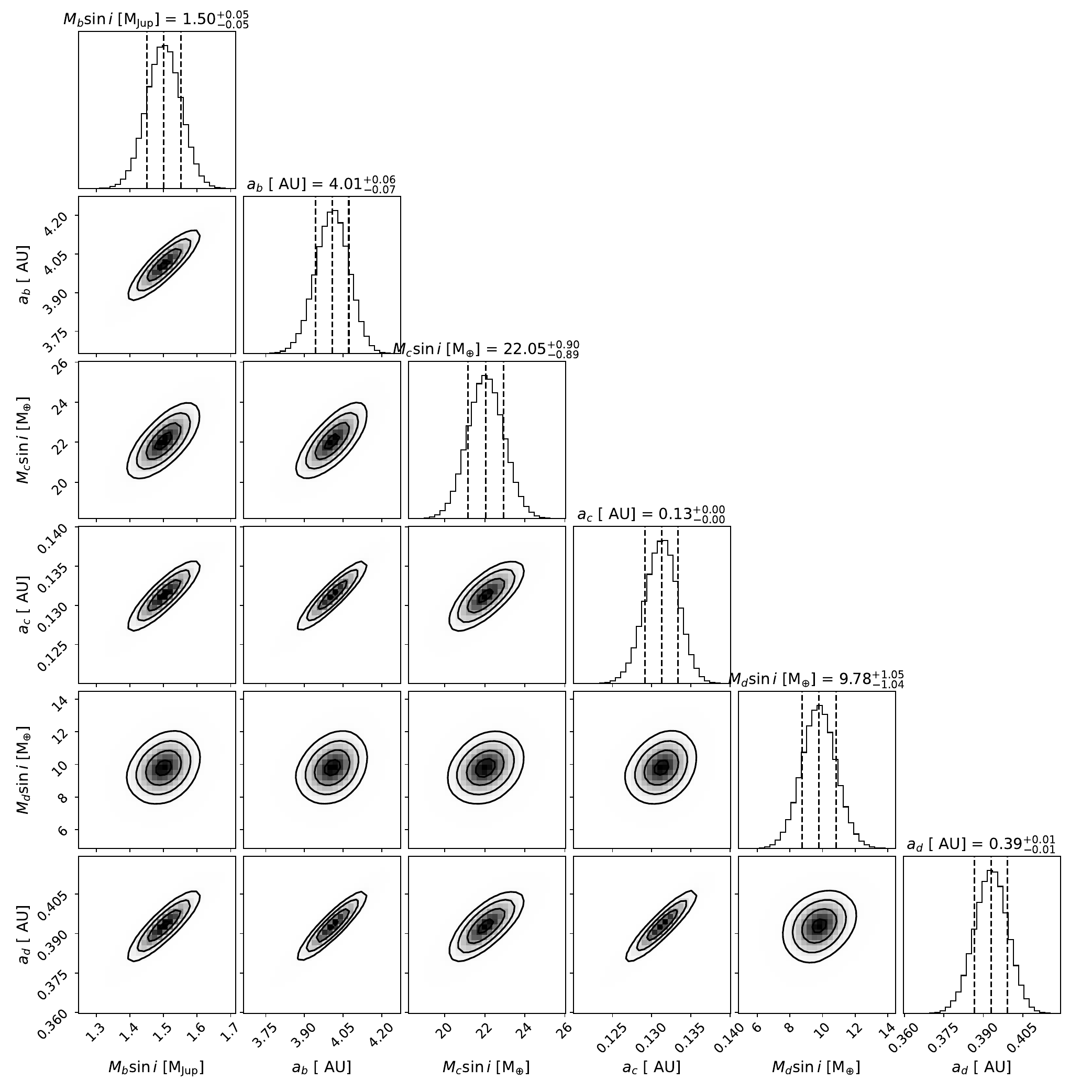}
    \caption{Same as Figure \ref{fig:HD190360_corner_plot}, but for the derived planet parameters $M_{\rm p} \sin i$ and $a$. The complete figure set (49 images) is available for all systems with RV detections.}
    \label{fig:HD190360_derived_plot}
\end{figure*}

\begin{figure*}[b!]
    \centering
    \includegraphics[width=0.9\textwidth]{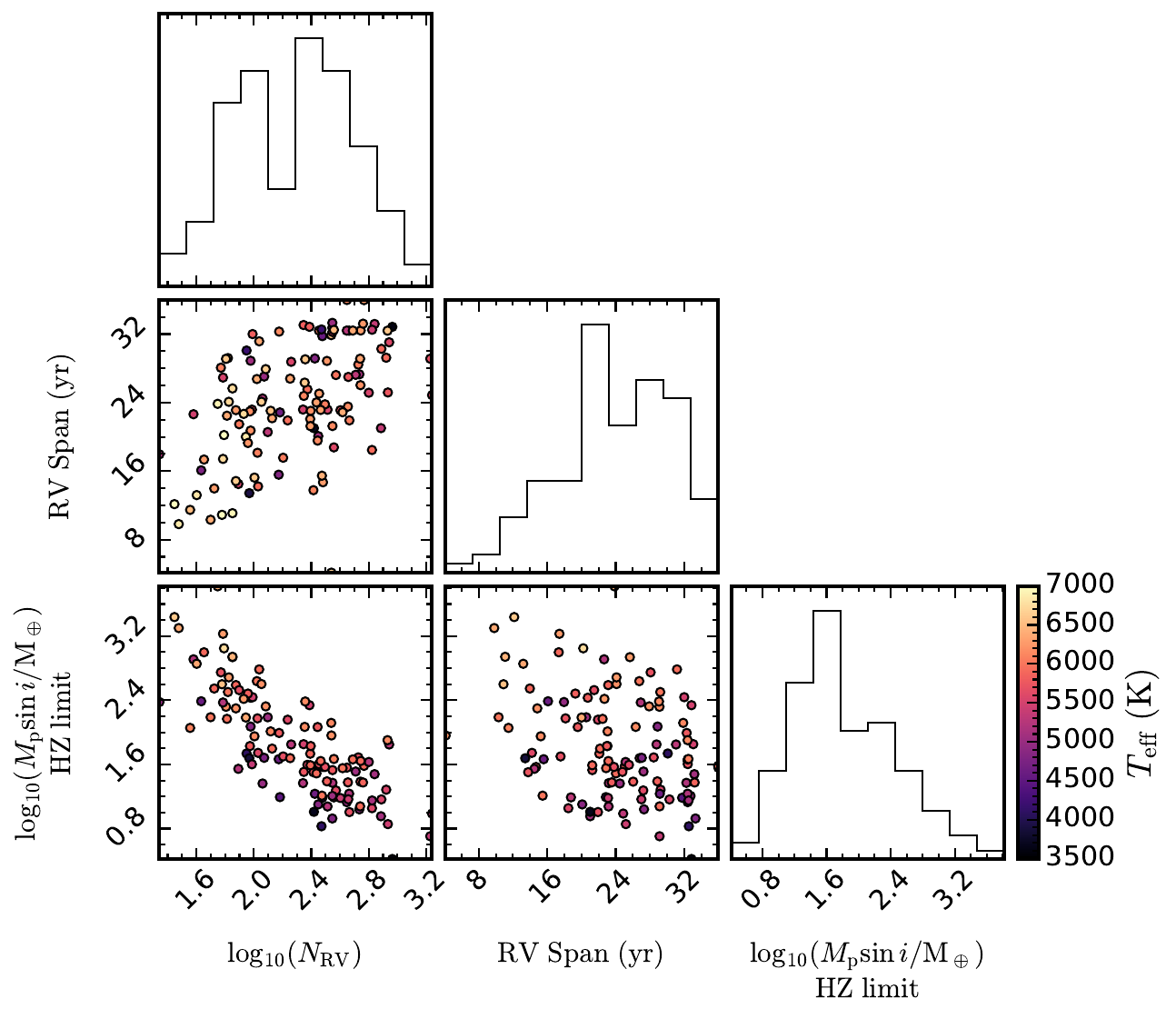}
    \caption{Comparison of the lower $M_{\rm p} \sin i$ sensitivity limits in the HZ with the number of RV observation epochs and the span of the RV observations for each target. The color scale corresponds to stellar effective temperature.}
    \label{fig:correlations}
\end{figure*}

Here we include large figures and tables that were not appropriate for the main text. Figures \ref{fig:HD190360_corner_plot} and \ref{fig:HD190360_derived_plot} show the MCMC posterior distributions of the RV model parameters for HD~190360 (shown in Figure \ref{fig:HD190360_summary} in the main text). Figure \ref{fig:correlations} shows the correlations between the minimum mass sensitivity limits in the MHZ, the number of RV observations, and the baseline of RV observations for each target. Table \ref{tab:targets} contains information for each star in our sample with at least one RV measurement, including stellar effective temperature, the number of stellar activity indicators measured, the number of epochs of RV observations, and the start date and time span of the RV observations. Table \ref{tab:observatories} details the observatories and instruments which collected the data used in this work. Table \ref{tab:example_rvs} shows a subset of the data used in this work (for HD~190360 as an example), including the RVs and stellar activity data. Table \ref{tab:rvsearch_signals} lists all of the Keplerian RV signals detected with \rvsearch, including best-fit parameters such as $P$, $K$, $e$, $\omega$, $a$, and $M_{\rm p} \sin i$, in addition to the detection FAP and the classification of each signal. Table \ref{tab:rvsearch_trends} lists trends detected by \rvsearch\ and the associated best-fit trend terms, $\dot{\gamma}$ and $\Ddot{\gamma}$. Table \ref{tab:undetected_planet_claims} shows the properties of the known planets (and disputed planets) in the sample which were not detected in this work. Table \ref{tab:sensitivity_limits} lists all of the derived RV sensitivity limits corresponding to the IHZ, MHZ, and OHZ in terms of $M_{\rm p} \sin i$ and RV semi-amplitude $K$. Lastly, Table \ref{tab:tab_neglected_stars} lists the target stars for which we found no public RV observations.

\newpage

\clearpage
\begin{longrotatetable}
\begin{deluxetable*}{lccp{4.5cm}ccllcp{1.5cm}}
\tablecolumns{10}
\tablecaption{Summary of targets}
\label{tab:targets}
\tablehead{
  \colhead{HD name} & 
  \colhead{Alt.\ name} & 
  \colhead{$T_{\rm eff}$ (K)} & 
  \colhead{Insts.} & 
  \colhead{\#\ RV} & 
  \colhead{\#\ SAI\tablenotemark{$^a$}} & 
  \colhead{First} & 
  \colhead{Last} & 
  \colhead{Span (yr)} & 
  \colhead{RV Refs.}
}
\startdata
HD 166 & V439 And & 5491 & HIRES-post, Hamilton, Levy & 98 & 53 & 1987-06-13 & 2019-06-13 & 32.0 & 19, 31 \\
HD 693 & 6 Cet & 6190 & HARPS-pre & 16 & 64 & 2005-08-19 & 2011-07-22 & 5.9 & 26 \\
HD 1581 & $\zeta$ Tuc & 5932 & CES-LC, CES-VLC, HARPS-post, HARPS-pre, UCLES & 550 & 1547 & 1992-11-03 & 2021-12-19 & 29.1 & 15, 26, 36 \\
HD 2151 & $\beta$ Hyi & 5806 & CES-LC, CES-VLC, HARPS-pre, UCLES & 249 & 232 & 1992-11-04 & 2015-11-19 & 23.0 & 15, 26, 36 \\
HD 3651 A & 54 Psc A & 5203 & HIRES-post, HIRES-pre, HRS, Hamilton, Levy, Tull & 461 & 168 & 1987-09-09 & 2020-02-01 & 32.4 & 19, 11, 31 \\
HD 4391 & GJ 1021 & 5887 & CES-LC, CES-VLC, HARPS-pre & 53 & 65 & 1993-09-16 & 2007-09-07 & 14.0 & 15, 26 \\
HD 4614 A & $\eta$ Cas A (Achird) & 5907 & HIRES-post, HIRES-pre, Hamilton, Levy, SOPHIE & 273 & 120 & 1995-09-09 & 2019-09-17 & 24.0 & 19, 16, 31, 7 \\
HD 4628 & GJ 33 & 5007 & HARPS-post, HARPS-pre, HIRES-post, Hamilton, Levy & 444 & 398 & 1987-09-08 & 2020-02-05 & 32.4 & 26, 19, 31 \\
\enddata
\tablecomments{Data have been binned to nightly intervals prior to counting. The stellar properties shown are taken from the SPORES-HWO Catalog \citep{Harada+2024ApJS}, which were inherited from the NASA ExEP Mission Star List \citep{Mamajek+Stapelfeldt_2024}.}
\tablenotetext{a}{SAI = Stellar Activity Indicators.}
\tablerefs{
(1) \citet{1999ApJ...526..916B};
(2) \citet{2001AA...374..675E};
(3) \citet{2001AA...375..205N};
(4) \citet{2003AA...410.1051N};
(5) \citet{2004AA...414..351N};
(6) \citet{2004AA...415..391M};
(7) \citet{2004PASP..116..693M};
(8) \citet{2006AJ....132.2206B};
(9) \citet{2006ApJ...646..505B};
(10) \citet{2007ApJ...654..625W};
(11) \citet{2009ApJS..182...97W};
(12) \citet{2010ApJ...715.1203M};
(13) \citet{2012ApJ...759...19E};
(14) \citet{2013AA...551A..90M};
(15) \citet{2013AA...552A..78Z};
(16) \citet{2014ApJS..210....5F};
(17) \citet{2015AA...584A..72M};
(18) \citet{2016ApJ...818...34E};
(19) \citet{2017AJ....153..208B};
(20) \citet{2018AA...619A...1B};
(21) \citet{2018AJ....155...24Z};
(22) \citet{2019AA...625A..17D};
(23) \citet{2019AA...630A..50B};
(24) \citet{2019AA...632A..37B};
(25) \citet{2019MNRAS.490.5002F};
(26) \citet{2020AA...636A..74T};
(27) \citet{2020AA...642A..31D};
(28) \citet{2020AA...643A.112S};
(29) \citet{2021AJ....161...26C};
(30) \citet{2021AJ....161..134H};
(31) \citet{2021ApJS..255....8R};
(32) \citet{2022AJ....163...19R};
(33) \citet{2022AJ....163..171Z};
(34) \citet{2023AA...671A..10S};
(35) \citet{2023AA...674A.114B};
(36) \citet{2023AJ....165..176L};
(37) \citet{2023AJ....166...46B};
(38) \citet{2025AA...693A.297N};
(39) \citet{2025AJ....169....1G};
(40) \citet{2025AJ....170...52G};
}
\end{deluxetable*}
\end{longrotatetable}

\clearpage
\begin{longrotatetable}
\begin{deluxetable*}{p{4cm}p{6cm}p{1.5cm}p{1cm}p{3.5cm}p{0.5cm}c}
\tablecolumns{8}
\tablefontsize{\scriptsize}
\tablecaption{Summary of observatory characteristics}
\label{tab:observatories}
\tablehead{
  \colhead{Observatory} & 
  \colhead{Instrument name} & 
  \colhead{$\lambda$ range} & 
  \colhead{\tablenotemark{$^\dagger$}Calib.} & 
  \colhead{Telescope} & 
  \colhead{$D$ (m)} & 
  \colhead{Ref.}
}
\startdata
\hline \textbf{Calar Alto Observatory}, \textit{Almer\'{i}a, Spain} & CARMENES (Calar Alto high-Resolution search for M dwarfs with Exoearths with Near-infrared and optical \'{E}chelle Spectrographs) & 0.52-1.71 $\mu$m & FP, ThNe, UAr, UNe & 3.5-m Telescope & 3.5 & 13 \\
\hline \textbf{Cerro Tololo Inter-American Observatory (CTIO)}, \textit{Mt.\ Cerro Tololo, Coquimbo, Chile} & CHIRON (CTIO HIgh ResolutiON spectrometer) & 410-870 nm & I$_2$, ThAr & Small and Moderate Aperture Research Telescope System (SMARTS) & 1.5 & 14 \\
 & CTIO-ES (CTIO \'{E}chelle Spectrometer) & 401-730 nm & I$_2$ & Small and Moderate Aperture Research Telescope System (SMARTS) & 1.5 & 22 \\
\hline \textbf{Kitt Peak National Observatory (KPNO)}, \textit{Kitt Peak, Arizona, USA} & NEID (NN-EXPLORE Exoplanet Investigations with Doppler Spectroscopy) & 380–930 nm & FP, LFC, ThAr & Wisconsin-Indiana-Yale-NOIRLab (WIYN) 3.5-m Telescope & 3.5 & 18 \\
\hline \textbf{La Silla Observatory}, \textit{Coquimbo, Chile} & CES-LC (Coud\'{e} \'{E}chelle Spectrometer + Long Camera) & 536.7-541.2 nm & I$_2$ & Coud\'{e} Auxiliary Telescope (CAT) & 1.4 & 1 \\
 & CES-VLC (Coud\'{e} \'{E}chelle Spectrometer + Very Long Camera) & 537.6-541.2 nm & I$_2$ & ESO 3.6-m Telescope & 3.6 & 1 \\
 & CORALIE (CORALIE) & 388-681 nm & ThAr & Leonhard Euler Swiss Telescope & 1.2 & 9 \\
 & HARPS (High Accuracy Radial Velocity Planet Searcher) & 378-691 nm & ThAr & ESO 3.6-m Telescope & 3.6 & 11 \\
\enddata
\tablenotetext{^\dagger}{FP: Fabry–P\'{e}rot ; I$_2$: Iodine; LFC: Laser Frequency Comb; ThAr: Thorium-Argon; ThArNe: Thorium-Argon-Neon; ThNe: Thorium-Neon; ThUXe: Thorium-Uranium-Xenon; UAr: Uranium-Thorium; UNe: Uranium-Neon}
\tablerefs{(1) \citet{1982SPIE..331..232E}; (2) \citet{1987PASP...99.1214V}; (3) \citet{1990SPIE.1235..562D}; (4) \citet{1994PASP..106.1285B}; (5) \citet{1994SPIE.2198..362V}; (6) \citet{1995PASP..107..251T}; (7) \citet{1996AAS..119..373B}; (8) \citet{1998SPIE.3355..387T}; (9) \citet{2000AA...354...99Q}; (10) \citet{2000SPIE.4008..534D}; (11) \citet{2003Msngr.114...20M}; (12) \citet{2008SPIE.7014E..0JP}; (13) \citet{2010SPIE.7735E..13Q}; (14) \citet{2010SPIE.7735E..4GS}; (15) \citet{2010SPIE.7735E..53C}; (16) \citet{2012SPIE.8446E..1VC}; (17) \citet{2014PASP..126..359V}; (18) \citet{2016SPIE.9908E..7HS}; (19) \citet{2017ApJ...836..142G}; (20) \citet{2020AJ....159..238B}; (21) \citet{2021AA...645A..96P}; (22) \url{https://www.ctio.noirlab.edu/~atokovin/echelle/FECH-overview.html}}
\end{deluxetable*}
\end{longrotatetable}

\clearpage
\begin{longrotatetable}
\begin{deluxetable*}{cCCCCCCCCCCCCCcc}
\tablecolumns{16}
\tablecaption{Example (unbinned) data for HD 190360}
\label{tab:example_rvs}
\tablehead{
  \colhead{HD} & 
  \colhead{BJD} & 
  \colhead{RV} & 
  \colhead{$\sigma_{\rm RV}$} & 
  \colhead{$S$-ind} & 
  \colhead{$\sigma_S$} & 
  \colhead{$H$-ind} & 
  \colhead{$\sigma_H$} & 
  \colhead{EW$_{\rm H \alpha}$} & 
  \colhead{$\sigma_{\rm EWH\alpha}$} & 
  \colhead{$R_{\rm HK}^\prime$} & 
  \colhead{$\sigma_{\rm RHK^\prime}$} & 
  \colhead{BIS} & 
  \colhead{FWHM} & 
  \colhead{Inst.} & 
  \colhead{Bibcode} \\
  \colhead{} & 
  \colhead{$\rm (d)$} & 
  \colhead{$\rm (m\,s^{-1})$} & 
  \colhead{$\rm (m\,s^{-1})$} & 
  \colhead{} & 
  \colhead{} & 
  \colhead{} & 
  \colhead{} & 
  \colhead{$\rm (\AA)$} & 
  \colhead{$\rm (\AA)$} & 
  \colhead{} & 
  \colhead{} & 
  \colhead{$\rm (m\,s^{-1})$} & 
  \colhead{$\rm (m\,s^{-1})$} & 
  \colhead{} & 
  \colhead{}
}
\startdata
190360 & 2455486.886290 & -0.98 & 1.27 & 0.1375 & -1 & 0.0322 & -1 & -1 & -1 & -1 & -1 & 0 & -1 & HIRES-post & 2017AJ....153..208B \\
190360 & 2455399.932050 & -2.32 & 1.17 & 0.1364 & -1 & 0.0323 & -1 & -1 & -1 & -1 & -1 & 0 & -1 & HIRES-post & 2017AJ....153..208B \\
190360 & 2455169.758110 & -14.17 & 1.46 & 0.1484 & -1 & 0.0322 & -1 & -1 & -1 & -1 & -1 & 0 & -1 & HIRES-post & 2017AJ....153..208B \\
190360 & 2455170.693010 & -16.84 & 1.10 & 0.1383 & -1 & 0.0322 & -1 & -1 & -1 & -1 & -1 & 0 & -1 & HIRES-post & 2017AJ....153..208B \\
190360 & 2455635.155120 & -5.50 & 1.07 & 0.1346 & -1 & 0.0321 & -1 & -1 & -1 & -1 & -1 & 0 & -1 & HIRES-post & 2017AJ....153..208B \\
190360 & 2455285.154770 & -8.04 & 1.05 & 0.1338 & -1 & 0.0322 & -1 & -1 & -1 & -1 & -1 & 0 & -1 & HIRES-post & 2017AJ....153..208B \\
190360 & 2455313.134240 & -2.61 & 1.19 & 0.1348 & -1 & 0.0321 & -1 & -1 & -1 & -1 & -1 & 0 & -1 & HIRES-post & 2017AJ....153..208B \\
190360 & 2455313.134900 & -2.59 & 1.21 & 0.1363 & -1 & 0.0322 & -1 & -1 & -1 & -1 & -1 & 0 & -1 & HIRES-post & 2017AJ....153..208B \\
\enddata
\tablecomments{Null stellar activity indicators are given as ``$-1$'' (or ``$0$'' in the case of BIS). Only the first 20 rows of data are shown here. The full table is available in machine-readable form.}
\end{deluxetable*}
\end{longrotatetable}

\clearpage
\begin{longrotatetable}
\begin{deluxetable*}{llCCCCCCCLl}
\tablecolumns{11}
\tablefontsize{\scriptsize}
\tablecaption{Signals identified by \texttt{RVSearch}}
\label{tab:rvsearch_signals}
\tablehead{
  \colhead{HD name} & 
  \colhead{Signal} & 
  \colhead{$P$ (d)} & 
  \colhead{$T_{\rm c} - 2450000$ (JD)} & 
  \colhead{$e$} & 
  \colhead{$\omega$ (rad)} & 
  \colhead{$K$ (m\,s$^{-1}$)} & 
  \colhead{$M_{\rm p} \sin i$ (M$_\oplus$)} & 
  \colhead{$a$ (AU)} & 
  \colhead{$\log {\rm FAP}$} & 
  \colhead{$^\dagger$Class.}
}
\startdata
HD 1581 & 0.01 & 937.0^{+18.0}_{-15.0} & 5737^{+81}_{-60} & 0.24^{+0.16}_{-0.15} & -1.98^{+0.73}_{-0.65} & 0.757^{+0.096}_{-0.093} & 11.8^{+1.6}_{-1.5} & 1.941 \pm 0.042 & -7.9 & SA (10) \\
HD 1581 & 0.02 & 15.6544^{+0.0048}_{-0.0074} & 5199.08^{+1.2}_{-0.77} & 0.16^{+0.16}_{-0.11} & 2.2^{+2.2}_{-1.3} & 0.732 \pm 0.098 & 2.99 \pm 0.41 & 0.1268^{+0.0022}_{-0.0023} & -7.35 & SA-R (10) \\
HD 3651 A & b & 62.2478^{+0.0024}_{-0.0025} & 4414.1^{+0.78}_{-0.81} & 0.617 \pm 0.015 & -2.041^{+0.035}_{-0.034} & 15.45^{+0.36}_{-0.35} & 68.6 \pm 3.4 & 0.2935^{+0.0066}_{-0.0069} & -78.2 & KC (17) \\
HD 3651 A & 0.02 & 5157.0^{+220.0}_{-230.0} & 6738^{+680}_{-1300} & 0.62^{+0.29}_{-0.26} & -1.2^{+0.34}_{-0.35} & 3.37^{+4.2}_{-0.88} & 65.0^{+15.0}_{-11.0} & 5.57^{+0.2}_{-0.21} & -6.3 & SA (22) \\
HD 4614 A & 0.01 & 91.018^{+0.13}_{-0.093} & 7203.0^{+2.6}_{-1.9} & 0.53^{+0.11}_{-0.14} & 0.42^{+0.25}_{-0.26} & 2.31^{+0.45}_{-0.38} & 14.4^{+2.1}_{-2.0} & 0.4051^{+0.0062}_{-0.0064} & -3.93 & FA  \\
HD 9826 A & b & 4.617118^{+9.5e-06}_{-9.4e-06} & 3335.89 \pm 0.01 & 0.0087^{+0.0074}_{-0.0059} & 1.41^{+0.82}_{-1.4} & 69.87^{+0.48}_{-0.5} & 227.7^{+8.7}_{-8.9} & 0.0607^{+0.0011}_{-0.0012} & -99.9 & KC (8) \\
HD 9826 A & c & 241.201 \pm 0.037 & 4652.2 \pm 1.2 & 0.274 \pm 0.0094 & -2.142^{+0.04}_{-0.039} & 55.72^{+0.61}_{-0.62} & 653.0 \pm 26.0 & 0.849 \pm 0.016 & -88.3 & KC (8) \\
HD 9826 A & d & 1281.4^{+0.73}_{-0.71} & 5937.3^{+7.2}_{-7.0} & 0.3139^{+0.0095}_{-0.0096} & -1.885 \pm 0.032 & 68.27^{+0.8}_{-0.76} & 1378.0^{+54.0}_{-55.0} & 2.585^{+0.048}_{-0.05} & -99.9 & KC (8) \\
\enddata
\tablecomments{The full table is available in machine-readable form.}
\tablenotetext{^\dagger}{FA: False Alarm; KC: Known Companion (“-S” indicates stellar); PC: Planet Candidate; SA: Stellar Activity (“-R” indicates rotation); LPS: Long Period Signal; SRC: Source Requiring Confirmation.}
\tablerefs{
(1) \citet{1995ApJ...438..269B};
(2) \citet{2001ApJ...555..410B};
(3) \citet{2006Natur.441..305L};
(4) \citet{2007AA...462..769P};
(5) \citet{2009ApJ...693.1084W};
(6) \citet{2010ApJ...708.1366V};
(7) \citet{2010MNRAS.403..731G};
(8) \citet{2011AA...525A..78C};
(9) \citet{2011AA...534A..58P};
(10) \citet{2011arXiv1107.5325L};
(11) \citet{2013AA...551A..90M};
(12) \citet{2015AA...584A..72M};
(13) \citet{2015ApJ...814...12V};
(14) \citet{2015MNRAS.446.1493B};
(15) \citet{2018AA...619A...1B};
(16) \citet{2019ApJ...886..132G};
(17) \citet{2019MNRAS.484.5859W};
(18) \citet{2020AJ....159..235L};
(19) \citet{2020Sci...368.1477J};
(20) \citet{2021AA...654A.104U};
(21) \citet{2021AJ....162..181L};
(22) \citet{2021ApJS..255....8R};
(23) \citet{2021NatAs...5..775D};
(24) \citet{2022AJ....163..218H};
(25) \citet{2022AJ....163..223H};
(26) \citet{2023AA...671A..10S};
(27) \citet{2023AA...674A.114B};
(28) \citet{2023AJ....166...46B};
(29) \citet{2024AJ....167..243B};
(30) \citet{2024ApJS..274...35I};
}
\end{deluxetable*}
\end{longrotatetable}

\clearpage
\startlongtable
\begin{deluxetable*}{lCC}
\tablecolumns{3}
\tablefontsize{\scriptsize}
\tablecaption{Trends identified by \texttt{RVSearch}}
\label{tab:rvsearch_trends}
\tablehead{
  \colhead{HD name} & 
  \colhead{$\dot{\gamma}$ (m\,s$^{-1}$\,d$^{-1}$)} & 
  \colhead{$\ddot{\gamma}$ (m\,s$^{-1}$\,d$^{-2}$)}
}
\startdata
HD 2151 & 0.00076 \pm 0.00022 & - \\
HD 4614 A & 0.01689 \pm 0.00033 & - \\
HD 16895 A & -0.0055^{+0.0009}_{-0.00092} & - \\
HD 19373 & -0.00073^{+0.00017}_{-0.00016} & - \\
HD 20010 A & -0.0303 \pm 0.0015 & - \\
HD 20630 & -0.0057 \pm 0.0016 & - \\
HD 20766 & -0.00244^{+0.00066}_{-0.00067} & - \\
HD 20807 & -0.00057^{+0.00019}_{-0.0002} & - \\
\enddata
\tablecomments{This table is available in machine-readable form.}
\end{deluxetable*}

\clearpage
\begin{longrotatetable}
\begin{deluxetable*}{lcccCCCCcp{7cm}}
\tablecolumns{10}
\tablefontsize{\scriptsize}
\tablecaption{Previously reported planets not detected by \texttt{RVSearch}}
\label{tab:undetected_planet_claims}
\tablehead{
  \colhead{HD name} & 
  \colhead{Pl.} & 
  \colhead{Det.\ meth.} & 
  \colhead{Disc.\ yr.} & 
  \colhead{$P$ (d)} & 
  \colhead{$e$} & 
  \colhead{$M_{\rm p} \sin i$ (M$_\oplus$)} & 
  \colhead{$a$ (AU)} & 
  \colhead{Ref.} & 
  \colhead{Notes}
}
\startdata
HD 10700 ($\tau$ Cet) & e & RV & 2017 & 162.87 & 0.18 & 3.93 & 0.538 & 3 & Disputed planet candidate. \\
 & f  &  RV  &  2017  &  636.13  &  0.16  &  3.93  &  1.334  &  3  &  Disputed planet candidate. \\
 & g  &  RV  &  2017  &  20.0  &  0.06  &  1.75  &  0.133  &  3  &  Disputed planet candidate. \\
 & h  &  RV  &  2017  &  49.41  &  0.23  &  1.83  &  0.243  &  3  &  Disputed planet candidate. \\
HD 20794 (82 Eri) & e & RV & 2017 & 147.02 & 0.29 & 4.77 & 0.509 & 2 & Disputed planet candidate. \\
 & f  &  RV  &  2025  &  647.6  &  0.45  &  5.82  &  1.3541  &  11  &  RV precision limited by possible unmodeled instrumental drift in the ESPRESSO data. \\
HD 33564 & b & RV & 2005 & 388.0 & 0.34 & 2892.1 & 1.1 & 5 & Original ELODIE data used for discovery are not public. \\
HD 86728 A (20 LMi A) & b & RV & 2025 & 31.1515 & 0.067 & 9.43 & 0.1916 & 7 & Signal was detected but did not meet the FAP=0.01 significance cutoff. \\
\enddata
\tablecomments{RV: Radial Velocity, Astr: Astrometry, Im: Imaging}
\tablerefs{ (1) \citet{2023AJ....166...46B}; (2) \citet{2017AA...605A.103F}; (3) \citet{2017AJ....154..135F}; (4) \citet{2022ApJS..262...21F}; (5) \citet{2005AA...444L..21G}; (6) \citet{2010MNRAS.403..731G}; (7) \citet{2025AJ....169....1G}; (8) \citet{2007ApJ...654..570L}; (9) \citet{2024Natur.633..789M}; (10) \citet{2004AA...415..391M}; (11) \citet{2025AA...693A.297N}; (12) \citet{2023AA...678A.107P}; (13) \citet{2011ApJ...727..103T}; (14) \citet{2015ApJ...814...12V} }
\end{deluxetable*}
\end{longrotatetable}

\clearpage
\begin{longrotatetable}
\begin{deluxetable*}{lC|CCLL|CCLL|CCLL}
\tablecolumns{14}
\tablecaption{RV sensitivity limits in the conservative HZ}
\label{tab:sensitivity_limits}
\tablehead{
  \colhead{} & 
  \colhead{} & 
  \multicolumn{4}{c}{IHZ (inner edge of the HZ)} & 
  \multicolumn{4}{c}{\tablenotemark{$^\dagger$}MHZ (middle of the HZ)} & 
  \multicolumn{4}{c}{OHZ (outer edge of the HZ)} \\
  \cline{3-14}
  \colhead{HD name} & 
  \colhead{$T_{\rm eff}$} & 
  \colhead{$a_{\rm IHZ}$} & 
  \colhead{$P_{\rm IHZ}$} & 
  \colhead{$M_{\rm p}\sin i$} & 
  \colhead{$K_{\rm RV}$} & 
  \colhead{$a_{\rm MHZ}$} & 
  \colhead{$P_{\rm MHZ}$} & 
  \colhead{$M_{\rm p}\sin i$} & 
  \colhead{$K_{\rm RV}$} & 
  \colhead{$a_{\rm OHZ}$} & 
  \colhead{$P_{\rm OHZ}$} & 
  \colhead{$M_{\rm p}\sin i$} & 
  \colhead{$K_{\rm RV}$} \\
  \colhead{} & 
  \colhead{(K)} & 
  \colhead{(au)} & 
  \colhead{(yr)} & 
  \colhead{($\rm M_\oplus$)} & 
  \colhead{($\rm m\,s^{-1}$)} & 
  \colhead{(au)} & 
  \colhead{(yr)} & 
  \colhead{($\rm M_\oplus$)} & 
  \colhead{($\rm m\,s^{-1}$)} & 
  \colhead{(au)} & 
  \colhead{(yr)} & 
  \colhead{($\rm M_\oplus$)} & 
  \colhead{($\rm m\,s^{-1}$)}
}
\startdata
HD 166 & 5491 & 0.81 & 0.75 & 246.7~[414.9] & 24.1~[39.3] & 0.99 & 1.01 & 274.4~[470.5] & 24.3~[40.1] & 1.39 & 1.70 & 324.8~[563.7] & 24.2~[41.0] \\
HD 1581 & 5932 & 1.11 & 1.11 & 10.7~[22.5] & 0.8~[1.7] & 1.35 & 1.49 & 11.9~[24.0] & 0.8~[1.7] & 1.87 & 2.43 & 13.5~[26.2] & 0.8~[1.7] \\
HD 2151 & 5806 & 1.85 & 2.42 & 59.0~[110.1] & 3.5~[6.0] & 2.26 & 3.26 & 67.5~[138.3] & 3.7~[6.6] & 3.14 & 5.37 & 91.2~[226.3] & 4.3~[9.1] \\
HD 3651 A & 5203 & 0.74 & 0.69 & 15.0~[31.6] & 1.7~[3.2] & 0.91 & 0.94 & 17.2~[34.6] & 1.7~[3.2] & 1.29 & 1.59 & 20.8~[39.4] & 1.7~[3.2] \\
HD 4391 & 5887 & 0.95 & 0.94 & 315.4~[556.7] & 24.5~[41.3] & 1.16 & 1.26 & 352.1~[620.8] & 24.9~[42.2] & 1.61 & 2.06 & 431.6~[857.0] & 26.0~[45.2] \\
HD 4614 A & 5907 & 1.11 & 1.09 & 25.8~[69.3] & 2.1~[4.8] & 1.35 & 1.46 & 30.7~[93.2] & 2.3~[5.7] & 1.87 & 2.39 & 41.5~[150.8] & 2.6~[8.2] \\
HD 4628 & 5007 & 0.56 & 0.48 & 11.4~[22.4] & 1.5~[2.7] & 0.69 & 0.65 & 13.5~[27.3] & 1.6~[2.9] & 0.99 & 1.11 & 18.4~[36.1] & 1.8~[3.3] \\
HD 7570 & 6110 & 1.39 & 1.53 & 90.7~[166.3] & 6.2~[10.2] & 1.68 & 2.05 & 102.7~[193.6] & 6.4~[10.3] & 2.33 & 3.32 & 128.3~[247.0] & 6.8~[11.0] \\
\enddata
\tablecomments{Values in brackets (``[ ]'') denote the sensitivity limit defined by the 90\% completeness contour (opposed to the 50\% completeness contour). The full table is available in machine-readable form.}
\tablenotetext{\dagger}{Computed in instellation flux space.}
\end{deluxetable*}
\end{longrotatetable}

\clearpage
\startlongtable
\begin{deluxetable*}{lLLLLL}
\tablecolumns{6}
\tablefontsize{\scriptsize}
\tablecaption{Stars with no public RV data}
\label{tab:tab_neglected_stars}
\tablehead{
  \colhead{HD name} & 
  \colhead{R.A.} & 
  \colhead{Dec.} & 
  \colhead{$T_{\rm eff}$ (K)} & 
  \colhead{$\log L_\star / {\rm L_\odot}$} & 
  \colhead{$\log R_{\rm HK}^\prime$}
}
\startdata
HD 739 & 2.933421 & -35.13311 & 6495.0 & 0.4814 & -4.715 \\
HD 4813 & 12.53162 & -10.64433 & 6208.0 & 0.2506 & -4.865 \\
HD 7788 A & 18.94204 & -68.87594 & 6436.0 & 0.5911 & -4.272 \\
HD 17206 & 41.27583 & -18.57262 & 6330.0 & 0.4178 & -4.461 \\
HD 23754 & 56.71203 & -23.24972 & 6685.0 & 0.7049 & -4.855 \\
HD 33262 A & 76.37773 & -57.4727 & 6158.0 & 0.1662 & -4.426 \\
HD 33564 & 80.63971 & 79.23115 & 6354.0 & 0.5171 & -4.979 \\
HD 35296 & 81.1061 & 17.38354 & 6131.0 & 0.2499 & -4.378 \\
HD 64379 & 118.0653 & -34.70544 & 6525.0 & 0.4043 & -4.363 \\
HD 74576 & 130.8251 & -38.88238 & 4992.0 & -0.4982 & -4.351 \\
HD 82885 A & 143.9146 & 35.81013 & 5518.0 & -0.1057 & -4.638 \\
HD 90589 & 156.0988 & -74.03161 & 6905.0 & 0.7128 & -4.685 \\
HD 90089 A & 157.7696 & 82.55865 & 6758.0 & 0.5124 & -4.55 \\
HD 105452 A & 182.1034 & -24.72888 & 6990.0 & 0.6487 & -4.214 \\
HD 114837 A & 198.5631 & -59.10324 & 6238.0 & 0.4722 & -4.665 \\
HD 126660 A & 216.2992 & 51.85074 & 6280.0 & 0.6054 & -4.361 \\
HD 155886 & 258.8366 & -26.6017 & 5132.0 & -0.4848 & -4.57 \\
HD 155885 & 258.8374 & -26.60283 & 5144.0 & -0.4788 & -4.559 \\
HD 156897 A & 260.2516 & -21.11294 & 6756.0 & 0.6161 & - \\
HD 160032 & 265.0993 & -49.41558 & 6620.0 & 0.6579 & -4.665 \\
HD 160915 & 265.8575 & -21.68319 & 6404.0 & 0.4315 & -4.698 \\
HD 165341 B & 271.3644 & 2.498947 & 4475.0 & -0.7843 & -4.71 \\
HD 165185 & 271.5988 & -36.01979 & 5892.0 & 0.0186 & -4.406 \\
\enddata
\end{deluxetable*}

\newpage


\section{Systems with Updated Planet Parameters}\label{appx:signals_known_planets}

As discussed in the main text, at least 70 exoplanets (of which at least 9 have been disputed) have previously been detected around 30 of the 164 stars in the sample (\secref{sec:discussion:architectures}). Of the known planets, we detected a total of 51 (\secref{sec:results:rvsearch}) orbiting 21 different stars. The updated masses and orbital properties of these planets are shown in Table \ref{tab:rvsearch_signals}. We also detected signals associated with two known stellar companions (\secref{app:sec:hd165499} and \secref{app:sec:hd212330}). While some of these systems were discussed in the main text (\secref{sec:hd190360} to \ref{sec:hd219623}), here we briefly discuss our results for the rest of these systems individually.

\subsection{HD 3651 A (54 Psc A)}
HD 3651 A (54 Psc A; HIP 3093 A; HR 166 A; GJ 27 A; TIC 434210589) is a K0.5V star \citep{Keenan+1989ApJS} at a distance of $d = 11.108 \pm 0.006 \,{\rm pc}$ \citep{GaiaDR3_2022yCat}. This star hosts an eccentric planet (54~Psc~b; $e\approx0.645$) with an orbital period of $P\approx62.25\,{\rm d}$ and a minimum mass of $M_{\rm p} \sin i \approx 72.5 \,{\rm M_\oplus}$ \citep{2019MNRAS.484.5859W}. The planet 54~Psc~b has been studied extensively in the literature using observations from various instruments, including Hamilton/Shane at Lick Observatory \citep{Fischer+2003ApJ}, HIRES/Keck \citep{2006ApJ...646..505B}, and the HRS/HET at McDonald Observatory \citep{2009ApJS..182...97W}. Here, we robustly detected 54~Psc~b at $P\approx62.25\,{\rm d}$ with 461 RVs from the HIRES, HRS, Hamilton, Levy, and Tull instruments spanning 30 years. We found a Keplerian orbital solution for this planet consistent with previous studies (see Table \ref{tab:rvsearch_signals}). We also report an additional Keplerian signal with a period of $P=5157\,{\rm d}$, which has previously been attributed to long-period magnetic stellar activity \citep{2021ApJS..255....8R}. We also note the presence of a signal in the $S$-index GLS periodogram at $P\approx3500\,{\rm d}$, which may also be due to a magnetic activity cycle \citep[e.g.,][]{2024ApJS..274...35I}.

\subsection{HD 9826 A ($\upsilon$ And A)}\label{sec:upsAnd}
HD 9826 A (Titawin; $\upsilon$ And A; HIP 7513 A; HR 458 A; GJ 61 A; TIC 189576919) is an F8V star in a binary system \citep{Gray+2001AJ} at a distance of $d = 13.49 \pm 0.03 \,{\rm pc}$ \citep{vanLeeuwen2007AA}. At least three giant planets are known to orbit $\upsilon$~And~A, which were initially detected with RV observations obtained at the Lick and Whipple Observatories \citep{Butler+1997ApJ, 1999ApJ...526..916B}. The orbits and masses of the three planets, $\upsilon$~And~b, c, and d, were subsequently refined with additional RVs from ELODIE/OHP \citep{2004AA...414..351N}, Tull/HJS spectrograph \citep{2007ApJ...654..625W}, and HRS/HET \citep{2010ApJ...715.1203M}, in addition to astrometry from the \textit{Hubble Space Telescope} (\textit{HST}) \citep{2010ApJ...715.1203M}. Evidence for a fourth planet was later reported by \citet{2011AA...525A..78C}, who detected a 3848.9-day signal (corresponding to a $M_{\rm p} \sin i = 1.06 \,{\rm M_{Jup}}$ planet) using a novel detection algorithm and archival RV data. We report six Keplerian signals using a total of 853 RVs collected using the AFOE, ELODIE, Hamilton, HIRES, HRS, Levy, SOPHIE, and Tull instruments across 30 years. Three of the signals found by \rvsearch\ match the known planets. Two additional signals with orbital periods of $3150\,{\rm d}$ and $174.35\,{\rm d}$ found by \rvsearch\ were previously attributed to stellar activity and instrumental systematics, respectively, by \citet{2021ApJS..255....8R}. Lastly, we note that \rvsearch\ identified a long-period signal with an orbital period of 32,946 days, although this signal does not have sufficient orbital phase coverage to be properly classified.

\subsection{HD 10647 ($q^1$ Eri)}\label{sec:q1Eri}
HD 10647 ($q^1$ Eri; HIP 7978; HR 506; GJ 3109; TIC 229137615) is an F9V star \citep{Keenan+1989ApJS} at a distance of $d = 17.35 \pm 0.01 \,{\rm pc}$ \citep{GaiaDR3_2022yCat} with one known planet ($q^1$~Eri~b) and an asymmetric debris disk \citep{Stencel+1991ApJS, Liseau+2008AA, Lovell+2021MNRAS}. The planet $q^1$~Eri~b is an eccentric ($e\approx0.15$) giant ($M_{\rm p} \sin i \approx 299\,{\rm M_\oplus}$) on a $P\approx989.2\,{\rm d}$ orbit \citep{2013AA...551A..90M}. It was initially detected using 28 RVs from UCLES/AAT, and was later re-observed using CORALIE \citep{2013AA...551A..90M}. We recovered the planet again here using 226 RVs from the HARPS, UCLES, CES-LC, CES-VLC, and CORALIE instruments across 26 years. The derived planet parameters are consistent with those previously reported in \citet{2013AA...551A..90M}.

\subsection{HD 17051 ($\iota$ Hor)}
HD 17051 ($\iota$ Hor; HIP 12653; HR 810; GJ 108; TIC 166853853) is an F9V Fe+0.3 star \citep{Gray+2006AJ} at a distance $d = 17.36 \pm 0.01 \,{\rm pc}$ \citep{GaiaDR3_2022yCat}. The star has one known giant ($M_{\rm p} \sin i \approx 677\,{\rm M_\oplus}$) planet (HR~810~b) with an orbital period of $P\approx312\,{\rm d}$ \citep{2001ApJ...555..410B}. The planet HR~810~b was discovered with RV observations from the ESO Coud\'{e} Auxiliary Telescope (CAT) Spectrograph at La Silla \citep{Kurster+2000AA} and UCLES/AAT at the Australian National University \citep{2001ApJ...555..410B}. We recovered the planet here at a period of $P\approx308\,{\rm d}$ using 228 HARPS, CES-LC, CES-VLC, CORALIE, and UCLES RVs with an orbital solution consistent with previous studies. We also identified a signal with $P = 383.0 \pm 2.2 \,{\rm d}$. We classified this signal as a likely systematic False Alarm, as its orbital period is close to a sidereal year and there are corresponding sub-threshold peaks in the H-index, $R^\prime_{\rm HK}$, BIS, and FWHM periodograms and a significiant peak in the window function periodogram.

\subsection{HD 20794 (82 Eri)}\label{sec:82Eri}
HD 20794 (82 Eri; HIP 15510; HR 1008; GJ 139; TIC 301051051) is a G6V star \citep{Keenan+1989ApJS} with evidence of a circumstellar debris disk \citep{Wyatt+2012MNRAS, Montesinos+2016AA} at a distance of $d = 6.041 \pm 0.003 \,{\rm pc}$ \citep{GaiaDR3_2022yCat}. The star hosts a multi-planet system of at least 2 low-$M_{\rm p} \sin i$ planets, HD~20794~b and HD~20794~d, with orbital periods of $P_b\approx18.3\,{\rm d}$ and $P_d\approx90.3\,{\rm d}$ and masses of $M_{{\rm p},b} \sin i \approx 2.7\,{\rm M_\oplus}$ and $M_{{\rm p},d} \sin i \approx 4.8\,{\rm M_\oplus}$ \citep{2011AA...534A..58P}. Here, we detected both known planets using 858 RVs collected by ESPRESSO, HARPS, PFS, and UCLES. Interestingly, we note the presence of a weakly-significant peak in the $H$-index periodogram consistent with the orbital period of planet d at $\sim$90 days, suggesting that perhaps a more detailed look into the stellar activity patterns of the host star near this period is warranted. We also detected a significant RV signal at $364.2\,{\rm d}$ that is not associated with any known planets or planet candidates, which we classified this signal as a false alarm due to its consistency with the sidereal year---there were also significant peaks at this period in the $R^\prime_{\rm HK}$, BIS, FWHM, and window function periodograms.

We also note that this system has been suggested to host two additional low-$M_{\rm p} \sin i$ planets, HD~20794~c \citep[$P_c\approx40.1\,{\rm d}$;][]{2011AA...534A..58P} and HD~20794~e \citep[$P_e\approx147.0\,{\rm d}$;][]{2017AA...605A.103F}, which were not detected in this analysis. Though the planetary signals originally reported at 18 and 90 days (corresponding to planets b and d) were ``very significant,'' \citet{2011AA...534A..58P} cautioned that their detection of a possible planet c at 40 days was ``less apparent'' in the phase-folded RV plot and that the ``orbital period is close to the supposed rotational period of the parent star,'' casting some doubt as to whether the signal was real or not. An independent reanalysis of the HARPS data using a more sophisticated treatment of correlated noise later found only ``weak evidence'' for the 40-day signal \citep{2017AA...605A.103F}, and a more recent analysis of data from HARPS/ESO, UCLES/AAT, and PFS/Magellan also failed to detect the reported signal at 40 days \citep{2023AJ....165..176L}. Our analysis of the combined ESPRESSO, HARPS, PFS, and UCLES data further suggest that the 40-day signal is not real. Recent analysis by \citet{2025AA...693A.297N} confirmed that the 40-day signal is due to stellar rotation. Additionally, \citet{2017AA...605A.103F} reported two more potential new detections at 147 and 330 days. Neither of these signals were detected by \citet{2023AJ....165..176L}, and we did not find evidence of them in this study either. \citet{Cretignier+2023AA} also claimed a $10\sigma$ detection of a $<1\,{\rm m\,s}^{-1}$ signal at 650 days using the HARPS data. This was subsequently confirmed by \citet{2025AA...693A.297N}, who notably did not find evidence of the 147-day signal (making this a disputed planet). Although we did not detect the 147-day and 650-day signals, given the more sophisticated data reduction and analysis of stellar variability and correlated noise by the \citet{2017AA...605A.103F} and \citet{2025AA...693A.297N} studies, we cannot rule them out as potential candidate planets. Lastly, we note that an apparent instrumental drift in the ESPRESSO RVs may have limited the sensitivity of our analysis of this system, as correcting instrumental red noise is beyond the scope of this work.

\subsection{HD 22049 ($\epsilon$ Eri)}
HD 22049 (Ran, $\epsilon$ Eri; HIP 16537; HR 1084; GJ 144; TIC 118572803) is a K2V star \citep{Keenan+1989ApJS} at a distance of $d = 3.220 \pm 0.001 \,{\rm pc}$ \citep{GaiaDR3_2022yCat}. The star's young age \citep[200–800 Myr;][]{Fuhrmann2004AN, Mamajek+2008ApJ} and prominent debris disk \citep[e.g.,][]{Greaves+1998ApJ, Backman+2009ApJ} make $\epsilon$~Eri a particularly interesting system. This star hosts at least one known planet ($\epsilon$~Eri~b) on a $P = 2671\,{\rm d}$ (7.3 yr) orbit with a minimum mass of $M_{\rm p} \sin i = 209.8\,{\rm M_\oplus}$ \citep{2021AJ....162..181L}. The planet was first announced by \citet{Hatzes+2000ApJ}, who analyzed six independent RV data sets taken at four ground-based telescopes. Subsequently, it has been extensively observed and characterized in various studies spanning the last least three decades, both from the ground \citep[e.g.,][]{Mawet+2019AJ} and from space \citep[e.g.,][]{2006AJ....132.2206B, 2021AJ....162..181L}. With our data set of 693 total RVs from HIRES, HARPS, CES-LV, CES-VLC, EXPRES, Hamilton, Levy, and Tull, we detected $\epsilon$~Eri~b at a period of $2686.0 \pm 18.0 \,{\rm d}$, consistent with previous values. A significant $S$-index signal in the activity GLS periodogram at $P \approx 1100\,{\rm d}$ is consistent with a the known magnetic activity cycle of this star \citep{Metcalfe+2013ApJ}.

\subsection{HD 39091 ($\pi$ Men)}
HD 39091 ($\pi$ Men; HIP 26394; HR 2022; GJ 9189; TIC 261136679) is a G0V star \citep{Gray+2006AJ} with evidence of a debris disk \citep{Sibthorpe+2018MNRAS} at a distance of $d = 18.29 \pm 0.01 \,{\rm pc}$ \citep{GaiaDR3_2022yCat}. The star hosts a multi-planet system consisting of at least three planets. The planet $\pi$~Men~b was initially published by \citet{Jones+2002MNRAS} based on observations from the UCLES spectrograph. The second planet, $\pi$~Men~c, was the first planet discovered by the \textit{Transiting Exoplanet Survey Satellite} (TESS) and is consistent with a super-Earth on a 6.3-day orbit \citep{Gandolfi+2018AA, Huang+2018ApJ}. The third planet, $\pi$~Men~d, was identified with EPRV observations from ESPRESSO \citep{2022AJ....163..223H} with a period of $P_{\rm d} = 125 \,{\rm d}$, and later confirmed by \citet{2023AJ....165..176L}. In this work, we detected the giant planet $\pi$~Men~b at $P = 2088.82 \pm 0.31 \,{\rm d}$ using 415 HARPS, ESPRESSO, CORALIE, PFS, and UCLES RVs. Our orbital solution is consistent with recent estimates by \citet{2022ApJS..262...21F}, \citet{2022AJ....163..223H}, and \citet{2023AJ....165..176L}. We also detected planet c with a precise period of $P = 6.26815 \pm 0.00045$ d and planet d at an orbital period of $120.3 \pm 0.1$ d.

\subsection{HD 69830}\label{sec:hd69830}
HD 69830 (HIP 40693; HR 3259; GJ 302; TIC 307624961) is a G8+V star \citep{Gray+2006AJ} with a debris disk \citep{Beichman+2005ApJ, Trilling+2008ApJ} at a distance of $d = 12.579 \pm 0.006 \,{\rm pc}$ \citep{GaiaDR3_2022yCat}. The star hosts three known planets, which were initially detected after two years of HARPS RV observations \citep{2006Natur.441..305L}. The planets all have minimum masses consistent with sub-Neptunes and orbital periods between 8 and 200 days. In this work, we recovered the planets HD~69830~b, HD~69830~c and HD~69830~d with periods $P_{\rm b} = 8.6690 \pm 0.0005 \,{\rm d}$, $P_{\rm c} = 31.613 \pm 0.009 \,{\rm d}$, and $P_{\rm d} = 201.1 \pm 0.5 \,{\rm d}$. The periods and masses are consistent with recently published values \citep{2021ApJS..255....8R, 2023AJ....165..176L}. We also detected a signal at $P = 379\,{\rm d}$ consistent with a signal identified by \citet{2021ApJS..255....8R}, which we classified as FA due to the strong peak in the window function at this period.

\subsection{HD 75732 A (55 Cnc A)}\label{sec:55Cnc}
HD 75732 A (Copernicus; 55 Cnc A; HIP 43587 A; HR 3522 A; GJ 324 A; TIC 332064670) is a well-known K0IV-V star \citep{Gray+2003AJ} at a distance of $d = 12.587 \pm 0.007 \,{\rm pc}$ \citep{GaiaDR3_2022yCat}. This star hosts 5 planets, which have been abundantly observed and characterized over many years \citep[see, e.g.,][and references therein]{Fischer+2008ApJ, 2012ApJ...759...19E, 2018AA...619A...1B}. In this work, we analyzed 876 RVs spanning 31 years from the ELODIE, HARPS-North, HIRES, HRS, Hamilton, Levy, SOPHIE, and Tull spectrographs, and recovered 55~Cnc~b, c, d, e and f with masses and orbits consistent with previous studies. We also detected a signal at $P = 10585 \pm 2400\,{\rm d}$, which is consistent with the second harmonic of the the known stellar activity cycle period around 13 years \citep[e.g.,][]{2015MNRAS.446.1493B}. 

\subsection{HD 95128 (47 UMa)}\label{sec:47Uma}
HD 95128 (Chalawan; 47 UMa; HIP 53721; HR 4277; GJ 407; TIC 21535479) is a G1-V Fe-0.5 star \citep{Keenan+1989ApJS} at a distance of $d = 13.89 \pm 0.02 \,{\rm pc}$ \citep{GaiaDR3_2022yCat}. This star hosts a well-studied planetary system with three long-period planets. The first planet, 47~UMa~b was published by \citet{Butler+1996ApJ} based on data from the Hamilton spectrograph at Lick Observatory. It was later suggested that the system hosted a second \citep{Fischer+2002ApJ, 2004AA...414..351N, 2007ApJ...654..625W, 2009ApJS..182...97W} and, later, third \citep{2010MNRAS.403..731G, 2021ApJS..255....8R} planet based on additional RV measurements and more sophisticated analysis techniques. Using 585 RVs from the HIRES, ELODIE, HRS, Hamilton, Levy, and Tull instruments, we detected 47~UMa~b and 47~UMa~c at periods of $P_{\rm b} = 1076.1 \pm 1.2 \,{\rm d}$ and $P_{\rm b} = 2265 \pm 15 \,{\rm d}$, which are consistent with previous measurements. We did not directly detect 47~UMa~d, which has an orbital period of $\sim$14,000 days (about 38 years). Planet d was detected by \citet{2010MNRAS.403..731G}, who developed a hybrid MCMC model that could constrain the long orbital period of the planet despite sampling data from only a fraction of the orbital period (Table \ref{tab:undetected_planet_claims}). In this work, we detected a long-period signal at about 17,650 d, which is likely associated with 41~UMa~d. However, because our regular MCMC analysis was unable to converge on a single, we did not classify this as a detection of planet d. Lastly, we also detected a FA signal at 358.3 d, which was close to the sidereal year and had corresponding peaks in the $H$-index and window function GLS periodograms.

\subsection{HD 95735 (Lalande 21185)}\label{sec:Lalande21185}
HD 95735 (Lalande 21185; HIP 54035; GJ 411; TIC 166646191) is an M2V star \citep{Kirkpatrick+1991ApJS} at a distance of $d = 2.5461 \pm 0.0002 \,{\rm pc}$ \citep{GaiaDR3_2022yCat}. A planet candidate with $M_{\rm p} \sin i = 2.99 \,{\rm M_\oplus}$ was initially detected orbiting this star on a 12.95-day orbit using 157 RVs from SOPHIE/OHP \citep{2019AA...625A..17D}. This planet (GJ~411~b) was later confirmed by archival HIRES data and new RVs from the CARMENES spectrograph at Calar Alto Observatory \citep{2020AA...643A.112S}. With additional RV observations from APF, a second planet in the system (GJ~411~c) with $M_{\rm p} \sin i = 13.6 \,{\rm M_\oplus}$ was later confirmed at a period of about 2900 days \citep{2022AJ....163..218H}. The authors also reported a possible third planet with a period near 215 days and $M_{\rm p} \sin i = 3.89 \,{\rm M_\oplus}$. In this work, we analyzed 762 RVs collecting using the CARMENES, HIRES, Hamilton, and Levy instruments. We recovered GJ~411~b and GJ~411~c at $P_{\rm b} = 12.938 \pm 0.002 \,{\rm d}$ and $P_{\rm c} = 2801 \pm 81 \,{\rm d}$ and masses consistent with those reported in the literature. However, we note that the sparse RV phase coverage of planet c and presence of significant power at long periods in the GLS periodograms of FWHM, BIS, and EW$_{\rm H \alpha}$ bring into question the planetary nature of this signal. We recommend further analysis to rule out stellar activity as the culprit. We also identified a signal at 55.32 d, corresponding to the measured stellar rotation of HD 95735 reported in \citet{2022AJ....163..218H}.

\subsection{HD 115617 (61 Vir)}
HD 115617 (61 Vir; HIP 64924; HR 5019; GJ 506; TIC 422478973) is a G6.5V star \citep{Keenan+1989ApJS} with a circumstellar debris disk \citep{Trilling+2008ApJ, Tanner+2009ApJ, Wyatt+2012MNRAS} at a distance of $d = 8.53 \pm 0.01 \,{\rm pc}$ \citep{GaiaDR3_2022yCat}. This star hosts a multi-planet system discovered with observations from HIRES and UCLES, which consists of three planets with orbital periods of approximately 4.2, 38, and 124 days \citep{2010ApJ...708.1366V}. In this work, with 773 RVs from the HARPS, HIRES, PFS, UCLES, Hamilton, and Levy instruments, we recovered all three planets at orbital periods of $P_{\rm b} = 4.21501 \pm 0.00009 \,{\rm d}$, $P_{\rm c} = 38.086 \pm 0.005 \,{\rm d}$, and $P_{\rm d} = 123.15 \pm 0.16 \,{\rm d}$ . These parameters are consistent with recent estimates from \citet{2023AJ....165..176L}. We also detected a long-period RV signal at 4026.3 days, which is similar to the LPS identified for this system by \citet{2023AJ....165..176L}. While this signal may correspond to a magnetic activity cycle, additional data and are required to confirm its true nature. 

\subsection{HD 136352 ($\nu^2$ Lupi)}
HD 136352 ($\nu^2$ Lupi; HIP 75181; HR 5699; GJ 582; TIC 136916387) is a G2-V star \citep{Gray+2006AJ} at a distance of $d = 14.74 \pm 0.01 \,{\rm pc}$ \citep{GaiaDR3_2022yCat}. Three planets were detected in this system based on HARPS RV observations \citep{Udry+2019AA}, and were later detected in transit with observations from TESS \citep{Kane+2020AJ} and CHEOPS \citep{2021NatAs...5..775D}. In this work we used 463 RVs spanning 22 years from the HARPS, HIRES, PFS, and UCLES instruments. We recovered all three known planets with parameters consistent with recently published values \citep[e.g.,][]{2023AJ....165..176L}.

\subsection{HD 141004 ($\lambda$ Ser)}
HD 141004 ($\lambda$ Ser; HIP 77257; HR 5868; GJ 598; TIC 296740796) is a G0-V star \citep{Keenan+1989ApJS} at a distance of $d = 11.92 \pm 0.02 \,{\rm pc}$ \citep{GaiaDR3_2022yCat}. \citet{2021ApJS..255....8R} reported a detection of a sub-Neptune planet ($M_{\rm p} \sin i = 13.6 \pm 1.5 \,{\rm M_\oplus}$) with an orbital period of 15.5 d. In this work, we analyzed 553 RVs spanning 32 years collected using the HIRES, Hamilton, and Levy spectrographs. We detected the planet at $P_{\rm b} = 15.509 \pm 0.002$ d with properties consistent with those reported in \citet{2021ApJS..255....8R}.

\subsection{HD 143761 ($\rho$ CrB)}
HD 143761 ($\rho$ CrB; HIP 78459; HR 5968; GJ 9537; TIC 458494003) is a G0+Va Fe-1 star \citep{Keenan+1989ApJS} at a distance of $d = 17.51 \pm 0.02 \,{\rm pc}$ \citep{GaiaDR3_2022yCat}. A giant planet was initially detected at a period of about 39 days from RV data obtained with the AFOE instrument at Whipple Observatory \citep{Noyes+1997ApJ}. A second lower-mass planet was later reported based on additional RV measurements from HIRES and APF \citep{Fulton+2016ApJ}. Recently, \citet{2023AJ....166...46B} obtained EPRV observations of this system using EXPRES/LDT. In addition to refining the masses and orbits of $\rho$~CrB~b and c, \citet{2023AJ....166...46B} reported two new planets in the system with orbital periods of $P_{\rm d} \approx 282.2 \,{\rm d}$ ($M_{\rm p,d} \sin i \approx 21.6 \, {\rm M_\oplus}$) and $P_{\rm e} \approx 12.949 \,{\rm d}$ ($M_{\rm p,d} \sin i \approx 3.79 \, {\rm M_\oplus}$). Our analysis of 552 RVs spanning 26 years recovered the planets discovered by \citet{Noyes+1997ApJ} and \citet{Fulton+2016ApJ} with periods of $P _{\rm b} = 39.8434 \pm 0.0004 \,{\rm d}$ and $P_{\rm c} = 102.58 \pm 0.04 \,{\rm d}$. We also detected a signal at $267.8 \pm 1.7 \,{\rm d}$ with $M_{\rm p,d} \sin i = 13.5 \pm 2.3\, {\rm M_\oplus}$, corresponding to $\rho$~CrB~d identified in \citet{2023AJ....166...46B}. The final residual periodogram of the RVs shows a fourth signal at 12.9 days, which corresponds to planet e \citep{2023AJ....166...46B}. However, likely due to the small amplitude of this planet's RV signal ($\approx 1$ m/s), the FAP of the 12.9-day signal was 0.27\%---too high to be considered a bona fide detection in this work (Table \ref{tab:undetected_planet_claims}).

\subsection{HD 160691 ($\mu$ Ara)}\label{sec:muAra}
HD 160691 (Cervantes; $\mu$ Ara; HIP 86796; HR 6585; GJ 691; TIC 362661163) is a G3IV-V star \citep{Gray+2006AJ} at a distance of $d = 15.60 \pm 0.02 \,{\rm pc}$ \citep{GaiaDR3_2022yCat}. HD~160691 is a closely-studied multi-planet system which was first discovered by the Anglo-Australian Planet Search using the UCLES spectrograph \citep{2001ApJ...555..410B, Jones+2002MNRAS}. Subsequent ground-based observing campaigns with instruments including HARPS and CORALIE confirmed four planets in the system \citep{Santos+2004AA, McCarthy+2004ApJ, 2007AA...462..769P, Gozdziewski+2007ApJ}, and more recently \textit{HST} Fine Guidance Sensor astrometry was used to further refine the masses and orbits of the planets \citep{Benedict+2022AJ}. In this work we analyzed 351 RVs collected over 21 years using the HARPS, PFS, and UCLES instruments. We detected all four of the known planets at periods of $P_{\rm b} = 644.93 \pm 0.27 \,{\rm d}$, $P_{\rm c} = 4241 \pm 145 \,{\rm d}$, $P_{\rm d} = 9.6392 \pm 0.0008 \,{\rm d}$, and $P_{\rm e} = 308.54 \pm 0.22 \,{\rm d}$, consistent with previously published results \citep{2007AA...462..769P}.

\subsection{HD 165499 ($\iota$ Pav)} \label{app:sec:hd165499}
HD 165499 ($\iota$ Pav; HIP 89042; HR 6761; GJ 9616; TIC 303704858) is a G0V star \citep{Gray+2006AJ} at a distance of $d = 17.75 \pm 0.04$~pc \citep{Gaia_DR2_2018AA}. HD~165499 is a binary star system, initially identified by \citet{2023AA...674A.114B}. Using a combination of CORALIE RVs and proper motion anomalies between HIPPARCOS \citep{perryman1997} and \textit{Gaia} epochs, they find that the binary has an orbital period of $P\approx 6.2$ years. The binary companion's orbit has a modeled semi-major axis of $8.6 \pm 0.3$ AU and mass of $141 \pm 10\, {\rm M_{J}}$. In this work, we used 65 CORALIE RVs spanning 22.5 years and clearly detected this binary companion. We reported a period of $8296 \pm 25 \,{\rm d}$ ($8.6 \pm 0.5$ AU) and minimum mass of $139.8 \pm 14.6\, {\rm M_{Jup}}$, consistent with \citet{2023AA...674A.114B}. We did not detect any significant stellar activity signals or additional planetary signals. However, this system is noteworthy due to its binarity; multiple star systems pose substantial challenges for starlight suppression for coronagraphic observations. The ExEP Mission Star List \citep[EMSL][]{Mamajek+Stapelfeldt_2024} excluded unresolved spectroscopic binaries and binaries with angular separations $<3^{\prime\prime}$), using the Washington Visual Double Star Catalog \citep[WDS;][]{wds} and the Ninth Catalogue of Spectroscopic Binary Orbits \citep{pourbaix+04}. Given the small physical separation of this binary (resulting in a maximum angular separation $<$1$^{\prime\prime}$), it should have been excluded from the EMSL. We caution that this system may not be suitable target for HWO unless future coronagraphic instruments can effectively mask starlight from a close-by off-axis companion.

\subsection{HD 189567}
HD 189567 (HIP 98959; HR 7644; GJ 776; TIC 352402781) is a G2V star \citep{Gray+2006AJ} at a distance of $d = 17.932 \pm 0.008 \,{\rm pc}$ \citep{GaiaDR3_2022yCat}. This star hosts two known planets at periods of 14.3 days and 33.7 days and minimum masses of 8.8\,\Mearth\ and 7.2\,\Mearth\ \citep{Mayor+2011arXiv, 2021AA...654A.104U}. We detected both planets with orbits and masses consistent with previous estimates based on 260 RVs from HARPS. We also found evidence for a long-term quadratic trend in the RVs. 

\subsection{HD 212330 A} \label{app:sec:hd212330}
HD 212330 A (HIP 110649 A; HR 8531 A; GJ 857 A; TIC 259291108) is a G2IV-V star \citep{Gray+2006AJ} at a distance of $d = 20.34 \pm 0.04$~pc \citep{GaiaDR3_2022yCat}. This star is in a binary system with an orbital period of $\sim$16,000 days ($\sim$44 years) and a true companion mass of $663 \pm 77\ {\rm M_{J}}$ \citep[$\approx$0.63 $\rm M_\odot$;][]{kane+19,2023AA...674A.114B}. In this work, we used 79 CORALIE RVs and detected the binary companion with orbital parameters consistent with those reported in \citet{2023AA...674A.114B}. The orbital parameters alongside the distance to this system result in a maximum angular separation less than $1^{\prime\prime}$. As was the case for HD 165499 (\secref{app:sec:hd165499}), this small angular separation is not consistent with the EMSL binary criteria set by \citep{Mamajek+Stapelfeldt_2024}, so the star should have been excluded from the list. Interestingly, this star is flagged as a binary on the EMSL (WDS designation 22249-5748), but the WDS angular separation and $\Delta$mag are given as 23.6$^{\prime\prime}$ and 7.68 mag, respectively, which are allowed by the EMSL criteria. It is possible that this star is a hierarchical multi-star system with a wide-separation component at 23.6$^{\prime\prime}$ and a much closer inner component. This may be consistent with the long-term quadratic trend observed in the CORALIE RVs, in addition to the $\sim$16,000-day orbit of the inner stellar companion. Nonetheless we caution that this system may not be suitable target for HWO depending on how well future coronagraphs are able to handle off-axis stellar companions.


\section{Non-detections in Systems with Previously Reported Planets}\label{appx:undetected_KPs}
Our RV search did not recover 19 of the 70 previously-reported planets in the sample. At least 9 of these non-detections were of planets that have been disputed by subsequent papers (e.g., \secref{sec:tauCeti}) or were not detailed in the original discovery paper (). In some cases, auxiliary observations from either imaging or astrometry were used in the original study, which are beyond the scope of this paper (). In other cases, the discovery data may not have been published (e.g., \secref{sec:HD33564}) or the statistical significance of a potential RV signal was not adequate to be classified as a detection in this work (\secref{sec:HD86728}). A summary of the undetected planets is given in Table \ref{tab:undetected_planet_claims}, and in the following subsections we briefly discuss the systems where we did not detect any planetary signals that were previously identified in the literature. Other non-detections in systems where at least one additional planet \textit{was} successfully detected are discussed in \secref{sec:results} or Appendix \ref{appx:signals_known_planets}.

\subsection{HD 10700 ($\tau$ Cet)}\label{sec:tauCeti}
HD 10700 ($\tau$ Cet; HIP 8102; HR 509; GJ 71; TIC 419015728) is a G8V star \citep{Keenan+1989ApJS} with a debris disk \citep[e.g.,][]{Greaves+2004MNRAS, MacGregor+2016ApJ, Sibthorpe+2018MNRAS} at a distance of $d = 3.652 \pm 0.002 \,{\rm pc}$ \citep{GaiaDR3_2022yCat}. Notably, $\tau$~Cet has the single highest number of RV observations in our sample ($N_{\rm RV} = 1692$), yielding the highest sensitivity to HZ planets of the Sun-like stars in the sample. Previously, \citet{Tuomi+2013AA} identified up to five possible planets in this system, of which two (e and f) were also later detected by \citet{2017AJ....154..135F}. The authors of the latter study used a ``differential radial velocity'' model that accounted for wavelength-dependent noise in the RV data to identify four planet candidates with $M_{\rm p} \sin i < 4 \,{\rm M_\oplus}$ (RV amplitudes down to $0.3 \,{\rm m\,s}^{-1}$) and orbital periods between 20 and 637 days (see Table \ref{tab:undetected_planet_claims}). However, the existence of any planets in this system has been called into question in recent work. \citet{korolik+23} used long-baseline interferometric data and high-precision spectroscopic observation to measure $\tau$ Ceti's stellar inclination, and found a near-polar orbit. This architecture would mean that the planets reported in \citet{2017AJ....154..135F} may not remain dynamically stable, calling into question whether there are planets orbiting this star. We detected a single Keplerian signal with our analysis, which we classified as a FA due to its small amplitude and proximity to the 1/3 annual harmonic. We did not identify any sub-threshold peaks in the RV residuals that correspond to any of the putative planets reported in \citet{2017AJ....154..135F}. We attribute our non-detection to the more sophisticated treatment of wavelength-dependent noise in previous studies and the disputed nature of these planets. 

\subsection{HD 33564} \label{sec:HD33564}
HD 33564 (HIP 25110; HR 1686; GJ 196; TIC 142103211) is an F7V star \citep{Gray+2003AJ} at a distance of $d = 20.79 \pm 0.03 \,{\rm pc}$ \citep{GaiaDR3_2022yCat}. A single massive ($M_{\rm p} \sin i = 9.1 \,{\rm M_{J}}$) planet, HD~33564~b, has been reported in this system with an orbital period of 388 days \citep{2005AA...444L..21G} using only 15 RV measurements from the ELODIE spectrograph spanning one orbital period. The ELODIE data analyzed by \citet{2005AA...444L..21G} are not publicly available apart from the points in their Figure 1. To the best of our knowledge, HD~33564 has not been observed by any of the instruments included in this analysis. The limited RV data analyzed in the discovery paper and lack of follow-up observations make HD 33564 b a dubious planet detection.

\subsection{HD 86728 A} \label{sec:HD86728}
HD 86728 A (20 LMi A; HIP 49081; HR 3951; GJ 376; TIC 172954294) is a G3V star at a distance of $d = 14.92 \pm 0.02 \,{\rm pc}$ \citep{bailer-jones+21}. A single planet on a circular orbit at $31.1503 \pm 0.0066$ d was detected using NEID in \citet{2025AJ....169....1G} and found to have a minimum mass of $M_{\rm p} \sin i = 9.16^{+0.56}_{-0.55} \,{\rm M_\oplus}$. We identified a sub-threshold peak at 31.2 days in 445 RVs from the HIRES, Hamilton, Levy, and NEID instruments. However, this peak did not pass the FAP=0.01 threshold, and is only statistically significant in the NEID RVs.

\subsection{HD 102365}
HD 102365 (HIP 57443; HR 4523; GJ 442 A; TIC 454082369) is a G2V star \citep{Gray+2006AJ} at a distance of $d = 9.319 \pm 0.008 \,{\rm pc}$ \citep{GaiaDR3_2022yCat}. The star has one known planet, HD~102365~b, which was discovered using 12 years of UCLES data \citep{2011ApJ...727..103T}. The mass ($M_{\rm p} \sin i = 16.0 \pm 2.6 \,{\rm M_\oplus}$) and orbit ($P = 122.1 \pm 0.3 \,{\rm d}$) of this planet were later confirmed and refined using additional RVs from HIRES, HARPS, and PFS \citep{2023AJ....165..176L}. Strangely, we did not detect HD~102365~b in this work despite analyzing 373 RVs from HARPS, HIRES, CES-LC, CES-VLC, PFS and UCLES. In the instrument-by-instrument GLS periodogram of the RVs, a significant $\sim$120 signal is clearly present. However, this peak only appears in the UCLES RV periodogram. This suggests that the $122$-day signal may be an instrumental effect associated with UCLES. It may also be that the lower-precision measurements from CES-LC and CES-VLC effectively dilute the UCLES signal to the point where it disappears from the periodogram. We recommend additional follow-up analysis of this signal to determine whether it is a bona fide planet or instrumental systematic.

\subsection{HD 115404 A}
HD 115404 A (HIP 64797 A; GJ 505 A; TIC 373765355) is a K2.5V (k) star \citep{Gray+2003AJ} at a distance of $d = 10.987 \pm 0.003 \,{\rm pc}$ \citep{GaiaDR3_2022yCat}. This star was recently suggested to host two planet candidates from a combined analysis of HIRES RVs and Hipparcos-Gaia astrometry \citep{2022ApJS..262...21F}. However, the original study reporting these planets does not detail the analysis of the RVs and astrometry for this system specifically. We were unable to detect either planet with only 95 RVs from ELODIE, HIRES, and SOPHIE. While the astrometry data may be necessary to recover these putative planets, we treat them as disputed planets in this work because the discovery paper does not discuss this system specifically and there have been no subsequent follow-up studies of the system to confirm the planets.

\subsection{HD 140901 A}
HD 140901 A (HIP 77358 A; HR 5864 A; GJ 599 A; TIC 179348425) is a G7IV-V star \citep{Gray+2006AJ} at a distance of $d = 15.246 \pm 0.008 \,{\rm pc}$ \citep{GaiaDR3_2022yCat}. Using both RV observations and astrometric data from Gaia and Hipparcos, \citet{2022ApJS..262...21F} reported two planet candidates with orbital periods of about 9.0 days and 7417.5 days ($\approx$20.3\,yr). The long-period planet was later confirmed by \citet{2023AA...678A.107P}, although the presence of the inner planet at 9 days remains controversial (and is flagged as such on the NASA Exoplanet Archive). We did not recover either of these signals in our analysis of 172 HARPS, PFS, and UCLES RVs spanning 21.9 years. We attribute these non-detections to the controversial nature of the inner planet, and the use of astrometric data in \citet{2023AA...678A.107P} to identify the longer-period planet.

\subsection{HD 147513}\label{sec:hd147513}
HD 147513 (HIP 80337; HR 6094; GJ 9559; TIC 350673608) is a G1V CH-0.4 star \citep{Gray+2006AJ} at a distance of $d = 12.89 \pm 0.01 \,{\rm pc}$ \citep{GaiaDR3_2022yCat}. This star hosts one known giant planet, which was initially detected using 30 RV measurements from CORALIE spanning about 4.6 years, but has not since been confirmed by follow-up observations. It was reported to have an orbital period of $P = 528.4 \rm 6.3 \,{\rm d}$ and a minimum mass of $M_{\rm p} = 1.21 \,{\rm M_{Jup}}$ \citep{2004AA...415..391M}. Interestingly, adding 46 nights of more recent HARPS RVs to the analysis in this study causes the planet's signal to vanish. With the combined RVs, we only detected a sub-threshold peak in the RV periodogram at 504.1 days, which did not meet the FAP=0.01 requirement. We suggest further analysis of the CORALIE and HARPS RVs of this system to determine whether the HD 147513 b is a real planet.

\subsection{HD 206860 (HN Peg)}
HD 206860 (HN Peg; HIP 107350; HR 8314; GJ 9751; TIC 301880196) is a G0IV-V star \citep{Gray+2003AJ} with a debris disk \citep{Trilling+2008ApJ, Ertel+2012AA} at a distance of $d = 18.13 \pm 0.01 \,{\rm pc}$ \citep{GaiaDR3_2022yCat}. This target has a directly-imaged T dwarf companion, HD~206860~b, at a physical separation of $795 \pm 15 \,{\rm AU}$, which was discovered with the \textit{Spitzer Space Telescope} \citep{2007ApJ...654..570L}. In this work, we analyzed 89 RVs from the Hamilton and Levy instruments spanning about 31 years. However, given the extremely long orbital period of HD~206860~b, the RV baseline was simply much too short to probe this directly-imaged planet's parameter space.

\subsection{HD 209100 ($\epsilon$ Ind)}
HD 209100 ($\epsilon$ Ind; HIP 108870; HR 8387; GJ 845 A; TIC 231698181) is a K4V(k) star \citep{Gray+2006AJ} at a distance of $d = 3.638 \pm 0.001 \,{\rm pc}$ \citep{GaiaDR3_2022yCat}. A combined analysis of Gaia-Hipparcos astrometry and archival RVs led to the discovery of a Jupiter-analog planet ($\epsilon$~Ind~b) in this system with a mass of roughly 3\,$\rm M_{Jup}$ on a slightly eccentric orbit with a period of about 45 years \citep{2019MNRAS.490.5002F}. This planet was recently imaged with the \textit{James Webb Space Telescope} using the MIRI coronagraphic imager \citep{2024Natur.633..789M}, which suggested that the planet has a high-metallicity, high carbon-to-oxygen atmosphere. We were unable to detect $\epsilon$~Ind~b here using only the 266 RV measurements from HARPS, CES-LC, CES-VLC, and UVES that span 29 years. This was due to the lack of additional information provided by imaging or astrometric observations in our analysis. However, we were able to detect a long-term quadratic trend in the RVs that is likely caused by the planet.


\section{Classifications of Other Signals Detected}\label{appx:other_signals}

Here we comment on systems with other RV signals detected by \rvsearch\ which have not yet been discussed in \secref{sec:results} or Appendices \ref{appx:signals_known_planets}-\ref{appx:undetected_KPs}. These systems have no previously-detected planets, so the signals discussed here are either related to stellar activity or instrumental false alarms.

\subsection{HD 1581 ($\zeta$ Tuc)}
HD 1581 ($\zeta$ Tuc; HIP 1599; HR 77; GJ 17; TIC 425935521) is an F9.5V \citep{Gray+2006AJ} star at a distance of $d = 8.61 \pm 0.01 \,{\rm pc}$ \citep{GaiaDR3_2022yCat}. This star has no known exoplanets, although an observed infrared excess suggests the presence of a circumstellar debris disk \citep{Trilling+2008ApJ}. We analyzed 550 RV observations of this target from HARPS, CES-LC, CES-VLC, and UCLES spanning a total of 29 years. We found two significant signals at $P_{.01} = 937.0 \pm 16.5 \,{\rm d}$ ($\sim$2.6\,yr) and $P_{.02} = 15.654 \pm 0.006 \,{\rm d}$. \citet{2011arXiv1107.5325L} predicted a stellar rotation rate for this target of $P_\text{rot} = 16.7 \pm 2.6 \,{\rm d}$, which matches the 15.65-day RV signal we detected. We have therefore designated signal HD~1581.01 as SA-R (indicating rotation-modulated stellar activity). \citet{2011arXiv1107.5325L} also reported a stellar activity cycle for HD~1581 with a period of $P_\text{cycle} = 1018_{-47}^{+51} \,{\rm d}$, based on 127 $\log R^\prime_{\rm HK}$ measurements from HARPS. This activity period is consistent with our second RV detection (HD~1581.01) within $2\sigma$; therefore, we categorized this signal as being due to stellar activity. This interpretation is supported by a peak in the FWHM GLS periodogram at $\approx 1000$ days. 

\subsection{HD 4614 A ($\eta$ Cas A)}
HD 4614 A (Achird; $\eta$ Cas A; HIP 3821 A; HR 219 A; GJ 34 A; TIC 445258206) is a F9V star \citep{Keenan+1989ApJS} at a distance of $d = 5.010 \pm 0.003$~pc \citep{GaiaDR3_2022yCat}. This target has no known planets, and we identified one RV signal at $91.02 \,{\rm d}$ using 273 observations from HIRES, SOPHIE, Hamilton, and Levy spanning 24 years. This signal was previously classified as an annual and/or instrumental systematic by \citet{2021ApJS..255....8R}. We classified this signal as a False Alarm due to the period being similar to the 1/4 annual harmonic and dubious RV phase curve.

\subsection{HD 14412}
HD 14412 (HIP 10798; HR 683; GJ 95; TIC 72748794) is a G8V star \citep{Gray+2006AJ} at a distance of $d = 12.835 \pm 0.005$~pc \citep{GaiaDR3_2022yCat}. There are no previously known planets orbiting this star. In this work, we used 400 RVs collected by HARPS, HIRES, UCLES, PFS, and Levy across 23.1 years. We identified one RV signal at $361.7 \,{\rm d}$, which we classified as a False Alarm due to its proximity to 1 year, as demonstrated by the strong peak near one year in the window function GLS periodogram for this target.

\subsection{HD 20807 ($\zeta$2 Ret)}
HD 20807 ($\zeta$2 Ret; HIP 15371; HR 1010; GJ 138; TIC 279649049) is a G1V star in a binary star system \citep{Keenan+1989ApJS} at a distance of $d = 12.039 \pm 0.009$~pc \citep{GaiaDR3_2022yCat}. There are no previously known planets in this system. In this work, we analyzed 323 RVs from CES-LC, CES-VLC, HARPS, UCLES, and PFS and identified one RV signal at $2943 \pm 130\,{\rm d}$. We observe moderate correlations between the RVs and the stellar activity data, and a peak in the $R_{\rm HK}^\prime$ periodogram consistent with the signal. This suggests that the RV signal is due to underlying stellar activity. Using HARPS spectra, \citet{2011arXiv1107.5325L} characterized the magnetic cycle of HD 20807 and found a period of $1133 \pm 700 \,{\rm d}$. Additional work by \citet{flores2021} identified a $2899 \pm 139\,{\rm d}$ ($\approx 7.9$ year) chromospheric activity cycle for this star using 430 spectra. We therefore classified the 2943-day signal found in our analysis as being due to stellar activity.

\subsection{HD 23249 ($\delta$ Eri)}
HD 23249 (Rana; $\delta$ Eri; HIP 17378; HR 1136; GJ 150; TIC 38511251) is a K0+IV star \citep{Keenan+1989ApJS} at a distance of $d = 9.09 \pm 0.02$~pc \citep{GaiaDR3_2022yCat}. This target has no previously known planets. In this work we used 512 RVs collected across 27 years using the CES-LC, CES-VLC, HARPS, HIRES, SOPHIE, UCLES, and Levy instruments. Using \rvsearch, we identified one RV signal at $595.3 \pm 4.2 \,{\rm d}$. Recently, \citet{2023AJ....165..176L} found a nearly identical signal using a subset of the data analyzed here, which they categorized as a SRC. We do not find evidence of significant stellar activity correlations or signals in the GLS periodograms, so we do not associate the signal with stellar activity. Due to the highly-eccentric dubious phase curve of the RVs and the lack of a well-defined peak in the RV periodogram, we deemed this signal unlikely to be physical and classified it as a False Alarm.

\subsection{HD 26965 A (40 Eri A)}
HD 26965 A (Keid, 40 Eri A; HIP 19849 A; HR 1325 A; GJ 166 A; TIC 67772871) is a K0.5V standard star \citep{Gray+2006AJ} at a distance of $d = 5.010 \pm 0.003 \,{\rm pc}$ \citep{GaiaDR3_2022yCat}. An RV signal near 43 days was first detected using data from HARPS, HIRES, PFS, and the CHIRON \citep{Diaz+2018AJ}, which the authors were unable to classify as either planetary or stellar activity due to the similarity between the star's rotation period and the period of the RV signal. A subsequent analysis by \citet{Ma+2018MNRAS} later suggested that the 43-day signal was indeed due to a planet after analyzing additional RV data. However, subsequent deep dives into the nature of this signal have revealed that is indeed most likely due to stellar activity and not a planet. Recently, it was shown that a combination of spots and convective blueshift suppression likely created this signal \citep{2024AJ....167..243B, Siegel+2024AJ....168..158S}. Here, we detected the RV signal at $P = 42.322 \pm 0.015\,{\rm d}$ using 666 RVs from the HARPS, HIRES, PFS, UCLES, Hamilton, and Levy spectrographs. Following the previous studies, we classified this signal as SA-R in Table \ref{tab:rvsearch_signals}. We also identified a signal with a period of $P = 37.326 \pm 0.021\,{\rm d}$, which has been attributed to differential stellar rotation. We classified this signal as SA-R. Lastly, we note the detection of a known magnetic activity cycle at $P\approx3000$ d in the residual RV periodogram (as a non-significant peak), as well as the $S$-index and $R_{\rm HK}^\prime$ GLS periodograms \citep{2011arXiv1107.5325L, 2019ApJ...886..132G, 2024ApJS..274...35I}.

\subsection{HD 32147}
HD 32147 (HIP 23311; HR 1614; GJ 183; TIC 213041474) is a K3+V star \citep{Gray+2003AJ} at a distance of $d = 8.844 \pm 0.002 \,{\rm pc}$ \citep{GaiaDR3_2022yCat} with no known planets. Our analysis of 352 RVs from HIRES, HARPS, PFS, Hamilton, and Levy spanning 33.3 years revealed a single Keplerian signal at $P_{.01} = 3127 \pm 120 \,{\rm d}$ ($\sim$8.6\,yr), which also appears as a prominent peak in the RV GLS periodogram. \citet{2019ApJ...886..132G} measured two activity cycles for this star with periods of $P_\text{cyc} = 9.33 \,{\rm yr}$ (3407.8\,d) and $P_\text{cyc} = 12.42 \,{\rm yr}$ (4536.4\,d). The former activity cycle period is consistent with our RV detection (HD~32147.01) to within $2\sigma$, so we classified it as stellar activity. This activity signal was also detected by \citet{2021ApJS..255....8R} at $P = 3444.0_{-81.0}^{+91.0} \,{\rm d}$ and by \citet{2024ApJS..274...35I} at a period of 9.53\,yr (3,480.8\,d).

\subsection{HD 38858}
HD 38858 (HIP 27435; HR 2007; GJ 1085; TIC 176521059) is a G2V star \citep{Gray+2003AJ} at a distance of $d = 15.210 \pm 0.007$~pc \citep{GaiaDR3_2022yCat}. There are no known planets orbiting this star. Previous work identified a stellar magnetic activity cycle with a period of $3460^{+2151}_{-570} \,{\rm d}$ \citep{2011arXiv1107.5325L}, which was also reported in \citet{2023AJ....165..176L} at $2893 \pm 150  \,{\rm d}$ and in \citet{2021ApJS..255....8R} at $3113.0 \pm 80  \,{\rm d}$. In this work, we used 223 RVs from HARPS, HIRES, and Levy and identified one RV signal at $417.4 \pm 3.8  \,{\rm d}$. This corresponds to the yearly alias of the previously-measured stellar magnetic activity cycle, and we therefore classified this signal as stellar activity. 

\subsection{HD 46588 A}
HD 46588 A (HIP 32439 A; HR 2401 A; GJ 9215 A; TIC 141523112) is an F8V star \citep{Gray+2003AJ} at a distance of $d = 18.2 \pm 0.02$~pc \citep{GaiaDR3_2022yCat}. This star has a common proper motion companion at the L/T transition with a projected physical separation of 1420 AU \citep{loutrel+11}. In this work we used 347 RVs from CARMENES and SONG spanning 4.1 years, and identified four RV signals using \rvsearch. These signals have orbital periods of $121.81 \pm 0.54\,{\rm d}$ (HD 46588.01), $181 \pm 1\,{\rm d}$ (HD 46588.02), $410 \pm 4\,{\rm d}$ (HD 46588.03), and $889^{+2000}_{-240}\,{\rm d}$ (HD 46588.04). Previous work searching for planets orbiting stars with wide brown dwarf companions identified three Keplerian signals in the RV data of HD 46588 A with periods of 127, 224, and 365 days \citep{2023AA...671A..10S}. Due to the strong correlation between the chromatic index and CARMENES RVs, \citet{2023AA...671A..10S} classified the 127-day peak as stellar activity due to the star's magnetic cycle, which was strengthened by the SONG H$\alpha$ data. They also attributed the 1-year peak to the SONG window function. Following \citet{2023AA...671A..10S}, we classified HD 46588.01 as stellar activity, but we did not see clear evidence of the 224-day peak reported in that study. Rather, the other 3 signals we detected are likely false alarms (FA) due to corresponding periodicity in the observational window function. The window function periodogram for this star is quite complex with multiple peaks. The signals we detected at 181, 410, and 889 days all correspond to significant peaks in the window function periodogram or the annual alias of a significant peak.

\subsection{HD 48682 (56 Aur)}
HD 48682 (56 Aur; HIP 32480; HR 2483; GJ 245; TIC 307754027) is an F9V star \citep{Gray+2003AJ} at a distance of $d = 16.61 \pm 0.02$~pc \citep{GaiaDR3_2022yCat}. There are no known planets in this system, and it may host an extended, cold debris disk \citep{hengst+20}. In this work, we analyzed 361 RVs collected using HIRES, Hamilton, and Levy spanning 32.5 years. We identified one significant signal in the RVs at $929.2 \pm 4.9 \,{\rm d}$, which was previously identified as stellar activity by \citet{2021ApJS..255....8R}. We note that RV phase plot for this signal is not well described by the model and the peak in the periodogram is not well defined. Following the \citet{2021ApJS..255....8R}, we classified this signal as SA.

\subsection{HD 76151}
HD 76151 (HIP 43726; HR 3538; GJ 327; TIC 62569281) is a G2V star \citep{Keenan+1989ApJS} at a distance of $d = 16.85 \pm 0.01$~pc \citep{GaiaDR3_2022yCat}. There are no known planets in this system, but it may have a circumstellar disk as indicated by infrared excess \citep{eiroa+13}. In this work, we used 150 HARPS, HIRES, Hamilton, and Levy RVs spanning 32.3 years to identify one Keplerian RV signal at $1850 \pm 24 \,{\rm d}$. This system was recently analyzed by \citet{2023AJ....165..176L}, but the authors did not report any significant RV signals. However, \citet{1995ApJ...438..269B} identified a stellar activity cycle period of $2.52 \pm 0.02$ years ($\sim 920\,{\rm d}$). The signal we identified at 1850 days is very close to the second harmonic of this known activity period. Additionally, the periodogram peak we observe is not well defined, and we measure significant correlations between the $H$-index and BIS activity indicators and the RVs. We therefore classified this signal as SA.

\subsection{HD 102870 ($\beta$ Vir)}
HD 102870 (Zavijava; $\beta$ Vir; HIP 57757; HR 4540; GJ 449; TIC 366661076) is an F9V star \citep{Morgan+1973ARA&A..11...29M} at a distance of $d = 10.93 \pm 0.03$~pc \citep{vanLeeuwen2007AA}. There are no known planets orbiting this target. In this work, we used 346 RVs collected using the HARPS, SOPHIE, Levy, Hamilton, and Tull instruments, and identified one Keplerian signal at $2167 \pm 43 \,{\rm d}$. We note that \citet{Seach+2022MNRAS.509.5117S} used Zeeman Doppler Imaging to measure the surface magnetic field of this star, and measured a rotation period of $\approx 9.1 \,{\rm d}$, which is inconsistent with the longer period signal we found with \rvsearch. We found a significant correlation between the Tull RVs and $S$-Index measurements ($r=0.29$, $p=0.0014$), which may suggest that the signal is related to stellar activity. However, there are no significant peaks in the activity GLS periodograms, except for a strong peak in the window function at $\approx 2000$ days. We interpret this window function periodicity as evidence that the 2167-day RV signal is due to an observational bias. Therefore, we classified this signal as a False Alarm.

\subsection{HD 114613}
HD 114613 (HIP 64408; HR 4979; GJ 9432; TIC 30293512) is a G4IV star \citep{Gray+2006AJ} at a distance of $d = 20.46 \pm 0.04 \,{\rm pc}$ \citep{GaiaDR3_2022yCat}. The star was observed as part of the Anglo-Australian Planet Search, and was initially found to host a planet candidate on a $\sim$10.5-year orbit based on RV measurements from UCLES \citep{Wittenmyer+2014ApJ}. Although the amplitude and period of the putative planet were refined by \citet{Luhn+2019AJ}, additional measurements from Keck later showed an activity correlation which subsequently led \citet{2020AJ....159..235L} to attribute the signal to activity-induced RV jitter (the current disposition of HD~114613~b on the NASA Exoplanet Archive at the time of writing is ``False Positive Planet''). 

In our analysis of 534 HIRES, HARPS, CES-LC, CES-VLC, UCLES, and PFS RVs, we detected an RV signal at $P=3851 \pm 34$ d, which is consistent with the signal (now known to be caused by stellar activity) first reported by \citet{Wittenmyer+2014ApJ}. We note that the recent study by \citet{2023AJ....165..176L} did not report this signal in their analysis of RVs from HARPS, HIRES, and PFS, but instead found signals at 6622 days, 73 days, and 1954 days (reported as stellar activity or SRCs). We did not detect any of these signals in this work, but did detect an additional signal at 4814 $\pm$ 260 d. Because of significant correlations between the RVs and activity data, and significant power at long periods ($\gtrsim2000$ d) in the periodograms for $H$-index, $R_{\rm HK}^\prime$, BIS, and FWHM, we determined the nature of this signal also to be stellar activity. The multitude of signals in previous studies of the RVs and activity indicator data for this system are likely the result of increased stellar jitter due to the evolved nature of HD 114613 \citep{2020AJ....159..235L}.

\subsection{HD 128620/128621 ($\alpha$ Cen A/B)}
HD 128620 (Rigil Kentaurus; $\alpha$ Cen A; HIP 71683; HR 5459; GJ 559 A; TIC 471011145) and HD 128621 (Toliman; $\alpha$ Cen B; HIP 71681; HR 5460; GJ 559 B; TIC 471011144) are a well-known binary system consisting of a G2V star \citep[$\alpha$ Cen A;][]{Gray+2006AJ} and a K1V star \citep[$\alpha$ Cen B;][]{Torres+2006AA...460..695T} at a distance of $d = 1.3319 \pm 0.0007$~pc \citep{Akeson+2021AJ....162...14A}. As one of the closest and brightest multi-star systems, this system has been studied extensively in the literature \citep[e.g.,][and references therein]{2018AJ....155...24Z}. There are no confirmed planets orbiting either star, although recent \textit{JWST}/MIRI imaging observations have suggested there may be a giant planet in the HZ of $\alpha$ Cen A \citep{aca1, aca2}, and past RV work has suggested an Earth-mass planet orbiting $\alpha$ Cen B \citep{acb1}. The $\alpha$ Cen B planet was later ruled out as a planet and attributed to a window function systematic and stellar activity \citep{acb2}. 

Here, we analyzed 278 RVs of $\alpha$ Cen A collected with CES-LC, CHIRON, and CTIO-ES and 632 RVs of $\alpha$ Cen B collected with CES-LC, CHIRON, CTIO-ES, and HARPS. We observed significant long-term trends in the RVs consistent with the stars' orbits around their common barycenter. We also identified two Keplerian signals in the $\alpha$ Cen B data at $P=333.9 \pm 1.3\,{\rm d}$ and $P=62.34 \pm 0.3\,{\rm d}$. Both signals are dubious given the multiple clustered peaks in both $\Delta$BIS periodograms and the non-monotonically increasing running-LS periodograms. The first signal is also consistent with an annual systematic, seen as a series of peaks in the window function GLS periodogram. We classified both signals as False Alarms. We note that the putative giant planet in $\alpha$ Cen A \citep{aca1, aca2} has a suggested mass below the RV sensitivity limit, which explains why we did not detect a signal associated with this planet candidate. For $\alpha$ Cen B, we identified one significant long-period signal (LPS) at $P\approx 12600$ days after modeling out the long-term RV trend associated with the stellar binary. This signal is consistent with significant power in the BIS, FWHM, and window function GLS periodograms, but given that the limited RV data cannot sample the full orbit, this signal is likely a systematic. 

\subsection{HD 140538 A ($\psi$ Ser A)}
HD 140538 A ($\psi$ Ser A; HIP 77052 A; HR 5853 A; GJ 9527 A; TIC 459427073) is a solar-twin G5V star \citep{Gray+2001AJ} with no known planets at a distance of $d = 14.793 \pm 0.008 \,{\rm pc}$ \citep{GaiaDR3_2022yCat}. We analyzed 160 RVs from HIRES, HARPS, and Levy spanning a total of 17.6 years. We found one significant RV signal at $P = 1449 \pm 17 \,{\rm d}$ ($\sim$4.0\,yr), which matches a very strong peak in the $S$-index GLS periodogram. This is a well-studied signal that has been attributed to a stellar magnetic activity cycle by multiple past studies \citep[e.g.,][]{Mittag+2019AA, 2021ApJS..255....8R, 2024ApJS..274...35I}. We therefore classified the signal as SA.

\subsection{HD 146233 (18 Sco)}
HD 146233 (18 Sco; HIP 79672; HR 6060; GJ 616; TIC 135656809) is a solar-twin G2Va standard star \citep{Keenan+1989ApJS} at $d = 14.14 \pm 0.01 \,{\rm pc}$ \citep{GaiaDR3_2022yCat} with no known exoplanets. Our analysis used 440 RV data points taken over a 23.5-year span from HIRES, HARPS, Levy, PFS, and UCLES. We detected one RV signal at $P = 2435 \pm 31 \,{\rm d}$ ($\sim$6.7\,yr), which is also present in the $R_{\rm HK}'$, BIS, and window function periodograms below the significance threshold. This signal has been identified previously as stellar activity, with literature values ranging from 6.27\,yr \citep[$\sim$2290\,d;][]{2024ApJS..274...35I} to $2803_{-392}^{+2663} \,{\rm d}$ \citep[$\sim$7.7\,yr][]{2011arXiv1107.5325L}. The activity period we measured is most consistent with the value reported by \citet{2021ApJS..255....8R} at $P = 2426_{-42}^{+60} \,{\rm d}$. Following these previous studies, we classified the signal as SA.

\subsection{HD 158633}
HD 158633 (HIP 85235; HR 6518; GJ 675; TIC 219880402) is a K0V star \citep{Cowley+1967AJ.....72.1334C} at a distance of $d = 12.792 \pm 0.004$~pc \citep{GaiaDR3_2022yCat}. This star hosts no known planets. We used 361 HIRES, SOPHIE, Hamilton, and Levy RVs to identify one significant RV signal at $366.5 \pm 1.6 \,{\rm d}$. This is consistent with the sidereal year and a significant peak in the window function periodogram. As in previous studies of this system \citep[e.g.,][]{2021ApJS..255....8R}, we classified this signal as a False Alarm.

\subsection{HD 185144 ($\sigma$ Dra)}
HD 185144 (Alsafi; $\sigma$ Dra; HIP 96100; HR 7462; GJ 764; TIC 259237827) is a K0V star \citep{Keenan+1989ApJS} with no known planets at a distance of $d = 5.764 \pm 0.002 \,{\rm pc}$ \citep{GaiaDR3_2022yCat}. Our analysis of this star included 728 RVs from HIRES, Hamilton, and Levy observed over a span of 24 years. We identified three significant RV signals at $P_{.01} = 2338 \pm 34 \,{\rm d}$ ($\sim$6.4\,yr), $P_{.02} = 363.34 \pm 0.69 \,{\rm d}$, and $P_{.03} = 121.55 \pm 0.22 \,{\rm d}$. The first signal appears as a very strong peak in the $S$-index GLS periodogram and we observed a strong correlation between the HIRES $S$-index data and the RVs. Following past studies of this signal that are consistent with our results \citep{2021ApJS..255....8R, 2024ApJS..274...35I}, we classified it as a stellar magnetic activity cycle. The second signal sits very close to the annual harmonic and is consistent with the peak in the window function periodogram. We therefore classified it as a False Alarm (\citet{2021ApJS..255....8R} detected a similar signal with $P = 347.11_{-0.82}^{+0.92} \,{\rm d}$, which they marked as ``annual and/or instrumental systematic'' ). The period of the final signal is almost exactly at the 1/3 annual harmonic and also has a corresponding significant peak in the window function periodogram. We classified this as a False Alarm. 

\subsection{HD 201091/201092 (61 Cyg A/B)}
HD 201091 (61 Cyg A; HIP 104214; HR 8085; GJ 820 A; TIC 165602000) and HD 201092 (61 Cyg B; HIP 104217; HR 8086; GJ 820 B; TIC 165602023) are a well-known binary system consisting of a K5V star \citep[61 Cyg A;][]{Keenan+1989ApJS} and a K7V star \citep[61 Cyg B;][]{Keenan+1989ApJS} at distances of $d_{\rm A} = 3.4966 \pm 0.0007$~pc and $d_{\rm B} = 3.4964 \pm 0.0004$~pc \citep{GaiaDR3_2022yCat}. Neither star hosts any known planets. In the RVs, we see opposite long-term trends for these stars associated with their mutual orbital motions. For 61 Cyg A, we used 300 RVs collected using HIRES, Hamilton, and Levy to identify one significant RV signal at $6918 \pm 40\,{\rm d}$, which we classified as systematic False Alarm due to a corresponding strong peak in the window function at this period. We note that past analysis of this star also detected an stellar activity cycle with a period of $\sim$2600 days \citep{2021ApJS..255....8R, 2024ApJS..274...35I}, which we see weak evidence for in the GLS periodograms for $S$-index and $H$-index. For 61 Cyg B, we analyzed 296 RVs collected using HIRES, Hamilton, and Levy and identified one significant RV signal at $49.012 \pm 0.029 \,{\rm d}$. Past work has attributed this signal to stellar rotation \citep{2021ApJS..255....8R}. 

\subsection{HD 207129}
HD 207129 (HIP 107649; HR 8323; GJ 838; TIC 147407292) is a G0V Fe+0.4 star \citep{Gray+2006AJ} with an imaged debris disk \citep{Krist+2010AJ, Lohne+2012AA} at a distance of $d = 15.56 \pm 0.01 \,{\rm pc}$ \citep{GaiaDR3_2022yCat}. The star hosts no known planets. We analyzed 179 RVs from HARPS, CES-LC, and CES-VLC spanning almost 27 years. We found one RV signal at $P = 1393 \pm 11 \,{\rm d}$ ($\sim$3.8\,yr), which is consistent to within $1\sigma$ of the literature value for this star's activity cycle period \citep[$P = 1520_{-139}^{+171} \,{\rm d}$;][]{2011arXiv1107.5325L}. Additionally, \citet{2023AJ....165..176L} analyzed this system using RV data from HARPS and UCLES, and found an RV signal at $P = 1964 \pm 49 \,{\rm d}$ and an $S$-index signal at $P\approx1898\,{\rm d}$), which they attributed to a magnetic activity cycle.

\end{document}